\documentclass[11pt,a4paper]{article}

\usepackage[a4paper,text={160mm,247mm},centering]{geometry}

\usepackage{setspace}
\usepackage[utf8]{inputenc}

\usepackage[english]{babel}
\usepackage{hyphenat}
\usepackage[nosort]{cite}

\newcommand{\rcite}[1]{ref.~\cite{#1}}
\newcommand{\rcites}[1]{refs.~\cite{#1}}

\usepackage{amsmath, amssymb, amsfonts, amsthm}
\usepackage{mathtools}
\usepackage[normalem]{ulem}

\theoremstyle{plain}

\newtheorem*{theorem*}{Theorem}

\usepackage{mathfixs}
\ProvideMathFix{autobold}
\ProvideMathFix{multskip}
\ProvideMathFix{frac,root,rootclose}

\allowdisplaybreaks[1]

\makeatletter
\def\mr@ignsp#1 {\ifx\:#1\@empty\else #1\expandafter\mr@ignsp\fi}%
\newcommand{\multiref}[1]{\begingroup
\xdef\mr@no@sparg{\expandafter\mr@ignsp#1 \: }%
\def\mr@comma{}%
\@for\mr@refs:=\mr@no@sparg\do{\mr@comma\def\mr@comma{,}\ref{\mr@refs}}%
\endgroup}
\renewcommand{\eqref}[1]{(\multiref{#1})}
\makeatother

\makeatletter
\newcommand{\namedref}[2]{\hyperref[#2]{#1~\ref*{#2}}}%
\newcommand{\namedreff}[2]{\hyperref[#2]{#1\,\ref*{#2}}}%
\newcommand{\secref}{\namedreff{\S}}
\newcommand{\appref}{\namedref{Appendix}}
\newcommand{\tabref}{\namedref{Table}}
\newcommand{\figref}{\namedref{Figure}}
\makeatother

\renewcommand{\theequation}{\thesection.\arabic{equation}}
\numberwithin{equation}{section}

\newcommand{\eqn}[1]{eq.~\eqref{#1}}
\newcommand{\Eqn}[1]{Equation~\eqref{#1}}
\newcommand{\eqns}[2]{eqs.~\eqref{#1} and~\eqref{#2}}
\newcommand{\Eqns}[2]{Eqs.~\eqref{#1} and~\eqref{#2}}

\providecommand{\href}[2]{#2}

\usepackage[font=small,labelfont=bf]{caption}

\usepackage{graphbox}

\usepackage[compile=false,fonts]{mpostinl}
\begin{mpostdef}
	input metaobj;
	
	def drawzigzag(expr pzz)=
	pair zz[];
	zz[1]:= point 0 of pzz ; zz[2]:= point 1 of pzz;
	nczigzag(zz[1])(zz[2]) "coilwidth(0.15xu)", "linewidth(1pt)", "linearc(.01xu)", "arrows(-)", "coilarmA(0)", "coilarmB(0)";
	enddef;
\end{mpostdef}

\begin{mpostdef}
	pair vpos[];
	pair epos[];
	pair ext[];
	pair exta[];
	path paths[];
	picture pic;
	picture pic[];
	picture savepic;
	path cpath;
	path hpath;
	xu:=1cm;
	yu:=1cm;
\end{mpostdef}

\begin{mpostdef}
	def pensize(expr s)=withpen pencircle scaled s enddef;
	def fillshape(expr p,ci,tb,cb)=
	fill p withcolor ci;
	draw p pensize(tb) withcolor cb;
	enddef;
	def filldot(expr z,s,ci)=
	fillshape(fullcircle scaled s shifted z, ci, 0.5pt, 0.0white);
	enddef;
	def filltrig(expr z,s,r,ci)=
	fillshape(((dir 60)--(dir 180)--(dir 300)--cycle) scaled s rotated r shifted z, ci, 0.5pt, 0.0white);
	enddef;
	def fillsqr(expr z,s,r,ci)=
	fillshape(((dir 45)--(dir 135)--(dir 225)--(dir 315)--cycle) scaled s rotated r shifted z, ci, 0.5pt, 0.0white);
	enddef;
	def drawcross(expr z,s,r,t,c)=
	draw ((-0.5,-0.5)--(+0.5,+0.5)) scaled s rotated r shifted z pensize(t) withcolor c;
	draw ((+0.5,-0.5)--(-0.5,+0.5)) scaled s rotated r shifted z pensize(t) withcolor c;
	enddef;
	
	def midarrow (expr p, t) =
	fill arrowhead subpath(0,arctime(arclength(subpath (0,t) of p)+0.5ahlength) of p) of p;
	enddef;
	
	def dprop expr p=draw p pensize(3pt) withcolor feyncolor enddef;
	def dpropf expr p=draw p pensize(3pt) withcolor feyncolorfaint enddef;
	
	def drawprop expr p=draw p pensize(1.5pt) enddef;
	def drawline expr p=draw p pensize(0.5pt) enddef;
	def drawpos expr p=drawcross(p, 5pt, 0, 1.0pt, 0.0white) enddef;
	def drawvertex expr p=filldot(p, 5pt, 0.5white+0.5red) enddef;
	def drawblackdotscale expr p=filldot(p, 0.4*xu, black) enddef;
	def drawreddotscale expr p=filldot(p, 0.4*xu, red) enddef;
	def drawwhitedotscale expr p=filldot(p, 0.4*xu, white) enddef;
	def drawblackdot expr p=filldot(p, 8pt, black) enddef;
	def drawwhitedot expr p=filldot(p, 8pt, white) enddef;
	def drawlinearrow expr p=draw p pensize(0.5pt); midarrow (p,0.5); enddef;
	def drawproparrow expr p=draw p pensize(1.5pt); midarrow (p,0.5); enddef;
	def drawproparrowdouble expr p=draw p pensize(1.5pt); midarrow (p,0.65); midarrow (reverse p,0.65); enddef;
	def drawwhitedotmr (expr p, sz)=filldot(p, sz*xu, white) enddef;
	def drawblackdotmr (expr p, sz)=filldot(p, sz*xu, black) enddef;
	
	def framed(expr p)=fill bbox p withcolor white; draw bbox p pensize(0.8pt); draw p; enddef;
	def framedd(expr p)=fill bbox p withcolor 0.94white; draw bbox p pensize(1pt); draw p; enddef;
\end{mpostdef}
\begin{mpostdef}
	color ecolor; ecolor:=(.72,.03,.06);
	color bcolor; bcolor:=(.07,.05,.66);
	color gcolor; gcolor:=(.0,.44,.0);
	color feyncolor; feyncolor:=(.06,.4,.01);
	color feyncolorfaint; feyncolorfaint:=(.12,.8,.02);
	color arrowcolor; arrowcolor:=(0,0,0);
	color faint; faint:=(.7,0.7,0.7);
	vsize := 0.16xu;
	rad:= 2xu;
	cpath:= for i=0 upto 35: 
	dir (i*10) scaled rad ..
	endfor cycle;
	def vert(expr pos)=
	fill fullcircle scaled vsize shifted point pos of cpath withcolor white;
	draw halfcircle scaled vsize rotated (90+10*pos) shifted point pos of cpath withcolor ecolor pensize(1.3pt);
	enddef;
	def vertthick(expr pos)=
	fill fullcircle scaled vsize shifted point pos of cpath withcolor white;
	draw halfcircle scaled vsize rotated (90+10*pos) shifted point pos of cpath withcolor ecolor pensize(2.0pt);
	enddef;
	def vertcol(expr pos,col)=
	fill fullcircle scaled vsize shifted point pos of cpath withcolor white;
	draw halfcircle scaled vsize rotated (90+10*pos) shifted point pos of cpath withcolor col pensize(1.3pt);
	enddef;
	def drawbdry(expr startpos,endpos)=
	draw subpath(startpos,endpos) of cpath withcolor ecolor pensize(1.3pt);
	enddef;
	def drawbdryblack(expr startpos,endpos)=
	draw subpath(startpos,endpos) of cpath withcolor black pensize(1.3pt);
	enddef;
	def drawbdrythick(expr startpos,endpos)=
	draw subpath(startpos,endpos) of cpath withcolor ecolor pensize(2.0pt);
	enddef;
	def drawbdrythickblack(expr startpos,endpos)=
	draw subpath(startpos,endpos) of cpath withcolor black pensize(2.0pt);
	enddef;
	def drawbdrydashed(expr startpos,endpos)=
	draw subpath(startpos,endpos) of cpath withcolor ecolor dashed withdots scaled 0.5 pensize(1.3pt);
	enddef;
	def drawbdrydashedthick(expr startpos,endpos)=
	draw subpath(startpos,endpos) of cpath withcolor ecolor dashed withdots scaled 0.5 pensize(2.0pt);
	enddef;
	def vlabel(expr pos, vscal, tt)=
	vert(pos);
	label(tt,point pos of cpath scaled vscal);
	enddef;
	
	def vlabelcol(expr pos, vscal, tcol, vcol, tt)=
	vertcol(pos, vcol);
	label(tt,point pos of cpath scaled vscal) withcolor tcol;
	enddef;
	
	def ppath(expr p,spos,epos)=
	subpath(arctime(spos*arclength(p))of p,arctime(epos*arclength(p)) of p) of p
	enddef;
\end{mpostdef}

\begin{mpostdef}
	picture nicearrow; currentpicture:= nullpicture;
	ahlengthsave:=ahlength; ahlength:=12pt;
	drawarrow (0xu,0xu){dir 20}..(2xu,0xu) pensize(3pt) withcolor arrowcolor;
	ahlength:=ahlengthsave;
	nicearrow:= currentpicture; currentpicture:= nullpicture;
\end{mpostdef}

\begin{mpostdef}
	eNum:= 2.718281828459045235;
	numeric startangle, lsdist, scalefactor, stepangle;
	path ls;
	
	vardef logspiralder(expr lspos, endpoint, maxangle, stepnumber, enddir) =
	startangle := ((angle(endpoint-lspos) - (maxangle mod 360)) mod 360); 
	stepangle := (1.0maxangle)/(1.0stepnumber);
	lsdist := length(endpoint-lspos); 
	scalefactor := mlog(lsdist+1.0)/(1.0*maxangle)/256;
	ls:=lspos for t = startangle step stepangle until (maxangle + startangle + 0.02): .. (((mexp(256((t-startangle)*scalefactor))-1.0) * cosd(t),(mexp(256((t-startangle)*scalefactor))-1.0)  * sind(t))+lspos) endfor;
	ls:= subpath(1,(length ls) - 1) of ls;
	ls:= ls .. {dir enddir}endpoint;
	ls
	enddef;
\end{mpostdef}

\begin{mpostfig}[label=figureone]
	pens:=0.7pt;
	drawarrow (-0.2xu, 0) -- (4.3xu, 0) pensize (1.0pt);  
	drawarrow (0, -0.2xu) -- (0, 3.2xu) pensize (1.0pt);  
	pair rpos; pair rrpos; 
	pair midpos, midposright, midposleft;
	path con, rA, rrA;
	path lspr, lsprr, lsprupper, lsprlower, lsprrupper, lsprrlower;

	rcirc:=0.9xu;
	rpos:=(3.4xu,2.1xu);
	rrcirc:=1.3xu;
	rrpos:=(1.0xu,1.2xu);
	draw fullcircle scaled (rcirc) shifted rpos pensize(pens);
	draw fullcircle scaled (rrcirc) shifted rrpos pensize(pens);
	rA:=fullcircle scaled (rcirc+0.11xu) shifted rpos;
	rrA:=fullcircle scaled (rrcirc+0.11xu) shifted rrpos;
	con = rpos--rrpos;
	midpos := ((rcirc/(rcirc + rrcirc))[rpos,rrpos]);
	midposright := ((1.06*rcirc/(rcirc + rrcirc))[rpos,rrpos]);
	midposleft := ((0.94*rcirc/(rcirc + rrcirc))[rpos,rrpos]);

	numeric appanglea, appangleb;
	appanglea:=76;
	appangleb:=appanglea + 180;
	lsprr:=logspiralder(rrpos, midpos, 560, 12, appanglea);
	draw lsprr pensize(pens);
	lspr:=logspiralder(rpos, midpos, 560, 12, appangleb);
	draw lspr pensize(pens);

	lsprupper:=logspiralder(rpos, midposright, 595, 12, appangleb);
	lsprlower:=logspiralder(rpos, midposleft, 528, 12, appangleb);
	lsprrupper:=logspiralder(rrpos, midposright, 528, 12, appanglea);
	lsprrlower:=logspiralder(rrpos, midposleft, 588, 12, appanglea);

	draw lsprupper cutbefore rA pensize (pens) withcolor bcolor;
	draw lsprlower cutbefore rA pensize(pens) withcolor bcolor;
	draw lsprrupper cutbefore rrA pensize (pens) withcolor bcolor;
	draw lsprrlower cutbefore rrA pensize (pens) withcolor bcolor;

	draw (rA cutafter lsprupper) pensize (pens) withcolor ecolor;
	draw (rA cutbefore lsprlower) pensize (pens) withcolor ecolor;
	draw (rrA cutbefore lsprrupper) pensize (pens) withcolor ecolor;
	draw (rrA cutafter lsprrlower) pensize (pens) withcolor ecolor;
	pic[4]:=currentpicture;
	currentpicture:=nullpicture;
			
	draw (-2xu,0xu) ..controls (-2xu,0.75xu) and (2xu,0.75xu) .. ( 2xu,0xu) dashed evenly pensize (1.0pt) withcolor bcolor;
	draw (0xu,-0.15xu) .. controls (0.1xu,-0.18xu) and (0.1xu,-0.92xu) .. ( 0xu,-0.95xu) dashed evenly pensize (1.0pt) withcolor ecolor;
	draw fullcircle xscaled 4xu yscaled 1.9xu pensize (1pt);
	draw (-1xu,0.03xu){dir -20}..(1xu,0.03xu) pensize (1.0pt);
	draw (-0.8xu,-0.02xu){dir 25}..(0.8xu,-0.02xu) pensize (1.0pt);
	draw (-2xu,0xu) ..controls (-2xu,-0.75xu) and (2xu,-0.75xu) .. ( 2xu,0xu) pensize (1.0pt) withcolor bcolor;
	draw (0xu,-0.15xu) .. controls (-0.1xu,-0.18xu) and (-0.1xu,-0.92xu) .. ( 0xu,-0.95xu) pensize (1.0pt) withcolor ecolor;
	pic[1]:=currentpicture;
	currentpicture:=nullpicture;
		
	draw (3xu,0xu) -- (4xu,0xu) pensize(1pt);
	drawzigzag((4xu,0xu) -- (5xu,0xu));
	draw (5xu,0xu) -- (6xu,0xu) pensize(1pt);
	drawzigzag ((6xu,0xu) -- (7xu,0xu));
	draw halfcircle xscaled (2.0xu) yscaled (1.8xu) shifted (5.5xu,0xu) pensize(1pt) withcolor bcolor;	
	draw halfcircle xscaled (2.0xu) yscaled (-1.8xu) shifted (5.5xu,0xu) pensize(1pt) withcolor bcolor dashed evenly;	
	draw fullcircle xscaled (1.5xu) yscaled (1.5xu) shifted (4.5xu,0xu) pensize(1pt) withcolor ecolor;	
	dotlabeldiam:=4pt;
	dotlabel.top (btex $ $ etex, (4xu,0xu));
	dotlabel.top (btex $ $ etex, (5xu,0xu));
	dotlabel.top (btex $ $ etex, (6xu,0xu));
	pic[2]:=currentpicture;
	currentpicture:=nullpicture;

	drawarrow (-1.1xu, -4.5xu) -- (3.4xu, -4.5xu) pensize (1pt);  
	drawarrow (-0.9xu, -4.7xu) -- (-0.9xu, -1.3xu) pensize (1pt);  
	label.bot(btex $\textrm{Re}(z)$ etex, (2.9xu, -4.5xu));
	label.lft(btex $\textrm{Im}(z)$ etex, (-0.9xu, -1.7xu));
	draw (0.2xu,-2.1xu) -- (2.2xu,-2.1xu) pensize (1pt) withcolor ecolor;
	draw (-0.9xu,-4.5xu) -- (1.1xu,-4.5xu) pensize (1pt) withcolor ecolor;
	draw (-0.9xu,-4.5xu) -- (0.2xu,-2.1xu) pensize (1pt) withcolor bcolor;
	draw (2.2xu,-2.1xu) -- (1.1xu,-4.5xu) pensize (1pt) withcolor bcolor;
	label.top(btex $\tau$ etex, (0.2xu, -2.1xu));
	label.bot(btex $1$ etex, (1.1xu, -4.5xu));
	label.top(btex $1+\tau$ etex, (2.2xu, -2.1xu));
	pic[3]:=currentpicture;
	currentpicture:=nullpicture;
			
	label.rt(btex (a) etex, (-2.5xu, 1.7xu));
	label.rt(btex (b) etex, (4.0xu, 1.7xu));
	label.rt(btex (c) etex, (-2.5xu, -1.8xu));
	label.rt(btex (d) etex, (4.0xu, -1.8xu));

	draw pic[1] scaled 1.0 shifted (0.5xu, 0.0xu);
	draw pic[2] scaled 1.0 shifted (2.5xu, 0.0xu);
	draw pic[3] scaled 1.0 shifted (0.0xu, -0.5xu);
	draw pic[4] scaled 1.0 shifted (5.5xu, -5.0xu);
\end{mpostfig}

\begin{mpostfig}[label=genustwo]
		pens:=0.7pt;
		drawarrow (-2.2xu, 0) -- (2.5xu, 0) pensize (1.0pt);  
		drawarrow (0, -2.0xu) -- (0, 2.3xu) pensize (1.0pt);  
		pair rpos; pair rrpos; 
		pair midpos, midposright, midposleft;
		path con, rA, rrA;
		path lspr, lsprr, lsprupper, lsprlower, lsprrupper, lsprrlower;
		
		rcirc:=0.9xu;
		rpos:=(1.4xu,1.5xu);
		rrcirc:=1.3xu;
		rrpos:=(-1.0xu,1.0xu);
		draw fullcircle scaled (rcirc) shifted rpos pensize(pens);
		draw fullcircle scaled (rrcirc) shifted rrpos pensize(pens);
		rA:=fullcircle scaled (rcirc+0.11xu) shifted rpos;
		rrA:=fullcircle scaled (rrcirc+0.11xu) shifted rrpos;
		con = rpos--rrpos;
		midpos := ((rcirc/(rcirc + rrcirc))[rpos,rrpos]);
		midposright := ((1.06*rcirc/(rcirc + rrcirc))[rpos,rrpos]);
		midposleft := ((0.94*rcirc/(rcirc + rrcirc))[rpos,rrpos]);
		
		numeric appanglea, appangleb;
		appanglea:=76;
		appangleb:=appanglea + 180;
		lsprr:=logspiralder(rrpos, midpos, 560, 12, appanglea);
		draw lsprr pensize(pens);
		lspr:=logspiralder(rpos, midpos, 560, 12, appangleb);
		draw lspr pensize(pens);
		
		lsprupper:=logspiralder(rpos, midposright, 595, 12, appangleb);
		lsprlower:=logspiralder(rpos, midposleft, 528, 12, appangleb);
		lsprrupper:=logspiralder(rrpos, midposright, 528, 12, appanglea);
		lsprrlower:=logspiralder(rrpos, midposleft, 588, 12, appanglea);
		
		draw lsprupper cutbefore rA pensize (pens) withcolor bcolor;
		draw lsprlower cutbefore rA pensize(pens) withcolor bcolor;
		draw lsprrupper cutbefore rrA pensize (pens) withcolor bcolor;
		draw lsprrlower cutbefore rrA pensize (pens) withcolor bcolor;

		draw (rA cutafter lsprupper) pensize (pens) withcolor ecolor;
		draw (rA cutbefore lsprlower) pensize (pens) withcolor ecolor;
		draw (rrA cutbefore lsprrupper) pensize (pens) withcolor ecolor;
		draw (rrA cutafter lsprrlower) pensize (pens) withcolor ecolor;
		
		rcirc:=1.2xu;
		rpos:=(1.4xu,-1.1xu);
		rrcirc:=1.0xu;
		rrpos:=(-1.2xu,-1.0xu);
		draw fullcircle scaled (rcirc) shifted rpos pensize(pens);
		draw fullcircle scaled (rrcirc) shifted rrpos pensize(pens);
		rA:=fullcircle scaled (rcirc+0.11xu) shifted rpos;
		rrA:=fullcircle scaled (rrcirc+0.11xu) shifted rrpos;
		midpos := ((rcirc/(rcirc + rrcirc))[rpos,rrpos]);
		midposright := ((1.06*rcirc/(rcirc + rrcirc))[rpos,rrpos]);
		midposleft := ((0.94*rcirc/(rcirc + rrcirc))[rpos,rrpos]);
		
		numeric appanglea, appangleb;
		appanglea:=56;
		appangleb:=appanglea + 180;
		lsprr:=logspiralder(rrpos, midpos, 560, 12, appanglea);
		draw lsprr pensize(pens);
		lspr:=logspiralder(rpos, midpos, 560, 12, appangleb);
		draw lspr pensize(pens);
		
		lsprupper:=logspiralder(rpos, midposright, 595, 12, appangleb);
		lsprlower:=logspiralder(rpos, midposleft, 528, 12, appangleb);
		lsprrupper:=logspiralder(rrpos, midposright, 528, 12, appanglea);
		lsprrlower:=logspiralder(rrpos, midposleft, 588, 12, appanglea);
		
		draw lsprupper cutbefore rA pensize (pens) withcolor bcolor;
		draw lsprlower cutbefore rA pensize(pens) withcolor bcolor;
		draw lsprrupper cutbefore rrA pensize (pens) withcolor bcolor;
		draw lsprrlower cutbefore rrA pensize (pens) withcolor bcolor;
		
		draw (rA cutafter lsprupper) pensize (pens) withcolor ecolor;
		draw (rA cutbefore lsprlower) pensize (pens) withcolor ecolor;
		draw (rrA cutbefore lsprrupper) pensize (pens) withcolor ecolor;
		draw (rrA cutafter lsprrlower) pensize (pens) withcolor ecolor;
		
		pic[2]:=currentpicture;
		currentpicture:=nullpicture;
		
		draw (-3xu,-0.15xu) .. controls (-2.9xu,-0.18xu) and (-2.9xu,-0.92xu) .. ( -3xu,-0.95xu) dashed evenly pensize (1.0pt) withcolor ecolor;
		draw (-4.2xu,0xu) ..controls (-4.2xu,0.6xu) and (-1.8xu,0.6xu) .. (-1.8xu,0xu) pensize (1.0pt) withcolor bcolor;
		draw (-4.2xu,0xu) ..controls (-4.2xu,-0.6xu) and (-1.8xu,-0.6xu) .. (-1.8xu,0xu) pensize (1.0pt) withcolor bcolor;
		draw (-4xu,0.03xu){dir -20}..(-2xu,0.03xu) pensize (1.0pt);
		draw (-3.8xu,-0.02xu){dir 25}..(-2.2xu,-0.02xu) pensize (1.0pt);
		draw (-3xu,-0.15xu) .. controls (-3.1xu,-0.18xu) and (-3.1xu,-0.92xu) .. ( -3xu,-0.95xu) pensize (1.0pt) withcolor ecolor;
		
		draw (0xu,-0.15xu) .. controls (0.1xu,-0.18xu) and (0.1xu,-0.92xu) .. ( 0xu,-0.95xu) dashed evenly pensize (1.0pt) withcolor ecolor;
		draw (-1.2xu,0xu) ..controls (-1.2xu,0.6xu) and (1.2xu,0.6xu) .. (1.2xu,0xu) pensize (1.0pt) withcolor bcolor;
		draw (-1.2xu,0xu) ..controls (-1.2xu,-0.6xu) and (1.2xu,-0.6xu) .. (1.2xu,0xu) pensize (1.0pt) withcolor bcolor;
		draw (-1xu,0.03xu){dir -20}..(1xu,0.03xu) pensize (1.0pt);
		draw (-0.8xu,-0.02xu){dir 25}..(0.8xu,-0.02xu) pensize (1.0pt);
		draw (0xu,-0.15xu) .. controls (-0.1xu,-0.18xu) and (-0.1xu,-0.92xu) .. ( 0xu,-0.95xu) pensize (1.0pt) withcolor ecolor;
		
		draw halfcircle xscaled 1.9xu yscaled 3.5xu rotated -90 ..(-1.5xu,0.7xu){left}..(-3xu,0.95xu){left} pensize (1pt);
		draw halfcircle xscaled 1.9xu yscaled 3.5xu rotated 90 shifted (-3xu,0xu) ..(-1.5xu,-0.7xu){right}..(0xu,-0.95xu){right} pensize (1pt);
		
		pic[1]:=currentpicture;
		currentpicture:=nullpicture;
		
		label.rt(btex (a) etex, (-3.5xu, 2.2xu));
		label.rt(btex (b) etex, (4.0xu, 2.2xu));
	
		draw pic[1] scaled 1.0 shifted (1.5xu,0.0xu);
		draw pic[2] scaled 1.0 shifted (7.0xu,0.0xu);
\end{mpostfig}

\begin{mpostfig}[label=concentricschottky]
	pens:=1.0pt;
	drawarrow (-3xu, 0) -- (3xu, 0) pensize (pens);  
	drawarrow (0, -3xu) -- (0, 3xu) pensize (pens);  

	vardef b_cycle_x(expr t) = 
		pi:=3.14159;
		(2.5)*((3/5)**t)*(cosd(90*t))
	enddef;

	vardef b_cycle_y(expr t) = 
		pi:=3.14159;
		(2.5)*((3/5)**t)*(sind(90*t))
	enddef;

	draw ((b_cycle_x(0)*1xu,b_cycle_y(0)*1xu) for i:=0.05 step 0.05 until 1.05:
		..(b_cycle_x(i)*1xu,b_cycle_y(i)*1xu) endfor) withcolor bcolor pensize (pens);
	
	draw fullcircle scaled 3xu withcolor ecolor pensize (pens);
	draw fullcircle scaled 5xu withcolor ecolor pensize (pens);
	
	draw (2.5xu,0.1xu) -- (2.5xu,-0.1xu) pensize (pens);
	draw (1.5xu,0.1xu) -- (1.5xu,-0.1xu) pensize (pens);
	label.bot(btex $1$ etex, (2.65xu,0xu));
	label.bot(btex $|q|$ etex, (1.7xu, 0xu));
\end{mpostfig}

\begin{mpostfig}[label=morphing]
	draw (-2xu,0xu) ..controls (-2xu,0.75xu) and (2xu,0.75xu) .. ( 2xu,0xu) dashed evenly pensize (1.0pt) withcolor bcolor;
	draw (0xu,-0.15xu) .. controls (0.12xu,-0.18xu) and (0.12xu,-0.92xu) .. ( 0xu,-0.95xu) dashed evenly pensize (1.0pt) withcolor ecolor;
	draw fullcircle xscaled 4xu yscaled 1.9xu pensize (1pt);
	draw (-1xu,0.03xu){dir -20}..(1xu,0.03xu) pensize (1.0pt);
	draw (-0.8xu,-0.02xu){dir 25}..(0.8xu,-0.02xu) pensize (1.0pt);
	draw (-2xu,0xu) ..controls (-2xu,-0.75xu) and (2xu,-0.75xu) .. ( 2xu,0xu) pensize (1.0pt) withcolor bcolor;
	draw (0xu,-0.15xu) .. controls (-0.12xu,-0.18xu) and (-0.12xu,-0.92xu) .. ( 0xu,-0.95xu) pensize (1.0pt) withcolor ecolor;
	pic[1]:=currentpicture;
	currentpicture:=nullpicture;
	
	draw (-2xu,0xu) ..controls (-2xu,0.75xu) and (2xu,0.75xu) .. ( 2xu,0xu) dashed evenly pensize (1.0pt) withcolor bcolor;
	draw halfcircle xscaled 4xu yscaled 1.9xu pensize (1pt);
	draw (-1.05xu,0.0xu){dir 18}..(1.05xu,0.0xu) pensize (1.0pt);
	draw halfcircle rotated -90 xscaled 0.048xu yscaled 0.12xu shifted (0xu,-0.56xu) pensize (1.0pt) withcolor ecolor;
	draw (-1.2xu,0.05xu){dir -20}..{dir 0}(0xu,-0.5xu){dir 0} .. {dir 20}(1.2xu,0.05xu) pensize (1.0pt);
	draw (-2xu,0xu) ..controls (-2xu,-0.75xu) and (2xu,-0.75xu) .. ( 2xu,0xu) pensize (1.0pt) withcolor bcolor;
	draw (-2xu,0xu){dir 270} .. {dir 0}(0xu,-0.62xu) .. {dir 90}(2xu,0xu) pensize (1.0pt) withcolor black;
	draw halfcircle rotated 90 xscaled 0.048xu yscaled 0.12xu shifted (0xu,-0.56xu) pensize (1.0pt) withcolor ecolor;
	pic[2]:=currentpicture;
	currentpicture:=nullpicture;
	
	hpath:= (-2xu,0xu) ..controls (-2xu,0.75xu) and (2xu,0.75xu) .. ( 2xu,0xu);
	draw ppath(hpath,0.0,0.97) dashed evenly pensize (1.25pt) withcolor faint;
	draw (-2xu,0xu) ..controls (-2xu,-0.75xu) and (2xu,-0.75xu) .. ( 2xu,0xu) pensize (1.25pt) withcolor black;
	draw halfcircle xscaled 4xu yscaled 4xu rotated 90 withcolor black pensize(1.25pt);
	draw halfcircle xscaled 4xu yscaled 4xu rotated -90 withcolor bcolor pensize(1.25pt);
	fillshape (fullcircle xscaled 0.2xu yscaled 0.08xu shifted (0xu,2xu),white,1.25pt,ecolor);
	fillshape (fullcircle xscaled 0.2xu yscaled 0.08xu shifted (0xu,-2xu),white,1.25pt,ecolor);
	pic[3]:=currentpicture;
	currentpicture:=nullpicture;
	
	draw pic[1] shifted (0.2xu,0xu);
	draw nicearrow scaled 0.4 shifted (2.7xu,0xu);
	draw pic[2] shifted (6xu,0xu);
	draw nicearrow scaled 0.4 shifted (8.4xu,0xu);
	draw pic[3] scaled 0.8 shifted (11.2xu,0xu);
\end{mpostfig}

\begin{mpostfig}[label=schottky-degeneration]
	interim ahlength := 1.8pt;
	drawarrow (-4.2xu, 0) -- (-1.4xu, 0) pensize(0.3pt);  
	drawarrow (-2.8xu, -1xu) -- (-2.8xu, 1.42xu) pensize(0.3pt);  
	
	transp:=0.2;
	pens:=0.5pt;
	
	pair rpos; pair rrpos;
	rcirc:=0.7xu;
	rpos:=(-3.5xu,0.5xu);
	rrcirc:=1xu;
	rrpos:=(-2xu,0.9xu);
	draw fullcircle scaled (rcirc) shifted rpos dashed evenly pensize(pens) withcolor 0.1[white,ecolor];
	draw fullcircle scaled (rcirc*0.833) shifted (-3.51xu,0.515xu) pensize(pens) withcolor 0.3[white,ecolor];
	draw fullcircle scaled (rcirc*0.666) shifted (-3.52xu,0.52xu) pensize(pens) withcolor 0.4[white,ecolor];
	draw fullcircle scaled (rcirc*0.5) shifted (-3.53xu,0.525xu) pensize(pens) withcolor 0.5[white,ecolor];
	draw fullcircle scaled (rcirc*0.333) shifted (-3.54xu,0.53xu) pensize(pens) withcolor 0.7[white,ecolor];
	draw fullcircle scaled (rcirc*0.166) shifted (-3.55xu,0.535xu) pensize(pens) withcolor 0.85[white,ecolor];
	fill fullcircle scaled (rcirc*0.03) shifted (-3.56xu,0.54xu) pensize(pens) withcolor ecolor;
	draw fullcircle scaled (rrcirc) shifted rrpos dashed evenly pensize(pens) withcolor 0.1[white,ecolor];
	draw fullcircle scaled (rrcirc*0.833) shifted (-2.015xu,0.885xu) pensize(pens) withcolor 0.3[white,ecolor];	
	draw fullcircle scaled (rrcirc*0.666) shifted (-2.03xu,0.87xu) pensize(pens) withcolor 0.4[white,ecolor];
	draw fullcircle scaled (rrcirc*0.5) shifted (-2.045xu,0.855xu) pensize(pens) withcolor 0.5[white,ecolor];
	draw fullcircle scaled (rrcirc*0.333) shifted (-2.06xu,0.84xu) pensize(pens) withcolor 0.7[white,ecolor];
	draw fullcircle scaled (rrcirc*0.166) shifted (-2.075xu,0.825xu) pensize(pens) withcolor 0.85[white,ecolor];	
	fill fullcircle scaled (rrcirc*0.03) shifted (-2.09xu,0.81xu) pensize(pens) withcolor ecolor;
		
	pair rposi; pair rrposi;
	rcirci:=0.8xu;
	rposi:=(-3.45xu,-0.5xu);
	rrcirci:=0.6xu;
	rrposi:=(-2.1xu,-0.6xu);
	
	draw fullcircle scaled (rcirci) shifted rposi pensize(pens) withcolor black;
	draw fullcircle scaled (rrcirci) shifted rrposi pensize(pens) withcolor black;
\end{mpostfig}

\begin{mpostfig}[label=algcurve-schottky] 
		draw (-4xu,0xu) -- (-2xu,0xu) pensize(1pt);
		drawzigzag((-2xu,0xu) -- (-1xu,0xu));
		draw (-1xu,0xu) -- (0xu,0xu) pensize(1pt);
		drawzigzag ((0xu,0xu) -- (2xu,0xu));
		draw (2xu,0xu) -- (5xu,0xu) pensize(1pt);
		drawzigzag ((5xu,0xu) -- (7xu,0xu));
		draw fullcircle xscaled (1.8xu) yscaled (1.5xu) shifted (-1.5xu,0xu) pensize(1pt) withcolor ecolor;	
		draw fullcircle xscaled (2.8xu) yscaled (1.5xu) shifted (1xu,0xu) pensize(1pt) withcolor ecolor;	
		draw halfcircle xscaled (3.8xu) yscaled (2xu) shifted (3.5xu,0xu) pensize(1pt) withcolor bcolor;	
		draw halfcircle xscaled (3.8xu) yscaled (-2xu) shifted (3.5xu,0xu) pensize(1pt) withcolor bcolor dashed evenly;	
		draw halfcircle xscaled (7.1xu) yscaled (3xu) shifted (2.05xu,0xu) pensize(1pt) withcolor bcolor;	
		draw halfcircle xscaled (7.1xu) yscaled (-3xu) shifted (2.05xu,0xu) pensize(1pt) withcolor bcolor dashed evenly;	
		dotlabeldiam:=4pt;
		dotlabel.top (btex $a_1$ etex, (-2xu,0xu));
		dotlabel.top (btex $a_2$ etex, (-1xu,0xu));
		dotlabel.top (btex $a_3$ etex, (0xu,0xu));
		dotlabel.top (btex $a_4$ etex, (2xu,0xu));
		dotlabel.top (btex $a_5$ etex, (5xu,0xu));
		drawdblarrow (1.5xu,-1.8xu) -- (1.5xu,-4.2xu);
		label (btex $a(z)-a_1=\sum\limits_{\gamma\in G}\left(\gamma(z)^2-\gamma(0)^2\right)$ etex, (4.2xu,-3xu));
		draw (-4xu,-6xu) -- (7xu,-6xu) pensize(1pt);
		draw (-1.8xu,-5.25xu){dir 30}..{dir -30}(4.8xu,-5.25xu) withcolor bcolor pensize(1pt);
		draw (0.1xu,-5.58xu){dir 40}..{dir -40}(2.9xu,-5.58xu) withcolor bcolor pensize(1pt);
		draw fullcircle scaled (1xu) shifted (-0.2xu,-6xu) pensize(1pt) withcolor ecolor;	
		draw fullcircle scaled (1xu) shifted (3.2xu,-6xu) pensize(1pt) withcolor ecolor;	
		draw fullcircle scaled (2xu) shifted (-2.5xu,-6xu) pensize(1pt) withcolor ecolor;	
		draw fullcircle scaled (2xu) shifted (5.5xu,-6xu) pensize(1pt) withcolor ecolor;	
		dotlabel.top (btex $z_1$ etex, (1.5xu,-6xu));
		dotlabel.ulft (btex $-z_2$ etex, (0.3xu,-6xu));
		dotlabel.urt (btex $z_2$ etex, (2.7xu,-6xu));
		dotlabel.ulft (btex $-z_3$ etex, (-0.7xu,-6xu));
		dotlabel.urt (btex $z_3$ etex, (3.7xu,-6xu));
		dotlabel.ulft (btex $-z_4$ etex, (-1.5xu,-6xu));
		dotlabel.urt (btex $z_4$ etex, (4.5xu,-6xu));
		dotlabel.ulft (btex $-z_5$ etex, (-3.5xu,-6xu));
		dotlabel.urt (btex $z_5$ etex, (6.5xu,-6xu));
\end{mpostfig}

\usepackage[dvipsnames,table]{xcolor}
\definecolor{mygreen}{rgb}{0,0.4,0}
\definecolor{myblue}{rgb}{0,0.0,0.4}
\definecolor{refrcolor}{rgb}{0,0.4,0}
\definecolor{cgreen}{rgb}{0,0.7,0}
\definecolor{ecolor}{rgb}{.52,.03,.06}
\definecolor{bgcolor}{rgb}{.96,.95,.80}
\definecolor{bgcolordark}{rgb}{.80,.80,.67}
\definecolor{faint}{rgb}{.80,.80,.80}

\usepackage{enumitem}

\usepackage{tabularx}

\makeatletter
\providecommand*{\shuffle}{%
  \mathbin{\mathpalette\shuffle@{}}%
}
\newcommand*{\shuffle@}[2]{%
  \sbox0{$#1\vcenter{}$}%
  \kern .15\ht0 
  \rlap{\vrule height .25\ht0 depth 0pt width 2.5\ht0}%
  \raise.1\ht0\hbox to 2.5\ht0{%
    \vrule height 1.75\ht0 depth -.1\ht0 width .17\ht0 %
    \hfill
    \vrule height 1.75\ht0 depth -.1\ht0 width .17\ht0 %
    \hfill
    \vrule height 1.75\ht0 depth -.1\ht0 width .17\ht0 %
  }%
  \kern .15\ht0 
}
\makeatother

\usepackage{xparse}

\ExplSyntaxOn
\NewDocumentCommand{\Gtargz}{m m}
{
 \Gt\left(\begin{smallmatrix}
 \Gtargz_print:n {#1} \\
 \Gtargz_print:n {#2}
 \end{smallmatrix};z\right)
}
\seq_new:N \l_Gtargz_list_seq
\cs_new_protected:Npn \Gtargz_print:n #1
{
  \seq_set_split:Nnn \l_Gtargz_list_seq { , } { #1 }
  \seq_use:Nn \l_Gtargz_list_seq { , & }
}
\ExplSyntaxOff

\ExplSyntaxOn
\NewDocumentCommand{\Gtargt}{m m}
{
 \Gt\left(\begin{smallmatrix}
 \Gtargt_print:n {#1} \\
 \Gtargt_print:n {#2}
 \end{smallmatrix};t\right)
}
\seq_new:N \l_Gtargt_list_seq
\cs_new_protected:Npn \Gtargt_print:n #1
{
  \seq_set_split:Nnn \l_Gtargt_list_seq { , } { #1 }
  \seq_use:Nn \l_Gtargt_list_seq { , & }
}
\ExplSyntaxOff

\ExplSyntaxOn
\NewDocumentCommand{\Gtargzt}{m m}
{
 \Gt\left(\begin{smallmatrix}
 \Gtargzt_print:n {#1} \\
 \Gtargzt_print:n {#2}
 \end{smallmatrix};z|\tau\right)
}
\seq_new:N \l_Gtargzt_list_seq
\cs_new_protected:Npn \Gtargzt_print:n #1
{
  \seq_set_split:Nnn \l_Gtargzt_list_seq { , } { #1 }
  \seq_use:Nn \l_Gtargzt_list_seq { , & }
}
\ExplSyntaxOff

\ExplSyntaxOn
\NewDocumentCommand{\Gtargxit}{m m}
{
 \Gt\left(\begin{smallmatrix}
 \Gtargxit_print:n {#1} \\
 \Gtargxit_print:n {#2}
 \end{smallmatrix};\xi|\tau\right)
}
\seq_new:N \l_Gtargxit_list_seq
\cs_new_protected:Npn \Gtargxit_print:n #1
{
  \seq_set_split:Nnn \l_Gtargxit_list_seq { , } { #1 }
  \seq_use:Nn \l_Gtargxit_list_seq { , & }
}
\ExplSyntaxOff

\ExplSyntaxOn
\NewDocumentCommand{\Gtargtt}{m m}
{
 \Gt\left(\begin{smallmatrix}
 \Gtargtt_print:n {#1} \\
 \Gtargtt_print:n {#2}
 \end{smallmatrix};t|\tau\right)
}
\seq_new:N \l_Gtargtt_list_seq
\cs_new_protected:Npn \Gtargtt_print:n #1
{
  \seq_set_split:Nnn \l_Gtargtt_list_seq { , } { #1 }
  \seq_use:Nn \l_Gtargtt_list_seq { , & }
}
\ExplSyntaxOff

\ExplSyntaxOn
\NewDocumentCommand{\Gtargzg}{m m}
{
 \Gt\left(\begin{smallmatrix}
 \Gtargzg_print:n {#1} \\
 \Gtargzg_print:n {#2}
 \end{smallmatrix};z|\SGroup\right)
}
\seq_new:N \l_Gtargzg_list_seq
\cs_new_protected:Npn \Gtargzg_print:n #1
{
  \seq_set_split:Nnn \l_Gtargzg_list_seq { , } { #1 }
  \seq_use:Nn \l_Gtargzg_list_seq { , & }
}
\ExplSyntaxOff

\ExplSyntaxOn
\NewDocumentCommand{\Gtargxig}{m m}
{
 \Gt\left(\begin{smallmatrix}
 \Gtargxig_print:n {#1} \\
 \Gtargxig_print:n {#2}
 \end{smallmatrix};\xi|\SGroup\right)
}
\seq_new:N \l_Gtargxig_list_seq
\cs_new_protected:Npn \Gtargxig_print:n #1
{
  \seq_set_split:Nnn \l_Gtargxig_list_seq { , } { #1 }
  \seq_use:Nn \l_Gtargxig_list_seq { , & }
}
\ExplSyntaxOff

\ExplSyntaxOn
\NewDocumentCommand{\Gtargtg}{m m}
{
 \Gt\left(\begin{smallmatrix}
 \Gtargtg_print:n {#1} \\
 \Gtargtg_print:n {#2}
 \end{smallmatrix};t|\SGroup\right)
}
\seq_new:N \l_Gtargtg_list_seq
\cs_new_protected:Npn \Gtargtg_print:n #1
{
  \seq_set_split:Nnn \l_Gtargtg_list_seq { , } { #1 }
  \seq_use:Nn \l_Gtargtg_list_seq { , & }
}
\ExplSyntaxOff

\newcommand{\SI}[1]{\Sel[#1]}

\ExplSyntaxOn
\NewDocumentCommand{\SIE}{m m}
{
\SelE\!\Big[\begin{smallmatrix}
 \SI_print:n {#1} \\
 \SI_print:n {#2}
 \end{smallmatrix}\Big]
}
\seq_new:N \l_SI_list_seq
\cs_new_protected:Npn \SI_print:n #1
{
  \seq_set_split:Nnn \l_SI_list_seq { , } { #1 }
  \seq_use:Nn \l_SI_list_seq { , & }
}
\ExplSyntaxOff

\usepackage[pdfencoding=auto,bookmarks=true,hyperfigures=true]{hyperref}%
\PassOptionsToPackage{unicode}{hyperref}
\providecommand{\hypersetup}[1]{}
\providecommand{\texorpdfstring}[2]{#1}
\hypersetup{plainpages=false}
\hypersetup{pdfpagemode=UseNone}
\hypersetup{bookmarksnumbered=true}
\hypersetup{pdfstartview=FitH}
\hypersetup{colorlinks=false}
\hypersetup{citebordercolor={0.7 0.7 1}}
\hypersetup{urlbordercolor={.4 .8 1}}
\hypersetup{linkbordercolor={1 .8 .6}}
\hypersetup{colorlinks=true, urlcolor=[rgb]{0.13,0.30,0.45}, linkcolor=[rgb]{0.13,0.30,0.45}, citecolor=[rgb]{0.55,0.0,0.05}}

\makeatletter
\let\@keywords\@empty
\let\@subject\@empty
\providecommand{\keywords}[1]{\gdef\@keywords{#1}}
\providecommand{\subject}[1]{\gdef\@subject{#1}}
\def\thetitle{\@title}
\def\theauthor{\@author}
\def\thesubject{\@subject}
\def\thedate{\@date}
\def\thekeywords{\@keywords}
\makeatother
\AtBeginDocument{
\hypersetup{pdftitle={\thetitle}}%
\hypersetup{pdfauthor={\theauthor}}%
\hypersetup{pdfsubject={\thesubject}}%
\hypersetup{pdfkeywords={\thekeywords}}%
}

\newif\ifnote 
\notetrue

\DeclareMathOperator{\id}{id}

\DeclareMathOperator{\ad}{ad}

\DeclareMathOperator{\Lie}{Lie}

\DeclareMathOperator*{\Res}{Res}

\DeclareMathOperator{\Gt}{\tilde{\Gamma}}

\DeclareMathOperator{\Sel}{S}
\DeclareMathOperator{\SelE}{S^E}

\newcommand{\SGroup}{G}

\ProvideMathFix{rfrac,vfrac}
\newcommand{\iunit}{{\mathring{\imath}}}
\newcommand{\eunit}{\mathinner{\mathrm{e}}}

\newcommand{\der}{\mathrm{d}}
\newcommand{\diff}[2][.]{\mathinner{\der#2\if #1.\else^{#1}\fi}}

\newcommand{\field}[1]{\mathbb{#1}}

\newcommand{\Integers}{\field{Z}}

\newcommand{\Complex}{\field{C}}

\newcommand{\grp}[1]{\mathrm{#1}}
\newcommand{\alg}[1]{\mathfrak{#1}}

\newcommand{\defeq}{\mathrel{:=}}
\newcommand{\eqdef}{\mathrel{=:}}

\newcommand{\val}{\vec{\alpha}}
\newcommand{\vbe}{\vec{\beta}}

\newcommand{\va}{\vec{a}}
\newcommand{\vb}{\vec{b}}
\newcommand{\vc}{\vec{c}}

\let\qed\relax\newcommand{\qed}
{\hfill\ensuremath{\Box}}

\newcommand{\SL}{\mathrm{SL}}

\newcommand{\te}{\textrm}

\newcommand{\dd}{\mathrm{d}}

\newcommand{\zC}{\mathbb C}

\newcommand{\zP}{\mathbb P}

\newcommand{\zR}{\mathbb R}
\newcommand{\zZ}{\mathbb Z}

\newcommand{\cF}{\mathcal{F}}       
\newcommand{\cG}{\mathcal{G}}

\newcommand{\abel}[0]{\mathfrak{u}}

\newcommand{\acyc}[0]{\mathfrak A}
\newcommand{\bcyc}[0]{\mathfrak B}

\newcommand{\genus}{\ensuremath{h}}
\newcommand{\Fth}{F^{\theta}}
\newcommand{\periodmatrix}{\tau}
\newcommand{\Atxt}{A}
\newcommand{\Btxt}{B}

\newcommand{\gkern}[1]{g^{(#1)}}

\newcommand{\skern}[1]{s^{(#1)}}

\newcommand{\conn}{K}
\newcommand{\kronfun}{F}
\newcommand{\skron}{S}
\newcommand{\funddom}{\cF}

\newcommand{\mindx}[1]{\mathbf{#1}}
\DeclareMathOperator{\RSurf}{\Sigma}

\title{\textbf{\texorpdfstring{Schottky–Kronecker forms \\ and hyperelliptic polylogarithms}{Schottky–Kronecker forms and hyperelliptic polylogarithms}}}
\author{Konstantin Baune\texorpdfstring{\textsuperscript{a}}{},
	Johannes Broedel\texorpdfstring{\textsuperscript{a}}{},
	Egor Im\texorpdfstring{\textsuperscript{a}}{},\texorpdfstring{\\}{}
        Artyom Lisitsyn\texorpdfstring{\textsuperscript{a}}{},
        Federico Zerbini\texorpdfstring{\textsuperscript{b}}{}}
\date{\today}

\begin{document}

\pdfbookmark[1]{Title Page}{title} 
\thispagestyle{empty}
\vspace*{1.0cm}
\begin{center}%
  \begingroup\LARGE\bfseries\thetitle\par\endgroup
\vspace{1.0cm}

\begingroup\large\theauthor\par\endgroup
\vspace{9mm}
\begingroup\itshape
$^{\te{a}}$Institute for Theoretical Physics, ETH Zurich\\Wolfgang-Pauli-Str.~27, 8093 Zurich, Switzerland\\[4pt]
$^{\te{b}}$Departamento de Matem\'aticas Fundamentales, UNED\\ Calle de Juan del Rosal~10, 28040 Madrid, Spain
\par\endgroup
\vspace*{7mm}

\begingroup\ttfamily
baunek@ethz.ch, jbroedel@ethz.ch, egorim@ethz.ch,\\
alisitsyn@ucdavis.edu, f.zerbini@mat.uned.es
\par\endgroup

\vspace*{2.0cm}

\textbf{Abstract}\vspace{5mm}

\begin{minipage}{13.4cm}
Elliptic polylogarithms can be defined as iterated integrals on a genus-one Riemann surface of a set of integration kernels whose generating series was already considered by Kronecker in the 19th century. In this article, we employ the Schottky parametrization of a Riemann surface to construct higher-genus analogues of Kronecker's generating series, which we refer to as \emph{Schottky–Kronecker forms}. Our explicit construction generalizes ideas from Bernard's higher-genus construction of the Knizhnik–Zamolodchikov connection.  

Integration kernels generated from the Schottky–Kronecker forms are defined as Poincar\'e series. Under technical assumptions, related to the convergence of these Poincar\'e series on the underlying Riemann surface, we argue that these integration kernels coincide with a set of differentials defined by Enriquez, whose iterated integrals constitute higher-genus analogues of polylogarithms. 

Enriquez' original definition is not well-suited for numerical evaluation of higher-genus polylogarithms. In contrast, the Poincar\'e series defining our integration kernels can be evaluated numerically for real hyperelliptic curves, for which the above-mentioned convergence assumptions can be verified. We numerically evaluate several examples of genus-two polylogarithms, thereby paving the way for numerical evaluation of hyperelliptic analogues of polylogarithms.
\end{minipage}
\end{center}
\vfill

\newpage
\setcounter{tocdepth}{2}
\tableofcontents

\newpage

\section{Introduction}
\label{sec:introduction}

Observables in high-energy particle physics, in particular scattering amplitudes or Wilson loops, are functions of a set of parameters comprising for example masses and momenta of external particles. These observables are usually calculated as solutions to differential equations, or -- interchangeably -- in terms of parameter integrals. It is convenient to express such integrals in terms of known special functions, such as the polylogarithm functions, whose study constitutes the main motivation of this article, as well as, for instance, the Bessel functions and the hypergeometric functions, but in simpler situations also the Gamma function, the trigonometric functions and the elliptic functions. The key to relate observables with special functions is to make use of the integral representation of the latter, which also contains information about the ``geometric setting'', encoded by the integration cycles and by the differential form integrated.

In recent years, it became clear that expressing special functions as iterated integrals is especially useful in physics, both because of the analytic and algebraic properties of the latter. A result of Chen implies that iterated integrals capture deep information about the topology of a manifold \cite{Chen}. Consequently, the information on observables formerly expressed through the corresponding differential equation can be framed more geometrically as properties of the manifold on which the integration path is located. 

In this article, we generically call \emph{polylogarithms} the iterated integrals over Riemann surfaces. The only strict requirement is that these iterated integrals must be homotopy invariant, i.e.~only depend on the homotopy class of the integration path, so that they define functions on the universal cover of the manifold, but we will mainly consider meromorphic (possibly multivalued\footnote{By saying that a function or differential is multivalued, we mean that it is defined on the universal cover of the manifold, rather than on the manifold itself.}) integration kernels with at most simple poles. Polylogarithms, also known in the mathematics literature as \emph{hyperlogarithms} to distinguish them from the classical polylogarithms $\mathrm{Li}_n$, are known to generate the smallest space of multivalued functions on a Riemann surface which contains all meromorphic functions and which is closed under taking primitives \cite{BrownThesis, EZ2}: this justifies their ubiquity in the computation of physics observables. 

A great advantage of this polylogarithmic formulation is that the space of functions generated by polylogarithms has an especially nice (Hopf-)algebraic structure (see \rcites{BrownThesis, Duhr:2012fh, EZ2} for example), which comes with useful tools like coproduct and coaction. This has been exploited for various calculations in the context of ${\cal N}{=}\,4$ super-Yang–Mills theory and string theories (see for example \rcites{Goncharov:2010jf,Broedel:2013tta,Caron-Huot:2020bkp}). 

Classically, polylogarithms are special functions defined by the series $\mathrm{Li}_k(z)=\sum_{n\geq 1}z^n/n^{k}$. It is well known that they can be represented as iterated integrals on the Riemann sphere, which has genus zero, and are therefore special cases of polylogarithms in the sense specified above. Elliptic integrals can also be formulated as elliptic polylogarithms: they are integrals of rational functions on an elliptic curve, which is a Riemann surface of genus one. Elliptic polylogarithms of different flavours have a prominent role in the evaluation of multi-loop Feynman integrals. In particular, they appear in the computation of the sunset and banana graphs, see e.g.~\cite{Laporta:2004rb,Bloch:2013tra,Adams:2015ydq,Adams:2014vja,Broedel:2017siw,Broedel:2019kmn,Primo:2017ipr,Broedel:2021zij}. In addition, elliptic polylogarithms provide a convenient language for expressing the results of one-loop open-string scattering amplitudes \cite{Broedel:2014vla,Broedel:2018izr,Broedel:2019gba}. Several situations, such as for example the evaluation of the Feynman integral associated to what is known as the multi-scale massive double box, require iterated integrals on Riemann surfaces of genus two and beyond \cite{Georgoudis:2015hca,Marzucca:2023gto}.

All this indicates that polylogarithms have altered the way we are presenting results for observables and in particular scattering amplitudes: the algebraic and analytic advantages of such a canonical language are compelling. As the construction of polylogarithms and the current understanding of their properties is different depending on the genus of the Riemann surface, let us collect the state of the art in the following list:  
\begin{itemize}
	\item \textbf{Genus zero.} Integrating rational functions on a Riemann surface of genus zero, i.e.~the Riemann sphere, leads to functions known as \emph{Goncharov polylogarithms} (or hyperlogarithms): they are iterated integrals of differentials $\frac{\dd z}{z-c}$, where $c\in\mathbb C$, over a path which starts at basepoint~$0$ and avoids the poles of the integrand everywhere, with the possible exception of the endpoints, where the integral may be divergent and must therefore be regularized. Being of genus zero, the Riemann sphere is topologically trivial, hence the only non-trivial topological information carried by Goncharov polylogarithms is encoded in the poles of the differentials. Goncharov polylogarithms for example comprise classical polylogarithms, harmonic polylogarithms and Nielsen polylogarithms \cite{Remiddi:1999ew,Nielsen1909,doi:10.1137/0517086}.

Numerical evaluation of genus-zero polylogarithms and their special values has been topic of numerous investigations and content of various software packages during the last two decades \cite{GiNaC:2001,GiNaC:2005,Moch:2001zr,Ablinger:2011te,Ablinger:2013cf}. Compared to the numerical evaluation of polylogarithms on higher-genus surfaces, genus-zero polylogarithms are under excellent numerical control.  
\item \textbf{Genus one.} Polylogarithms on genus-one Riemann surfaces are often called \emph{elliptic polylogarithms}. Different ways of parametrizing genus-one compact Riemann surfaces lead to different definitions of elliptic polylogarithms; we list here the three representations which are relevant for us. One approach is to represent such surfaces, via the so-called Jacobi uniformization, as the quotient of their Schottky cover $\mathbb C^\times$ by the Schottky group $\simeq \mathbb Z$ freely generated by one of their non-trivial homology cycles. Secondly, they can be seen, via the classical (Riemann) uniformization, as the quotient of their universal cover~$\mathbb C$ by a lattice $\Lambda\simeq\mathbb Z^2$ which is isomorphic to their fundamental group. Incidentally, at genus one this quotient corresponds also to the Jacobian variety, which is a complex torus of dimension equal to the genus of the surface, conveniently capturing the dependence of a Riemann surface on its complex structure. Finally, they can be viewed as elliptic curves in the algebraic sense, i.e.~the complex zeros of a genus-one smooth complex projective curve.

Historically, the first approach to define elliptic analogues of polylogarithms was by using the Jacobi uniformization and by generalising the series representation of classical polylogarithms \cite{Bloch, Zagier, Levin, BrownLevin}. They were later defined in \cite{BrownLevin} as iterated integrals on a complex torus using non-holomorphic single-valued integration kernels. For Feynman integrals application, however, it was desirable to work with meromorphic integration kernels. This led to considering iterated integrals $\tilde\Gamma$ of the meromorphic kernels $g^{(n)}(z-c) \dd z$, an infinite family of differentials with possible simple poles at $c\in\mathbb C/\Lambda$ and multivalued on the torus, i.e.~defined on its universal cover $\mathbb C$ \cite{Broedel:2014vla, Broedel:2017kkb}. The generating function of the kernels $g^{(n)}$ is the so-called Kronecker function \cite{Weil}; this function was already considered by Bernard to generalize the Knizhnik–Zamolodchikov (KZ) connection to genus one in the study of the Wess–Zumino–Witten (WZW) model \cite{Bernard:1987df}, as well as in the mathematics literature \cite{ZagierJacobi, CEE, LevinRacinet}. 

While the formulation of elliptic polylogarithms in terms of the Kronecker function proved to be very useful in string theory \cite{Broedel:2014vla}, it was desirable for the computation of observables in quantum field theory to define elliptic polylogarithms also in the algebraic setting. This was achieved in \cite{Broedel:2017kkb} using algebraic analogues of the $g^{(n)}$-kernels. Another algebraic approach using finitely many rational kernels with higher-order poles was recently proposed in \rcite{EZ3}. 

All these approaches are expected, or in some cases known, to (essentially) generate the same space of functions. Efficiency of numerical evaluation crucially depends on the chosen representation of elliptic polylogarithms. 
		
	\item \textbf{Genus two and higher.} Starting from genus two, the Jacobian variety and the Riemann surface cannot be identified with each other anymore: this makes the dependence on the complex structure of the surface more difficult to control, therefore causing immediate problems in manipulating and evaluating polylogarithms. More specifically, the Riemann uniformization identifies now a compact genus-\genus{} Riemann surface~$\Sigma$ with its so-called Fuchsian model, a quotient of the hyperbolic upper half-plane~$\mathfrak H$, i.e.~the universal cover of $\Sigma$, by a discrete (Fuchsian) subgroup of $\SL(2,\zR)$ isomorphic to the fundamental group of~$\Sigma$; on the other hand,~$\Sigma$ can be embedded into (but is not equal to) its Jacobian variety $\simeq \zC^\genus/\zZ^{2\genus}$, which is defined in terms of the complex structure of the surface, via Abel's map.
The Jacobi uniformization generalizes at higher genus as the identification of~$\Sigma$ with the quotient of its Schottky cover, which is a subset of the Riemann sphere, by a Schottky group, i.e.~a free subgroup of $\SL(2,\zC)$ on~$h$ generators which correspond to the~\genus{} \Btxt-cycles of the surface.

At least five different generalizations of the Kronecker function have been proposed in the literature. One can be extracted from Bernard's work on the higher-genus KZ connection for the WZW model \cite{Bernard:1988yv}, and is formulated in the Schottky-cover setting. A somewhat more abstract approach, leading to infinitely many multivalued meromorphic kernels with at most simple poles (generalising the $g^{(n)}$-kernels from genus one), was put forward by Enriquez to study braid groups in \cite{EnriquezHigher}. Recently, two families of rational kernels with higher-order poles were proposed in \cite{EZ1,EZ2} in an algebraic setting, with the explicit purpose of defining higher-genus polylogarithms, followed by an alternative string-theory motivated approach \cite{DHoker:2023vax} which generalizes the non-holomorphic single-valued kernels of \cite{BrownLevin}. Relations between some of these approaches are currently under investigation \cite{DESZ:WIP}. 

Unfortunately, it seems that none of these approaches to defining higher-genus polylogarithms can be directly exploited to obtain efficient numerical evaluations.  
\end{itemize}
The goal of this article is to construct higher-genus polylogarithms and their integration kernels in a framework in which the problem of their numerical evaluation can be tackled efficiently. The main tool we are going to work with is the Schottky parametrization of higher-genus Riemann surfaces. Inspired by Bernard's higher-genus KZ connection, we construct Poincaré series over the Schottky group, which we call \emph{Schottky–Kronecker forms}, which generalize at higher genus the genus-one Kronecker function. Under assumptions on the Riemann surface which ensure the convergence of the defining Poincaré series, we show that our construction reproduces Enriquez' connection.
Our higher-genus integration kernels can be expressed as averages over the Schottky group of genus-one kernels, and this property is the key to obtain quick numerical evaluations. A case for which the convergence assumption is satisfied is that of a real hyperelliptic curve, and we restrict our attention to this case. We study various examples, evaluate them and check their analytic properties, and we comment on convergence and numerical stability of the implementation. 

The article is organized as follows: in \secref{sec:polylogreview} we review the construction of polylogarithms at genus zero and one with a particular focus on the elliptic Kronecker function. In \secref{sec:higherlanguages} we recall basics of higher-genus surfaces, and we introduce the background on their Schottky parametrization needed for our construction. In \secref{sec:genustworeview} we describe other approaches to the construction of higher-genus analogues of the Kronecker functions and of elliptic polylogarithms, with a special focus on the work of Bernard and Enriquez on flat connections which generalize the KZ connection. The main results of this article are contained in \secref{sec:schottky_kronecker}. We construct a set of Schottky–Kronecker forms via formal averages over the Schottky group which generalize Bernard's construction. Relations, degenerations and in particular expansions into kernels are discussed, and we also compare our construction with that of Enriquez. Finally, in \secref{sec:practicabilities} we implement and employ the above results to numerically evaluate various polylogarithms. We also discuss the convergence of the Schottky averages which define our integration kernels, and give explicit examples of simple polylogarithms at genus two. We summarize and formulate various open questions in \secref{sec:summary}. 


\section{Polylogarithms on Riemann surfaces of genus zero and one}\label{sec:polylogreview}
In this section we review the construction of the iterated integrals on Riemann surfaces of genus zero and genus one which define classical and elliptic polylogarithms, respectively.


\subsection{Genus zero}
On the only compact Riemann surface of genus zero, the Riemann sphere $\zC\zP^1$, the integration kernels used to define polylogarithms are the (third-kind) differentials with a simple pole at a finite point $a\in\mathbb{C}$, a second simple pole at $\infty$ with opposite residue, and holomorphic elsewhere\footnote{Notice that one cannot have a single simple pole on a compact Riemann surface, because the Cauchy residue theorem would force the simple pole to have a vanishing residue.}, given by 
\begin{equation}
	\label{eqn:genuszerodiff}
	h(z-a)\,\dd z, \quad\quad\text{where} \,\,\, h(z)\coloneq \frac{1}{z}.
\end{equation}
The function $h(z)$ defined by \eqn{eqn:genuszerodiff} satisfies the partial fraction identity
\begin{align}
	h(z-z_1)\,h(z-z_2)&=\frac{1}{z-z_1}\frac{1}{z-z_2}\notag\\
	&=\frac{1}{z_1-z_2}\left(\frac{1}{z-z_1}-\frac{1}{z-z_2}\right)=h(z_1-z_2)\Big(h(z-z_1)-h(z-z_2)\Big).
\end{align}
This can be considered as a genus-zero version of the Fay identity for the genus-one Kronecker function to be discussed below in \eqn{eqn:KroneckerFayGenusOne}.

Employing the differentials \eqref{eqn:genuszerodiff}, one can define \emph{Goncharov polylogarithms}, also known as \emph{hyperlogarithms}, as the iterated integrals 
\begin{equation}
	\label{eqn:Goncharov}
	H(a_1, a_2, \ldots, a_n ; z)\coloneq \int_0^z \dd t\,h(t-a_1)\,H(a_2, \ldots, a_n ; t),\quad H(;z)=1\,.
\end{equation}
The points $a_i$ are called \textit{letters} and $n$ is called the \emph{length}. If a particular problem in physics can be described in terms of Goncharov polylogarithms, one needs to determine the corresponding alphabet containing a finite number of letters. In this article, we will also refer to Goncharov polylogarithms as classical\footnote{This terminology may be a bit confusing, as usually classical polylogarithms are just the (very) special cases $\mathrm{Li}_n(z)$ of Goncharov polylogarithms whose Taylor expansion is given by $\sum_{k\geq 1}z^k/k^n$.}, or genus-zero, polylogarithms.

In a string amplitude context, all Selberg integrals can be expanded into polylogarithms with $a_i\,{\in}\,\{0,1\}$, known as \emph{one-variable multiple polylogarithms} \cite{BrownSVHPL}. Goncharov polylogarithms can be expressed as specializations of multiple polylogarithms, which are functions of several complex variables \cite[Lemma 3.4.2]{Panzer:2015ida}.

Let us list some features of the Goncharov polylogarithms defined in \eqn{eqn:Goncharov}:  
\begin{itemize}
	\item \textbf{Regularization.} Integrals in \eqn{eqn:Goncharov} diverge logarithmically as soon as $a_n=0$. This problem can be remedied by setting
	\begin{equation}
		H(\underbrace{0,\ldots,0}_n;z)\coloneq\frac{1}{n!}\log^n(z)\,
	\end{equation}
	and requiring $w\to H(w;z)$ to be a homomorphism from the shuffle algebra of all words~$w$ in the letters $a_i$, which is already the case for all words corresponding to convergent integrals, by known properties of iterated integrals. This purely algebraic regularization is referred to as the \emph{shuffle regularization}, and is equivalent to the more geometric \emph{tangential base point regularization}, as explained for instance in \cite{EZ3}.
	\item \textbf{Algebraic structure.} Due to their definition as iterated integrals, polylogarithms obey \emph{shuffle product relations}
	\begin{equation}
		H(\va;z)\,H(\vb;z)=\sum_{\vc\,\in\,\va\,\shuffle\,\vb}H(\vc;z)\,,
	\end{equation}
	where $\va=(a_1,\ldots,a_r)$, $\vb=(b_1,\ldots,b_s)$ and $\va\shuffle\vb$ is the shuffle set of $\va$ and $\vb$, i.e.~the set of all words formed by using all letters from $\va$ and $\vb$ while keeping the relative orders of the letters from $\va$ and $\vb$ intact. One can actually show that the algebra of (regularized) Goncharov polylogarithms is isomorphic to the abstract shuffle algebra of all words in the chosen alphabet, which is known to be a Hopf algebra, graded by the length of the words \cite{BrownThesis}. Goncharov polylogarithms therefore also enjoy the same nice algebra structure, the grading being given by the weight; in particular, they are linearly independent.
	\item \textbf{Special values.} Evaluating at argument one Goncharov polylogarithms for the alphabet $\{0,1\}$, one obtains real numbers called \emph{multiple zeta values}:
	\begin{equation}
		\label{eqn:MZVs}
		\zeta(n_1\ldots,n_s)=(-1)^s H(\underbrace{0,\ldots,0}_{n_s-1},1,\ldots,\underbrace{0,\ldots,0}_{n_1-1},1;1)=\sum_{0<k_1<\ldots<k_s}\frac{1}{k_1^{n_1}\cdots k_s^{n_s}}.
	\end{equation}
	Also these values have to be regularized for $n_s=1$. 
 
	\item \textbf{Primitives.} Another important feature of polylogarithms is that they generate the closure under taking primitives of the space of rational functions \cite{BrownThesis}. The procedure of taking integrals within this space is completely algorithmic, which massively simplifies the calculation of some string amplitudes and Feynman integrals. 
\end{itemize}
%


\subsection{Genus one} \label{sec:genus-one-polylogs}
At genus one a compact Riemann surface $\RSurf$ has a more intricate geometric structure. The two non-trivial loops which generate the fundamental group are referred to as the \Atxt-cycle $\acyc_1$ and \Btxt-cycle~$\bcyc_1$. We pick a holomorphic differential\footnote{There is only one holomorphic one-form on a genus-one surface, up to constants.} $\omega_1$ which is normalized by setting its \Atxt-cycle period to be $\oint_{\acyc_1} \omega_1 = 1$. The corresponding \Btxt-cycle period defines the \emph{elliptic modulus} $\oint_{\bcyc_1} \omega_1 = \tau\in \mathbb H$, where $\mathbb H$ is the complex upper half-plane.

\begin{figure}
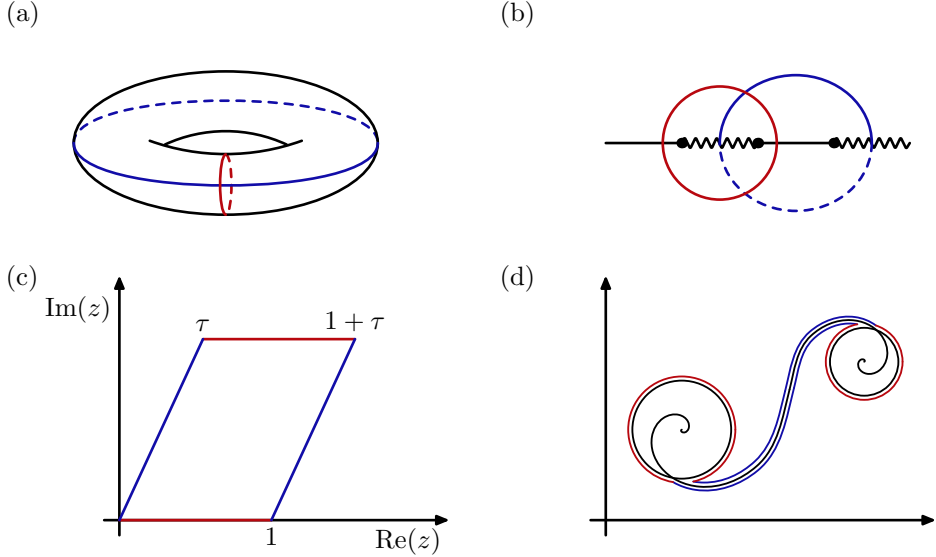

	\centering
	\mpostuse{figureone}
	\caption{Different representations of a compact Riemann surface of genus one. Picture (a) shows a torus with its \Atxt-cycle (red) and \Btxt-cycle (blue), (b) displays the corresponding algebraic curve with its branch cuts and \Atxt- and \Btxt-cycle, (c) is the fundamental domain in the universal cover $\mathbb C$ for the quotient $\mathbb C/\mathbb Z+\tau\mathbb Z$, and (d) is a picture of the Schottky parametrization to be discussed in \secref{sec:schottky_intro}.}
	\label{fig:genusone}
\end{figure}
Two different elliptic moduli $\tau_1$ and $\tau_2$ arise from the same Riemann surface, if they are related by a modular transformation
\begin{equation}
\tau_2=\frac{a\tau_1+b}{c\tau_1+d}\,,\quad\bigg(\begin{matrix}a & b\\ c & d\end{matrix}\bigg)\in\mathrm{SL}(2,\zZ)=\mathrm{Sp}(2,\zZ)\,,
\end{equation}
corresponding to a different choice of \Atxt- and \Btxt-cycles and evaluating the newly normalized holomorphic differential.

A compact genus-one Riemann surface can be viewed in different ways, as exemplified in \figref{fig:genusone}: 
\begin{enumerate}
	\item[(a)] \textbf{Complex manifold.} A compact genus-one Riemann surface is a one-dimensional compact complex manifold with fundamental group isomorphic to $\Integers^2$. To evaluate function-theoretic objects in this setting, one must specify a complex atlas, and we describe possible ways to do so in (b), (c) and (d).
	\item[(b)]  \textbf{Elliptic curve.} Any compact genus-one Riemann surface can be viewed as the (compactified) set of complex solution of an affine equation of the form 
	\begin{equation}
		\label{eqn:ellcurve}
		y^2=P_4(x)=\alpha\prod\limits_{i=1}^4(x-a_i)\,,
	\end{equation}
	i.e.~an elliptic curve. For applications, this algebraic viewpoint yields a more direct connection between Feynman integrals and elliptic polylogarithms.
	\item[(c)] \textbf{Complex torus.} Every compact genus-one Riemann surface can be viewed as a (one-dimensional) complex torus (i.e.~the complex manifold given by the quotient $\mathbb C/(\lambda_1\mathbb Z+\lambda_2\mathbb Z)$, for $\lambda_1,\lambda_2$ such that $\lambda_2/\lambda_1\notin \mathbb R$), for two very different reasons. The first is that every genus-$\genus$ compact Riemann surface is embedded via Abel's map into its Jacobian variety, which is an $\genus$-dimensional complex torus, and this map is an isomorphism at genus one. The second is that a genus-one compact Riemann surface has fundamental group isomorphic to $\mathbb Z^2$ and universal cover isomorphic to~$\mathbb C$, hence by the general theory of covering maps it must always be isomorphic to such a quotient. The complex numbers $\lambda_1,\lambda_2$ are the periods of~$\Sigma$, and can be obtained as the two integrals of a holomorphic one-form over the \Atxt- and \Btxt-cycle of the surface, respectively. As mentioned, we choose the holomorphic one-form in such a way that $\lambda_1=1$ and $\lambda_2=\tau\in\mathbb H$. This is the language used to describe the construction of the elliptic polylogarithms $\tilde\Gamma$, and so \secref{sec:eMPLs} will be phrased in this setting.
	
	\item[(d)] \textbf{Schottky cover uniformization.} In \secref{sec:schottky_intro}, we will describe the torus as a quotient of its Schottky cover in preparation for our discussion of higher genera. Essentially, in the Schottky language a torus is described as a sphere with a pair of circles on it that are identified with each other, thus generating a handle. 
\end{enumerate}

\subsubsection{Elliptic polylogarithms and the Kronecker form at genus one}\label{sec:eMPLs}

Now, we review the generalization of the genus-zero polylogarithms \eqref{eqn:Goncharov} for the elliptic curves. Following the logic of the classical polylogarithms, one could naively define the generalizations as the iterated integrals of the genus-one analogues of the third kind differentials \eqref{eqn:genuszerodiff}, which are meromorphic one-forms with at most simple poles. Unfortunately, unlike the genus-zero case, these differentials do not span a basis of the first de Rham cohomology of the Riemann surface minus their poles; it therefore follows from \cite{EZ2} that they cannot generate a space which is closed under taking primitives. One way to solve this problem and define elliptic polylogarithms is to make use of multivalued integration kernels, which can be defined by giving a formula for their generating series that involves the \emph{Kronecker function}
\begin{equation}
	\label{eqn:Kroneckergenusone}
	F(\xi,\alpha|\tau)=\frac{\theta'(0|\tau)\theta(\xi+\alpha|\tau)}{\theta(\xi|\tau)\theta(\alpha|\tau)} \,,
\end{equation}
where $\theta(\xi| \tau)$ is the only genus-one odd theta function. This is a classical function, introduced by Kronecker and further studied by Zagier \cite{Weil, ZagierJacobi}. It is meromorphic and symmetric in the two variables $\xi,\alpha\in\mathbb C$, with simple poles placed at all lattice points; the (holomorphic) dependence on the lattice is encoded by the dependence on $\tau\in \mathbb H$. It is not periodic, i.e.~single-valued on the torus, but only quasiperiodic\footnote{Since the quasiperiodicity is only in the direction of the \Btxt-cycle, it is single-valued on any cover that distinguishes between points that are separated by \Btxt-cycles, e.g.~a universal cover or a Schottky cover.}:
\begin{equation}\label{eqn:KroneckerGenusOneQuasiperiodicity}
	F(\xi+1,\alpha|\tau) = F(\xi,\alpha|\tau)\,, \quad F(\xi+\tau,\alpha|\tau) = e^{- 2 \pi \iunit \alpha} F(\xi,\alpha|\tau)\,.
\end{equation}
Because of the \Atxt-cycle periodicity in the variables $\xi,\alpha$, as well as the periodicity w.r.t.~$\tau\to \tau+1$, $F$ admits a triple Fourier expansion in an appropriate region of convergence, given by (see \cite{ZagierJacobi, BrownLevin})
\begin{align}
\label{eqn:kronecker_g1_z_w}
F(\xi, \alpha | \tau)=-2\pi \iunit\left(\frac{z}{1-z}-\frac{w}{1-w}+\sum_{m, n>0}\left(z^m w^{-n}-z^{-m} w^{n}\right) q^{m n}\right), 
\end{align}
where\footnote{The convention used here is different from that of \rcite{BrownLevin}, where $w_{\mathrm{BL}} = \exp(2\pi \iunit \alpha)$ is used. With our convention, the \Btxt-cycle quasiperiodicity in \eqn{eqn:KroneckerGenusOneQuasiperiodicity} reads $F(\xi+\tau,\alpha| \tau)=w \,F(\xi,\alpha| \tau)$.} 
\begin{equation}\label{eqn:Schottky_conventions}
	z \coloneq \exp(2\pi \iunit \xi)\,,\quad w \coloneq \exp(-2\pi \iunit \alpha)\,,\quad q \coloneq \exp(2\pi \iunit \tau)\,.
\end{equation}
The representation in \eqn{eqn:kronecker_g1_z_w}  is related to the Schottky parametrization to be introduced below, and will serve as a starting point for the generalization of the Kronecker function to higher genera, leading to a convenient approach in order to construct higher-genus polylogarithms, as explained in~\secref{sec:schottky_kronecker}.

A remarkable property of the Kronecker function is that it satisfies the \emph{Fay identity} 
\begin{equation}
	\label{eqn:KroneckerFayGenusOne}
	F(\xi_1, \alpha_1| \tau)\, F(\xi_2, \alpha_2| \tau)=F(\xi_1, \alpha_1\,{+}\,\alpha_2| \tau)\, F(\xi_2\,{-}\,\xi_1, \alpha_2| \tau)+F(\xi_2, \alpha_1\,{+}\,\alpha_2| \tau)\, F(\xi_1\,{-}\,\xi_2, \alpha_1| \tau)\,.
\end{equation}
The above identity is the simplest instance of the Fay trisecant equation, which is a relation between Riemann $\theta$-functions defined for every genus \genus{}. 

The Kronecker function can be lifted to a function $\Omega(\xi, \alpha| \tau)$ which is doubly-periodic in the variable $\xi$ and for which the Fay identity \eqref{eqn:KroneckerFayGenusOne} remains valid, by introducing a non-holomorphic prefactor:
\begin{equation}
	\label{eqn:gen1-periodic}
	\Omega(\xi, \alpha|\tau) \coloneq \exp \left(2\pi \iunit \alpha \frac{\operatorname{Im}(\xi)}{\operatorname{Im}(\tau)}\right) F(\xi, \alpha| \tau)\,.
\end{equation}

We discuss now how the Kronecker function serves as a generating function for integration kernels on the complex torus. Since $\theta(\alpha| \tau)$ has a simple zero at the origin, the function $\alpha\, F(\xi,\alpha| \tau)$ has a Taylor expansion at $\alpha=0$. We define a family of differentials $\gkern{n}(\xi| \tau)\,\dd \xi$, for $n\geq 0$, that will serve as integration kernels, out of this Taylor expansion:
\begin{equation}
	\label{eqn:genus1expansion}
  \alpha \, F(\xi, \alpha| \tau)\,\dd \xi \eqdef \sum_{n=0}^{\infty} \gkern{n}(\xi| \tau)\,\dd \xi\, \alpha^n\,.
\end{equation}
This generating function is the object which will be generalized in \secref{sec:schottky_kronecker} below. The integration kernels inherit the periodicity over \Atxt-cycles and quasiperiodicity over \Btxt-cycles from the Kronecker function (see \eqn{eqn:KroneckerGenusOneQuasiperiodicity}):
\begin{equation}
  \gkern{n}(\xi+1| \tau)=\gkern{n}(\xi| \tau)\,,\quad \gkern{n}(\xi+\tau| \tau) = \sum_{k=0}^n (-2\pi \iunit)^{k} \gkern{n-k}(\xi| \tau)\,.
\end{equation}
They are meromorphic, with possible simple poles at the lattice points and holomorphic elsewhere; in particular, the only differential which is not holomorphic at the origin is $g^{(1)}$, which has a simple pole at $\xi=0$ with residue $1$.

If one desires periodic (i.e.~single-valued) integration kernels, the periodic version of Kronecker function \eqref{eqn:gen1-periodic} can be similarly Taylor expanded to get a family of differentials $f^{(n)}(\xi| \tau)\,\dd \xi$ (which are however not holomorphic, but simply smooth), as discussed in refs.~\cite{BrownLevin,Broedel:2014vla,Broedel:2017kkb,Matthes:Thesis}.

The kernels $\gkern{n}$ in \eqn{eqn:genus1expansion} can, by definition, be written in terms of $\theta$ and its derivatives, but they also have a Fourier expansion inherited from \eqn{eqn:kronecker_g1_z_w} (see e.g.~\cite{Broedel:2014vla}); for example
\begin{align}\label{eqn:genusonekernels}
	\gkern{0}(\xi| \tau)=1\,,\quad\gkern{1}(\xi| \tau)=\frac{\theta'(\xi| \tau)}{\theta(\xi| \tau)}=\pi\cot(\pi \xi)+4\pi\sum_{m=1}^\infty\sin(2\pi m\xi)\sum_{n=1}^\infty q^{mn}\,.
\end{align}
Our numerical evaluations in \secref{sec:practicabilities} are based on the Fourier series of the genus-one kernels.

Elliptic polylogarithms can be defined as iterated integrals\footnote{These iterated integrals are not defined directly on the (punctured) complex torus, because the integration kernels are not single-valued, but rather on its covering given by the complex plane minus the points $a_i$. The integrals are then well-defined, because the meromorphicity of the differentials ensures homotopy invariance, and they define multivalued functions $\tilde\Gamma$ of $\xi\in\mathbb C$.} of the kernels $\gkern{n}(\xi|\tau)\,\dd \xi$ in the following way:  
\begin{equation}
	\label{eqn:ellpoly}
	\Gtargxit{n_1,\ldots,n_r}{a_1,\ldots,a_r}\defeq \int_0^\xi \dd t\, \gkern{n_1}\left(t-a_1|\tau\right) 
	\Gtargtt{n_2,\ldots,n_r}{a_2,\ldots,a_r},\quad \Gtargxit{}{}=1\,,
\end{equation}
where $a_1,\ldots , a_r\in\mathbb C$ and $n_1,\ldots ,n_r\geq 0$. In this definition, the sum $\sum_j n_j$ is called the weight and $r$ is called the length of a given elliptic polylogarithm. The notation $\tilde\Gamma$ was born to distinguish them from the iterated integrals $\Gamma$ of the doubly-periodic but non-holomorphic kernels $f^{(n)}(\xi|\tau)\,\dd \xi$.

The above elliptic polylogarithms have been subject to numerous investigations during the last few years. Let us list and comment on some of the properties:  
\begin{itemize}
	\item \textbf{Regularization.} Similarly to the genus-zero case, because of the simple pole at $\xi=0$ of the kernels $g^{(1)}$, the integrals in \eqn{eqn:ellpoly} diverge as soon as $(n_1,a_1)=(1,0)$. Elliptic polylogarithms are therefore also defined by regularizing the integral, either by the shuffle regularization or, equivalently, by the tangential basepoint regularization \cite{EZ3}. 
	\item \textbf{Algebraic structure.} Elliptic polylogarithms generate an algebra whose structure was completely described in \cite{EZ3} in terms of a shuffle algebra. In particular, it was shown that elliptic polylogarithms are linearly independent, as the only relations that they satisfy are the shuffle relations arising from their iterated integral nature. The calculation of symbols of elliptic polylogarithms and the construction of a coaction as discussed in \rcite{Broedel:2018iwv} should be related to the Hopf algebra structure inherited from this shuffle algebra description.
	\item \textbf{Special values.} In the same way as multiple zeta values are the special values defined as genus-zero polylogarithms evaluated at $1$, elliptic multiple zeta values are defined as elliptic polylogarithms evaluated at\footnote{These are actually called \Atxt-elliptic MZVs, because they describe the monodromy of the KZB connection along the \Atxt-cycle. One may also consider the special values at $\tau$, which are called \Btxt-elliptic MZVs, which are related to \Atxt-elliptic MZVs by a modular transformation.}~$1$ \cite{Enriquez:Emzv,Broedel:2014vla,Broedel:2015hia,Matthes:Thesis,ZerbiniThesis}. They are functions of the modular parameter $\tau$, and their modular dependence can be inferred from the fact that they can also be written as linear combinations of iterated integrals of Eisenstein series \cite{Enriquez:Emzv}.
	\item \textbf{Primitives.} Elliptic polylogarithms generate an algebra which is closed under taking primitives \cite{Broedel:2017kkb, EZ3}, similarly to their genus-zero counterparts, leading to the same benefits when evaluating one-loop string amplitudes or Feynman integrals.  Closure under taking primitives can be viewed as a fundamental property that we require for higher-genus generalizations of the polylogarithms \cite{EZ2}, and it is expected to hold for the iterated integrals generated by Enriquez' connection from \secref{ssec:Enriquezconn}. For this reason, when we generalize the Kronecker function to higher genera we will impose that it matches the requirements which uniquely determine Enriquez' connection.
\end{itemize}
%


\section{Schottky parametrization of higher-genus Riemann surfaces}\label{sec:higherlanguages}
In this section we introduce the background on higher-genus Riemann surfaces needed in the rest of the article. After a general introduction, we review a specific parametrization of the Riemann surface, known as the Schottky parametrization, which will be exploited in \secref{sec:schottky_kronecker} to construct higher-genus analogues of the genus-one integration kernels from the previous section.


\subsection{Basics of higher-genus Riemann surfaces}\label{sec:basicsRiemannsurf}
\begin{figure}
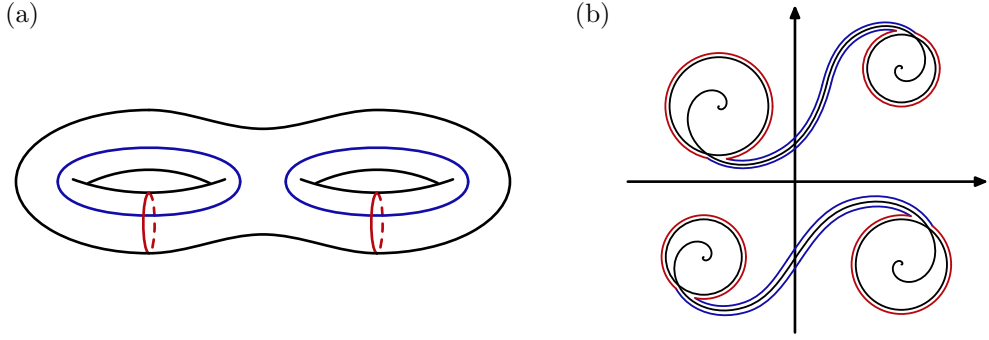

	\centering
	\mpostuse{genustwo}
	\caption{A Riemann surface of genus two with its \Atxt-cycles (red) and \Btxt-cycles (blue) represented by (a) a double torus and (b) a corresponding Schottky cover.}
	\label{fig:genustwo}
\end{figure}

The material of this section is standard, and our brief overview is simply meant to set the notation which will be used throughout the next sections. A more detailed treatment can be found for example in \cite{Farkas1992}.
For a compact Riemann surface $\RSurf$ of genus $\genus{}$, let 
\begin{equation}
  \label{eqn:homologybasis}
  \acyc_1,\bcyc_1,\ldots,\acyc_g,\bcyc_g
\end{equation}
be loops which generate the fundamental group of~$\RSurf$, such that the corresponding homology cycles satisfy intersection relations
\begin{equation}
	\acyc_i \# \acyc_j = 0 = \bcyc_i \# \bcyc_j\,, \qquad  \acyc_i \# \bcyc_j = \delta_{ij}\,,
\end{equation}
which make them into a symplectic basis of the first cohomology group, and which may be intuitively interpreted by having each $\bcyc_i$ wrap around a hole of the compact Riemann surface, and $\acyc_i$ wrapping around each handle. For a genus-two surface this is depicted in \figref{fig:genustwo}, showing the cycles on the double torus (as well as a genus-two Schottky parametrization, as discussed in \secref{sec:schottky_intro}).

The space of holomorphic differentials on the compact Riemann surface is of dimension $\genus{}$, for which we select a basis normalized by
\begin{equation}\label{eqn:holomorphic-basis}
	\oint_{\acyc_i} \omega_j = \delta_{ij}\,, \qquad \oint_{\bcyc_i} \omega_j = \tau_{ij}\,.
\end{equation}
where the \emph{period matrix} $\tau\coloneq (\tau_{ij})_{i,j}$, uniquely determined by the first condition for the \Atxt-cycle integrals in \eqn{eqn:holomorphic-basis}, is symmetric with positive definite imaginary part and contains the complex-geometric information of the surface. Two different period matrices $\tau$ and $\tau'$ describe the same Riemann surface if they are related by a modular transformation
\begin{equation}
	\tau' = (a \tau + b)(c \tau + d)^{-1}, \quad \begin{pmatrix} a & b \\ c & d \end{pmatrix} \in \mathrm{Sp}(2\genus{},\mathbb Z)\,.
\end{equation}
This transformation is a result of choosing a different symplectic basis of \Atxt- and \Btxt-cycles,
\begin{equation}
	\begin{pmatrix} \vec \bcyc' \\ \vec \acyc' \end{pmatrix}=\begin{pmatrix} a & b \\ c & d \end{pmatrix}\begin{pmatrix} \vec \bcyc \\ \vec \acyc \end{pmatrix} ,
\end{equation}
with correspondingly normalized holomorphic differentials $\vec \omega' = \vec \omega (c \tau + d)^{-1}$.

The Jacobian variety $J(\RSurf)$ is then defined by taking a quotient of $\Complex^\genus$ by the lattice 
generated by the \Atxt- and \Btxt-cycle periods of the holomorphic forms:
\begin{equation}
	J(\RSurf) \defeq \Complex^\genus / (\Integers^\genus + \periodmatrix \Integers^\genus)\,.
\end{equation}

Using the basis of holomorphic forms satisfying \eqn{eqn:holomorphic-basis} one can define \emph{Abel's map} $\abel$ from two copies of the universal cover of a genus-\genus{} Riemann surface to $\Complex^\genus$ as
\begin{equation}\label{eqn:abel-universal}
	\abel_i(p, p_0 | \periodmatrix) \defeq \int_{p_0}^p \omega_i\,,
\end{equation}
which restricts to a well-defined map from two copies of the Riemann surface to the Jacobian variety.
%

\subsection{Schottky parametrization} \label{sec:schottky_intro}

In this section we describe the Schottky parametrization \cite{Schottky1887} of compact Riemann surfaces. This approach is very geometrical: the starting point is the Riemann sphere $\overline{\zC}=\zC\cup\{\infty\}$, and each additional genus is represented by identifying a pair of disjoint Jordan curves\footnote{A Jordan curve is a simple (i.e.~not self-intersecting) closed curve in the plane. Jordan curves are homeomorphic to the unit circle.} in $\overline{\zC}$, which can be imagined as the ends of a handle spanning in between the circles. For a detailed review we refer to \rcite{Bobenko:2011,herrlich1schottky}. 

Let us start with some definitions. Let $C_1, \ldots, C_\genus, C_1^{\prime}, \ldots, C_\genus^{\prime}$ be a set of $2\genus$ mutually disjoint Jordan curves on a Riemann sphere $\overline{\mathbb{C}}$. The \emph{Schottky group} $\SGroup$ is defined as a Kleinian group, (i.e.~a discrete subgroup of $\grp{PSL}(2, \Complex)$) generated by $\genus$ \textit{loxodromic transformations} $\gamma_1,\ldots ,\gamma_\genus$ of the sphere that map the exterior of $C_i$ to the interior of $C_i^{\prime}$. We will call them \emph{Schottky generators}.

As for any loxodromic transformation, each Schottky generator $\gamma_i$ is parametrized by its two ``fixed points'', which are defined as $P_i\,{\coloneq} \lim_{n\to\infty} \gamma_i^{n}z$ and $P_i'\,{\coloneq} \lim_{n\to\infty} \gamma_i^{-n}z$ for some $z$ not being a fixed point, and by a multiplier parameter~$\lambda_i$ of absolute value $\ne 1$ related to the trace of $\gamma_i$ and adhering to the relation 
\begin{equation}
	\frac{\gamma_iz-P_i'}{\gamma_iz-P_i}=\lambda_i\frac{z-P_i'}{z-P_i}.
\end{equation}
Taking into account the equivalence of different Schottky groups related by a M\oe{}bius transformation, i.e.~an element of $\mathrm{PSL}(2,\mathbb{C})$, in total we have $3 \genus - 3$ parameters, which precisely matches the dimension of the moduli space of genus-$\genus$ compact Riemann surfaces.

The $2\genus$ curves $\lbrace C_i,C_i' \rbrace_{i=1}^h$ bound a connected region of the Riemann sphere denoted as $\funddom$ The union of the images of $\funddom$ under the Schottky group
\begin{equation}
  \Omega(\SGroup)\coloneq\bigcup_{\gamma \in \SGroup} \gamma(\funddom)
\end{equation}
is the \textit{region of discontinuity} of the action of $\SGroup$ on the sphere, and $\funddom$ is its \textit{fundamental domain}. The \emph{Schottky uniformization theorem} states that for any Riemann surface $\Sigma$ there exists a Schottky group $\SGroup$ such that 
\begin{equation}
	\Sigma = \Omega(\SGroup) / \SGroup\,.
\end{equation}
In other words, $ \Omega(\SGroup) \rightarrow \Sigma$ is an unramified cover of the Riemann surface $\Sigma$, and $\SGroup$ is the cover automorphism group, and these data provide a uniformization of the surface.  
In the following, the terms Schottky uniformization and Schottky parametrization are going to be used interchangeably. 

Evidently, the Schottky uniformization for a given Riemann surface is not unique: it is always possible to apply a M\oe{}bius transformation to the sphere and conjugate the Schottky generators by this transformation in order to obtain an equivalent uniformization.
The M\oe{}bius invariance will prove to be an important notion later in this article.

Although in principle there always exists a Schottky uniformization for any Riemann surface, in practice it is an open problem to find a Schottky group that uniformizes a given Riemann surface (called the inverse uniformization problem). Moreover, it is unknown if every Schottky group gives rise to a Riemann surface.

There exists, however, a subclass of the Schottky groups called \textit{classical Schottky groups}, for which one makes the specific choice for the Jordan curves $C_i, C_i^{\prime}$ to be circles. In what follows, we solely focus on this type of Schottky groups. Although it is not known whether any Riemann surface can be uniformized by a classical Schottky group, at least for real hyperelliptic curves\footnote{A hyperelliptic curve is a complex projective curve whose affine equation can be given in the form $y^2=P(x)$, with $P$ a polynomial of degree $d\geq 3$. Its complex points define a compact Riemann surface of genus $\lceil d/2 \rceil - 1$.
It is called real when the coefficients of $P$ are real.} this statement holds \cite{Seppala:2004}. Moreover, every classical Schottky group corresponds to a Riemann surface and particular statements about the convergence of the Poincar\'e series defined on the surface can be made (cf.~\secref{sec:Schottky_Poincare_series}).

Given a pair $\{C,C'\}$ of circles with radii $r_{C}$ and $r_{C^{\prime}}$ and centers $p_C$ and $p_{C^{\prime}}$, respectively, one can construct the corresponding M\oe{}bius transformation $\gamma_i$ by inverting at $C$, rescaling and successively shifting to $C^{\prime}$, i.e.
\begin{equation}
\gamma: z \longmapsto z-p_C \longmapsto \frac{-1}{z-p_C} \longmapsto \frac{-r_C r_{C^{\prime}}}{z-p_{C^{\prime}}} \longmapsto \frac{-r_C r_{C^{\prime}}}{z-p_C}+p_{C^{\prime}}\,.
\end{equation}

In the language of the Schottky uniformization it is easy to assign \Atxt- and \Btxt-cycles given a pair of circles. Conventionally, one takes the \Atxt-cycle to be the pair of circles (remember that they are identified on the Schottky cover) and the corresponding \Btxt-cycle can then be chosen along any line connecting the two boundaries. \figref{fig:genusone} (d) depicts this setup of \Atxt- and \Btxt-cycles for a genus-one surface Schottky uniformization, where the loxodrome is associated to the \Btxt-cycle. Similarly, \figref{fig:genustwo} (b) depicts a genus-two Schottky uniformization.

\subsubsection{Poincar\'e series}\label{sec:Schottky_Poincare_series}
In the language of the Schottky uniformization various objects on the underlying Riemann surface are expressed in terms of automorphic forms, which can be written as Poincar\'e series over (cosets of) the Schottky group. Classically, a Poincar\'e series of dimension $-2k$ is a series
\begin{equation} \label{eqn:poincare-series}
	\theta_{-2k}(z) = \sum_{\gamma \in \SGroup^*} (\gamma'(z))^k \psi(\gamma z )\,,
\end{equation}
where $\psi(z)$ is a meromorphic function on a complex domain $D$ and $\SGroup^*$ is an appropriate coset of a group~$\SGroup$ acting on $D$ such that the summation is performed over distinct terms only and $\psi(z)$ does not have poles in the \emph{singular set} of $\SGroup^*$, i.e.~the closure of the fixed points of all elements of $\SGroup^*$ excluding the identity.

Let us introduce, for $i\in\{1,\ldots ,h\}$, the subgroups $\SGroup_i\coloneq\langle\gamma_i\rangle\leq \SGroup$, where $\gamma_i$ are the Schottky generators of the Schottky group $\SGroup$, and consider the right and left cosets
\begin{align}\label{eqn:cosetdefinition}
    \SGroup \ / \ \SGroup_i = \{ \gamma_{j_1}^{n_1} \cdots \gamma_{j_k}^{n_k} : \gamma_{j_k} \neq \gamma_i \}\,, \nonumber\\
    \SGroup_i \setminus \SGroup = \{ \gamma_{j_1}^{n_1} \cdots \gamma_{j_k}^{n_k} : \gamma_{j_1} \neq \gamma_i \}\,.
\end{align}
Then, the holomorphic differentials of a Riemann surface can be expressed on the Schottky cover with Schottky group $\SGroup$ as the $(-2)$-dimensional Poincar\'e series 
\begin{align}\label{eqn:schottky-holomorphic-basis}
	\omega_i(z|\SGroup) 
	&= \frac{1}{2\pi \iunit}\sum_{\gamma \in \SGroup / \SGroup_i} \left(\frac{1}{z-\gamma P_i'} - \frac{1}{z-\gamma P_i}\right) \der z 
	\nonumber\\
	&= \frac{1}{2\pi \iunit}\sum_{\gamma \in \SGroup_i \setminus \SGroup } \left(\frac{1}{\gamma z-P_i'} - \frac{1}{\gamma z-P_i}\right) \der (\gamma z)\,,
\end{align}
where $P_i$ and $P_i'$ are the fixed points\footnote{Notice that due to the choice of the coset $\SGroup_i \setminus \SGroup$ the poles at $P_i$ and $P_i'$ are not problematic.} defined earlier, provided that the defining Poincar\'e series are absolutely convergent\footnote{For a generic Schottky group $\SGroup$ this series might be divergent. We will comment on this issue at the end of the section.}.
One way to recognize these differentials intuitively is by recalling the normalization conditions they must satisfy when integrating along the \Atxt-cycles. The circle corresponding to \Atxt-cycle $\acyc_j$ contains the fixed point $P_j$, as well as both $\gamma P_k$ and $\gamma P_k'$ for $\gamma = \gamma_j \cdots$. Integrating along the circle, we can use the Cauchy residue theorem, where the residues from $\gamma P_k$ and $\gamma P_k'$ cancel each other, while the only contribution that does not cancel out when integrating $\omega_j$ is from the pole at $P_j$, giving the desired unit normalization $\int_{\acyc_j}\omega_k=\delta_{jk}$. Additionally, one may notice that the genus-\genus{} holomorphic differentials from \eqref{eqn:schottky-holomorphic-basis} can be written as 
\begin{equation} \label{eqn:decomp_hol_diff}
	\omega_i(z| \SGroup) = \sum_{\gamma \in \SGroup_i \setminus \SGroup } \omega(\gamma z, \SGroup_i),
\end{equation}
where $\omega(z|\SGroup_i)$ is the only holomorphic differential of the genus-one surface\footnote{In the genus-one case the coset over which the summation is performed in \eqref{eqn:schottky-holomorphic-basis} only consists of the identity element.}
corresponding to the Schottky group $\SGroup_i$.

Thanks to this representation of the holomorphic differentials, Abel's map can be calculated on a pair of points $z$, $z_0$ on the Schottky cover by\footnote{We introduce this special notation for Abel's map on the Schottky cover in order to distinguish from the original definition on the universal cover in \eqn{eqn:abel-universal}.} 

\begin{equation}\label{eq:abel-map-schottky}
	\abel_i(z,z_0|\SGroup) = \int_{z_0}^z \omega_i = \frac{1}{2\pi \iunit} \sum_{\gamma \in \SGroup/\SGroup_i} \log \{z,\gamma P_i',z_0,\gamma P_i\}\,,
\end{equation}
where the definition of the cross-ratio 
\begin{equation}
	\{z_1,z_2,z_3,z_4\} = \frac{(z_1-z_2)(z_3-z_4)}{(z_1-z_4)(z_3-z_2)}	
\end{equation}
was used. Moreover, the corresponding period matrix, which we will denote $\tau(G)$, have entries
\begin{equation} \label{eq:period-matrix-schottky}
	\periodmatrix_{ij}(\SGroup) = \int_{\bcyc_j}\omega_i= \delta_{ij}\log\lambda_i+\sum_{\gamma\in\SGroup_j\backslash\SGroup/\SGroup_i}\log\{P_j', \gamma P_i', P_j, \gamma P_i\}\,.
\end{equation}
The invariance of Abel's map and of the period matrix under a M\oe{}bius transformation $\sigma=\left(\begin{smallmatrix}a&b\\c&d\end{smallmatrix}\right)\in\mathrm{PSL}(2,\mathbb{C})$ -- where we transform each point as $z \mapsto \sigma z$ and elements of the group as $\gamma \mapsto \sigma \gamma \sigma^{-1}$ -- directly follows from the corresponding property of the cross-ratio:
\begin{equation}
    \{\sigma z_1,\sigma z_2,\sigma z_3,\sigma z_4\} = \{z_1,z_2,z_3,z_4\} \implies \abel_i(\sigma z, \sigma z_0|\sigma \SGroup \sigma^{-1}) = \abel_i(z, z_0|\SGroup)\,.
\end{equation}

Similarly, the Poincar\'e series for the holomorphic differentials also exhibit M\oe{}bius invariance. In order to see this, one can study the fundamental building blocks and observe how they transform when we apply $\sigma$: 
\begin{equation} \label{eqn:transformation_properties}
\begin{aligned}
    \der(\sigma z) &= (cz+d)^{-2} \der z \,, \\
    (\sigma z - \sigma x)^n &= (cz+d)^{-n}(cx+d)^{-n} (z-x)^n.
\end{aligned}
\end{equation}
With these properties in mind, it can be verified that each term in the expansion of $\omega_i(z|\SGroup)$ is M\oe{}bius invariant. 
In what follows, we write the Kronecker function as a Poincar\'e series over the Schottky group and generalize this approach; in particular, we will require M\oe{}bius invariance, and we will use analogous M\oe{}bius invariant terms to obtain the desired independence from the choice of a Schottky cover.

Generally, the convergence of a Poincar\'e series over a Schottky group is not guaranteed, and the expression for the holomorphic differential \eqref{eqn:schottky-holomorphic-basis} is not always well-defined. However, as long as we restrict to classical Schottky groups, particular statements about the convergence of a $(-2)$-dimensional Poincar\'e series can be made. Specifically, the series turns out to be convergent if the configuration of the circles on the sphere is so-called \textit{circle decomposable} (also known as the Schottky criterion \cite{Bobenko:2011}). By means of the explicit construction (see \appref{sec:curve_to_schottky}) for a real hyperelliptic curve one can always find a classical circle-decomposable Schottky group such that the Poincar\'e series is convergent.

\subsubsection{Concentric Schottky cover}\label{sec:concentricSchottky}

\begin{figure}
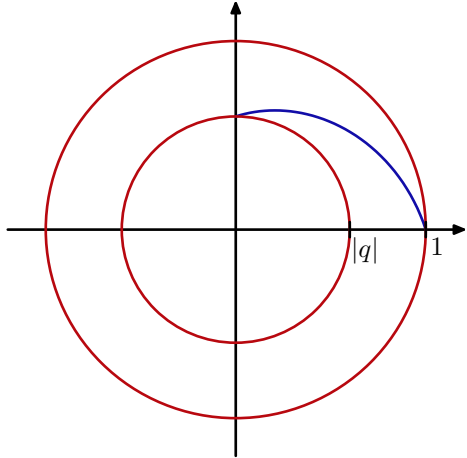

	\begin{center}
		\mpostuse{concentricschottky}
	\end{center}
	\caption{Schottky uniformization of a genus one Riemann surface, chosen such that the fixed points are at $0$ and $\infty$, yielding concentric circles. The red circles correspond to \Atxt-cycles, with a \Btxt-cycle connecting them drawn in blue.}
	\label{fig:concentricschottky}
\end{figure}

A useful example of a Schottky cover is the concentric Schottky cover at genus one. Using the M\oe{}bius invariance of the Schottky cover, we can choose to fix $P_1 = 0$ and $P_1' = \infty$, leaving us with the generator $\gamma_1 : z \mapsto qz$, with $q\in\mathbb C$, which corresponds to the element $\left(\begin{smallmatrix}\sqrt{q}&0\\0&1/\sqrt{q}\end{smallmatrix}\right)\in\mathrm{PSL}(2,\mathbb{C})$. The corresponding Schottky group will be denoted by $G_{\mathrm c}\coloneq \langle\gamma_1\rangle\simeq\mathbb Z$. The \Atxt-cycle of the underlying Riemann surface $\Omega(G_{\mathrm c})/G_{\mathrm c}$ can then be identified with the pair of circles of radii $1$ and~$|q|$, and the \Btxt-cycle connects the points $1$ and $q$; a fundamental domain isomorphic to the surface is the annulus bounded by the circles of radius $|q|$ and $1$, whereas the Schottky cover $\Omega(G_{\mathrm c})$ is $\mathbb C\setminus\{0\}$. The modulus $\tau(G_{\mathrm c})$ of the underlying Riemann surface, which can be computed using \eqn{eq:period-matrix-schottky} or through a direct integration, is such that $q=\exp(2\pi \iunit\tau)$. We refer the reader to \figref{fig:concentricschottky} for a picture. 

The normalized holomorphic differential on this Schottky cover is simply $\omega_1(z| G_{\mathrm c}) = \frac{\dd z}{2\pi \iunit z}$. One can pass from this Schottky cover to the torus, i.e.~the Jacobian variety, by exponentiation, writing $z = \exp(2\pi \iunit \xi)$ as in \eqn{eqn:Schottky_conventions}.
As a consequence, the addition of points we used on the genus one Jacobian variety corresponds to the multiplication of coordinates on the concentric Schottky cover. These properties serve as a starting point for translating the Kronecker function to Schottky covers for its generalization in \secref{sec:schottky_kronecker_def}, and is also useful as an alternative method for proving the Fay identity for the genus-one Kronecker function as shown later in \eqn{eqn:schottky-fay-proof}.

Similarly, one may choose to define a concentric Schottky cover for higher genera, choosing some arbitrary pair of fixed points to fix to $0$ and $\infty$. However, due to the presence of other generators in the Schottky group, choosing a concentric Schottky cover at higher genus does not lead to the same convenient properties.

\section{Higher-genus generalizations of the KZ connection}
\label{sec:genustworeview}
Homotopy invariant iterated 
integrals on a manifold can be systematically constructed out of a flat connection on a principal $\cG$-bundle over the manifold, for $\cG$ a Lie group. More specifically, suppose that one can write the connection as $\der+\conn$, with $K$ a differential one-form, valued in the Lie algebra of $\cG$, which satisfies the integrability condition (equivalent to the flatness of the connection)
\begin{equation}\label{eq:integrability}
  \der \conn + \conn \wedge \conn = 0.
\end{equation}
We will call $K$ a \emph{connection form}\footnote{In the literature, $K$ is also called a \emph{Maurer–Cartan element} \cite{EZ1, EZ2}.}. Eq.~\eqref{eq:integrability} implies that iterated integrals of $K$ over some path only depend on the endpoints of the path and on its homotopy class. This implies that the $\cG$-valued \emph{path-ordered exponential integral}
\begin{equation}
  \mathrm{P} \exp \int_{z_0}^z \conn\,\coloneq\, 1+\int_{z_0}^z \conn(t)+\int_{z_0}^z \conn(t_1)\int_{z_0}^{t_1}\conn(t_2)+\cdots , 
\end{equation}
where the (fixed) integration basepoint $z_0$ and the variable $z$ live in the universal cover, is well-defined, and serves as a generating series of homotopy invariant iterated integrals \cite{Hain:1985}. 

In this article, we are only interested in the case where the manifold is a Riemann surface\footnote{Multivariable polylogarithms can also be generalized by considering flat connections on configuration spaces of points on higher-genus Riemann surfaces, for example along the lines of \cite{EnriquezHigher,EZ1}.}, in which case homotopy invariant iterated integrals correspond to higher-genus analogues of the classical genus-zero polylogarithms, which arise from the path-ordered exponential integral of the Knizhnik–Zamolodchikov (KZ) connection. A property of the KZ connection which seems convenient for physics applications and therefore that we would like to hold, for a higher-genus generalization, is to be regular-singular, i.e.~the connection form~$K$ should have at most simple poles. 

Moreover, if the genus is strictly positive, one should notice a subtlety: unlike at genus zero, there is now a trade off between having a single-valued connection form and a holomorphic connection form\footnote{Another possibility, explored in \cite{LevinRacinet} at genus one and in \cite{EZ1, EZ2} at higher-genus, is to allow for higher-order poles, in which case it is possible to obtain a single-valued holomorphic connection form.}.  In the former case (see \cite{BrownLevin, DHoker:2023vax}), the connection is single-valued and smooth but not holomorphic. In the latter case, the connection form is holomorphic but cannot be single-valued: it is defined on some cover of the surface (see \cite{CEE, Broedel:2017kkb, EnriquezHigher, Bernard:1987df}).  In this article we focus on the holomorphic setting, as it seems more suitable for applications in particle physics \cite{Broedel:2017kkb}. At genus one, a connection form in this setting can be constructed (see \cite{CEE}) out of the Kronecker function of \secref{sec:genus-one-polylogs}, which is indeed holomorphic and multivalued. Therefore we formulate the problem of constructing a flat connection in terms of constructing a higher-genus generalization of the Kronecker function. In this section, we review known approaches towards a generalization of the KZ connection, with special focus on the approaches of Bernard and Enriquez which are, as we will see, closely related to the construction proposed in this article.


\subsection{The connection of Bernard} \label{sec:kzb_generalities}

The KZ connection on a genus-zero Riemann surface (sphere) was constructed in the context of Wess–Zumino–Witten models \cite{Knizhnik:1984nr}. This flat connection induces the 
differential equations (Ward identities) for the correlation functions of the model, and the iterated integrals which arise from it are the classical genus-zero polylogarithms.
The generalization of this approach for the model on Riemann surfaces 
of higher genera was achieved by Bernard \cite{Bernard:1987df,Bernard:1988yv}, who realized that, upon twisting affine currents with elements of a Lie group~$\cG$ on 
which the model is defined, it was possible to determine the Ward identities
similar to the KZ equations. The KZ connection in this case is then replaced by the 
KZB connection $\dd+K_{\mathrm{B}}$, where~$K_{\mathrm{B}}$ is defined as a Poincar\'e series over a Schottky group $\SGroup\leq \mathrm{PSL}(2,\mathbb C)$ that uniformizes the underlying surface (see \secref{sec:schottky_intro}), and depends on a twist homomorphism\footnote{More precisely, to obtain Bernard's connection one should apply the morphism $\mathrm{Ad}$ to $K_{\mathrm{B}}$, so that $K_{\mathrm{B}}$ is not valued in $\alg{g} \coloneq \Lie(\cG)$, but rather in the Lie algebra $\mathrm{ad}(\alg{g})$. Accordingly, the corresponding connection is not defined on a $\cG$-bundle, but rather on an $\mathrm{Ad}(\cG)$-bundle.} $g : \SGroup \to \cG$: 
\begin{equation} \label{eqn:Bernard_connection}
K_{\mathrm{B}}(z, x | g)  
= 
\sum_{\gamma \in \SGroup} \frac{\der (\gamma z)}{\gamma z - x} g^{-1}(\gamma).
\end{equation}
This is a one-form in the variable $z$, and a function in the auxiliary point $x$; both $z$ and $x$ live in the Schottky cover of the surface, and~$K_{\mathrm{B}}$ has a simple pole at $z=x$, as well as at every shift of $x$ by a \Btxt-cycle. 

If we specify to the case of genus one, it can be explicitly shown, upon an appropriate choice of the Lie group $\cG$ and of the twist homomorphism, that the connection form $K_{\mathrm{B}}$ can be written in terms of the Kronecker function \eqref{eqn:Kroneckergenusone} (see \secref{sec:schottky_kronecker_def}) and that it essentially coincides with the connection form~$K_{\mathrm{CEE}}$ constructed by Calaque–Enriquez–Etingof\footnote{Levin and Racinet independently introduced the same connection in \cite{LevinRacinet}.} \cite{CEE}. The latter can be used to generate the iterated integrals which define the elliptic polylogarithms $\tilde\Gamma$ of \cite{Broedel:2017kkb}.  


\subsection{The connection of Enriquez}\label{ssec:Enriquezconn}

Enriquez in~\cite{EnriquezHigher} defined a family of multivalued meromorphic differentials $\omega_{i_1\ldots i_sj}(z,x)$ on a compact Riemann surface $\Sigma$, where $s\geq 0$ and $i_1,\ldots ,i_s,j\in\{1,\ldots ,\genus\}$, which are one-forms in the variable $z$ but depend also on an auxiliary point $x$. These differentials have at most a simple pole at $z=x$ and are holomorphic elsewhere; moreover, they are single-valued along the \Atxt-cycles, and are therefore single-valued on the Schottky cover $\Omega(\Sigma)$ of $\Sigma$. We will therefore consider $z$ and $x$ as variables on $\Omega(\Sigma)$. The differential $\omega_j$ is the $j$-th holomorphic differential on $\Sigma$ from \secref{sec:basicsRiemannsurf}. At genus one, $i_1\cdots i_sj$ is just a string of $s+1$ ones, and $\omega_{i_1\ldots i_sj}(z,x)$ can be written in terms of the integration kernels $g^{(s)}$ from \secref{sec:genus-one-polylogs}.

For each $j$ one can consider the generating series
\begin{equation}\label{eq:diffenriq}
K_j(z,x)\coloneq \sum_{s\geq 0}\sum_{i_1,\ldots ,i_s=1}^\genus\omega_{i_1\ldots i_sj}(z,x)\,b_{i_1}\cdots b_{i_s}
\end{equation}
in terms of formal non-commutative variables $b_1,\ldots ,b_h$, which can be assembled to define a differential~$K_{\mathrm{E}}$ given by
\begin{equation}\label{eqn:KEnriquez}
K_{\mathrm{E}}(z,x)\coloneq \sum_{j=1}^\genus K_j(z,x)a_j,
\end{equation}
which is the only meromorphic multivalued differential with at most simple poles at $x$, holomorphic elsewhere, and single-valued along the \Atxt-cycles, which satisfies
\begin{enumerate}[label=(\roman*)]
	\item $\displaystyle K_{\mathrm{E}}(z+\mathfrak{B}_k,x)=e^{b_k}K_{\mathrm{E}}(z,x) \text{ for every }k=1,\ldots ,\genus\,,$\hfill\refstepcounter{equation}\textup{(\theequation)}\label{eqn:connection_constraint1}
	\item $\Res_{z=x}K_{\mathrm{E}}(z,x)=-\sum_{i=1}^\genus b_i a_i/2\pi \iunit\,.$\hfill\refstepcounter{equation}\textup{(\theequation)}\label{eqn:connection_constraint2}
\end{enumerate}

The differential $K_{\mathrm{E}}$ is the main building block of the construction by Enriquez in~\cite{EnriquezHigher} of a holomorphic flat connection on a non-trivial principal bundle over the configuration space ${\rm Conf}_n(\Sigma)$ of~$n$ points on~$\Sigma$, which extends to a meromorphic connection on $\Sigma^n$, with simple poles along the diagonals. Enriquez was motivated by the study of braid groups, which are fundamental groups of configuration spaces of Riemann surfaces, and more precisely of their pro-unipotent completion $\pi_1^{\rm un}({\rm Conf}_n(\Sigma))$, which is isomorphic to the fibre of the above mentioned bundle. Even though the definition of~$K_{\mathrm{E}}$ makes use of an auxiliary point $x$ on $\Sigma$, the flat connection turns out to be independent of~$x$.  
 
 In the genus-one case, Enriquez' connection reduces to the connection $\dd+K_{\mathrm{CEE}}$ from \cite{CEE} which, as discussed in the previous section, is closely related to the genus-one KZB connection. 

Here we are interested in the case of a single Riemann surface, i.e.~$n=1$. In this case, Enriquez' connection is essentially~$\dd+K_{\mathrm{E}}$, provided that one replaces $b_i$ by $\ad(b_i)$, so that~$K_{\mathrm{E}}$ is valued in the degree-completion of the Lie algebra on~$2\genus$ generators $a_1,\ldots ,a_\genus,b_1,\ldots ,b_\genus$, subject to the relation $\sum_{i=1}^\genus[a_i,b_i]=0$, which is the Lie algebra of $\pi_1^{\rm un}(\Sigma)$ (the generators $a_i$, $b_i$ correspond to the \Atxt- and \Btxt-cycles of the torus, respectively). The residue of~$K_{\mathrm{E}}$ at~$x$ has then residue proportional to $\sum_{i=1}^\genus[a_i,b_i]$, which in this Lie algebra equals zero, hence the connection is indeed holomorphic on the whole surface; this is consistent with the fact that Enriquez' connection should be independent of~$x$. 

Thinking of~$K_{\mathrm{E}}$ as a differential form valued in the degree-completion of the free Lie algebra on~$2\genus$ generators, with a true simple pole at~$x$, which is the approach taken here, corresponds geometrically to constructing a connection on a one-punctured Riemann surface, whose fundamental group is freely generated by \Atxt- and \Btxt-cycles.

The original definition of Enriquez of the differentials $\omega_{i_1\ldots i_sj}(z,x)$ is not constructive. Subsequently, recursive formulas which in principle allow to compute these differentials were given in \cite{EZ1}, but do not seem suitable for numerical computations.


\subsection{Other connections}

As previously mentioned, at genus-one the connections of Bernard and Enriquez essentially coincide with the connection $\dd+K_{\mathrm{CEE}}$ constructed by Calaque–Enriquez–Etingof\footnote{Levin and Racinet independently introduced the same connection in \cite{LevinRacinet}.}, where $K_{\mathrm{CEE}}\coloneq \mathrm{ad}(b)F(\xi,\mathrm{ad}(b))(a)\dd \xi$ is built out of the Kronecker function \eqref{eqn:Kroneckergenusone}. Levin–Racinet \cite{LevinRacinet} and Brown–Levin~\cite{BrownLevin} constructed two different gauge transformations $\der+J_{\mathrm{LR}}=g_{\mathrm{LR}}(\der+K_{\mathrm{CEE}})g_{\mathrm{LR}}^{-1}$ and $\der+J_{\mathrm{BL}}=g_{\mathrm{BL}}(\der+K_{\mathrm{CEE}})g_{\mathrm{BL}}^{-1}$ that are flat on the trivial bundle over the curve (whereas $\der+K_{\mathrm{CEE}}$ is a connection on a non-trivial principal bundle, because $K_{\mathrm{CEE}}$ is multi-valued). To do so, one must find a function $g$ such that $gK_{\mathrm{CEE}}$ is single-valued; Levin–Racinet achieved it by allowing for higher-order poles at the puncture, picking $g_{\mathrm{LR}}=\exp(g^{(1)}(z|\tau)\,\mathrm{ad}(b))$, whereas Brown–Levin achieved it by downgrading holomorphicity to smoothness, picking $g_{\mathrm{BL}}=\exp(2\pi \iunit\,\mathrm{Im}(z)/\mathrm{Im}(\tau)\mathrm{ad}(b))$, as already mentioned in \secref{sec:genus-one-polylogs}. The spaces of iterated integrals constructed out of these flat connections are very closely related\footnote{The space generated over the ring of rational functions on the elliptic curve by the iterated integrals of the Levin–Racinet flat connection is closed under taking primitives (this can be deduced from \cite{EZ2}). The spaces generated by the iterated integrals $\tilde\Gamma$ and $\Gamma$ of the Calaque–Enriquez–Etingof and Brown–Levin connection, respectively, coincide with the former space upon adding some extra functions arising from the gauge transformations (see \cite{EZ3} for more details about the $\tilde\Gamma$-space).}.

Both of the above-described constructions have a higher-genus analogue: the connection $\dd+K_{\mathrm{E}}$ of Enriquez was explicitly gauge conjugated by Enriquez–Zerbini \cite{EZ1} to a connection $\dd+J_{\mathrm{EZ}}$, where $J_{\mathrm{EZ}}$ is single-valued with a higher-order pole, thus generalizing Levin–Racinet, whereas the Brown–Levin connection was generalized by D'Hoker–Hidding–Schlotterer \cite{DHoker:2023vax}, who constructed a flat connection $\dd+J_{\mathrm{DHS}}$, with $J_{\mathrm{DHS}}$ single-valued and smooth, whose relation with the connection $\dd+K_{\mathrm{E}}$ of Enriquez is under investigation~\cite{DESZ:WIP}. The construction of these connections is explicit but recursive, and does not seem very suitable for numerical approximation. Similarly to the genus-one situation, all these connections produce iterated integrals which are closely related. In the case of $\dd+J_{\mathrm{EZ}}$, such iterated integrals generate a space over the ring of rational functions on the curve which is closed under taking primitives (this can be deduced from \cite{EZ2}). Moreover, in \cite{EZ2} it was shown that the same space can be generated also by the iterated integrals constructed out of ``simpler'' connections which only involve a finite set of meromorphic single-valued kernels with higher-order poles. All the iterated integrals arising from the above-mentioned connections can be considered as higher-genus analogues of polylogarithms.


\section{Higher-genus Kronecker forms in the Schottky parametrization} \label{sec:schottky_kronecker}

In this section we construct a M\oe{}bius-invariant version of Bernard's connection form~\eqref{eqn:Bernard_connection}.
Our construction, similarly to that of Bernard, makes use of the Schottky parametrization introduced and discussed in \secref{sec:schottky_intro}. More precisely, at genus $\genus{}\geq 1$ we explicitly construct a set of~$\genus{}$ higher-genus analogues of the Kronecker form as Poincar\'e series over elements of a Schottky group (see~\secref{sec:schottky_kronecker_def}), whose properties are explored and discussed in \secref{sec:properties_schottky_kronecker}. 

These Schottky–Kronecker forms serve as generating series of integration kernels, as explained in \secref{sec:expansionschottky}. Noticing the formal similarity with the expansion into kernels of the connection form of Enriquez in \eqn{eqn:KEnriquez}, we identify the two sets of kernels in \secref{sec:KZB} under the assumption that the defining Poincar\'e series are convergent. In particular, we argue that this is the case for real hyperelliptic curves. We define hyperelliptic polylogarithms in \secref{sec:hepolyschottky} as iterated integrals of these integration kernels. 

\subsection{Notation for the Kronecker form at higher genera}\label{sec:characterization-kronecker}

As we seek to find a generalization of a generating series for higher-genus integration kernels, it is convenient for us to introduce a notation for the Kronecker function accompanied by the differential $\der\xi$ as in \eqn{eqn:genus1expansion}. At the same time, we also want to consider it a function on the Schottky cover. At genus one we therefore introduce the \textit{Schottky–Kronecker form} $\skron$, which is defined for the concentric Schottky group~$\SGroup_\mathrm{c}$ defined in \secref{sec:concentricSchottky} by
\begin{equation} \label{eqn:redefinition_Kronecker_form}
	\skron(z,x,w | \SGroup_\mathrm{c}) \coloneq \kronfun(\xi-\chi,\alpha|\tau(\SGroup_\mathrm{c}))\,\der\xi\,,
\end{equation}
where the Kronecker function~$F$ on the right-hand side was defined in \eqn{eqn:Kroneckergenusone}. Here $\tau(\SGroup_\mathrm{c})$ is the modulus of the torus corresponding to the concentric Schottky group $\SGroup_\mathrm{c}$, and $z$ and~$w$ are related to $\xi$ and $\alpha$ according to the conventions \eqref{eqn:Schottky_conventions} used before. Additionally, here we have introduced an auxiliary point $\chi\in\mathbb C$, whose image on the Schottky cover is given by $x = \exp(2 \pi \iunit \chi)$.

As we will generalize the Kronecker form to higher genera, $\SGroup_\mathrm{c}$ will be a general Schottky group $\SGroup$ and the relations between $z,x$ and $\xi,\chi$ will be replaced by Abel's map:
\begin{equation}
	\xi = \abel(z, z_0|\SGroup)\,,
	\quad
	\chi = \abel(x, z_0| \SGroup)\,,
\end{equation}
whereas the $w$ will be promoted to a vector $\vec{w}$ of formal non-commutative variables. The Kronecker form will then be generalized to a set of $\genus$ forms $\{\skron_j(z,x,\vec{w}|\SGroup)\}_{j=1}^\genus$ at genus \genus{}.

The choices made for the definition and notation of the Schottky–Kronecker form were made with the following considerations:
\begin{itemize}
	\item \textbf{Translation invariance.} 
	At genus zero and one, the definition of polylogarithms takes advantage of the translation invariance on the universal cover. The pole of the integration kernel is usually placed at the origin, and can be moved to the desired location simply by subtracting the coordinate of the corresponding point. At higher genera, there does not exist a cover where such a simple translation property can be used to control the location of the pole. Consequently, the definition of the Schottky–Kronecker form in \eqn{eqn:redefinition_Kronecker_form} keeps track of the location of the pole for the higher-genus analogues of the kernels as the separate variable $x$.
	\item \textbf{Functions versus forms.} Since meromorphic differentials are defined to be independent of chart, their corresponding component functions transform under coordinate transformations,
	\begin{equation}
		\omega = f(z)\, \dd z = f(z')\, \dd z' \implies f(z') = f(z) \frac{\dd z}{\dd z'}\,.
	\end{equation}
	At genus one, one nonetheless works with the Kronecker \emph{function}, using only the universal cover where $\dd z$ is a global differential. However, should we choose to use a different choice of coordinates, such as a Schottky cover, one runs into two problems: the local chart changes when moving around a cycle, i.e.~$\dd (\gamma z) \neq \dd z$, and the choice of the underlying cover itself is not unique. Consequently, the definition of the Schottky–Kronecker form in \eqn{eqn:redefinition_Kronecker_form} includes the differential, allowing us to identify it as a chart-independent object.
\item \textbf{Characterization through quasiperiodicity and residue.} The Schottky–Kronecker form at genus one is quasiperiodic in the \Btxt-cycle as in \eqn{eqn:KroneckerGenusOneQuasiperiodicity}, has a simple pole\footnote{The definition of the residue for a one-form $\omega(z)$ at a point $x$ is $\Res_{z=x} \omega(z) = \frac{1}{2 \pi \iunit} \oint_{x} \omega(z)$, where the integration contour is taken to be a sufficiently small circle around~$x$.} at $z=x$, and is holomorphic at any other point of the fundamental domain $\cF$ of the Schottky cover. We would like the higher-genus Schottky–Kronecker forms $S_j$ to have similar properties, and more precisely we would like to match the properties \eqref{eqn:connection_constraint1} and~\eqref{eqn:connection_constraint2} satisfied by Enriquez' higher-genus connection form, or rather by its components~$K_j$. The quasiperiodicity condition which we will require is therefore
	\begin{equation}\label{eqn:KroneckerQuasiper}
		\skron_j(\gamma_k z, x, \vec w|\SGroup) = w_k \,\skron_j(z,x,\vec w|\SGroup)\,,\qquad j,k=1,\ldots,\genus\,,
	\end{equation}
which matches Enriquez' condition upon identifying $w_j = e^{b_j}$. Moreover, the Schottky–Kronecker forms $\skron_j(z,x,\vec w|G)$ will be required to have a simple pole in $\cF$ at $z=x$ with residue
	\begin{equation}\label{eqn:KroneckerResidue}
		\underset{z=x}{\Res} \, \skron_j(z,x,\vec w|\SGroup) = 1\,.
	\end{equation}

	One can show that there exist $\genus$ linearly independent Kronecker forms that satisfy the quasiperiodicity and residue requirement \cite{Lisitsyn:2024}, hence the corresponding index $j$. 
	
	\item \textbf{Working with formal variables.} As with the quasiperiodicity requirement above, we would like to match our construction with the connection of Enriquez, hence our Schottky–Kronecker forms should depend on non-commuting variables $b_i$. These serve as the power-counting variables which will be used to find an expansion into integration kernels in \secref{sec:expansionschottky}.

	The Kronecker function at genus one included a pole in its power-counting variable $\alpha$ that was cancelled in \eqn{eqn:genus1expansion} in order to perform the expansion. A similar property will be true for the Schottky–Kronecker forms at higher genera, as they will contain poles in the power-counting variables. More precisely, the $j$th Schottky–Kronecker form~$\skron_j$ will have a pole in the variable~$b_j$, in order to match the residue~\eqref{eqn:connection_constraint2} of Enriquez' connection form.
\end{itemize}

\subsection{The Schottky–Kronecker forms} \label{sec:schottky_kronecker_def}
As mentioned in \secref{sec:kzb_generalities}, on a genus-one Riemann surface Bernard's connection form \eqref{eqn:Bernard_connection} is closely related to the Kronecker function \eqref{eqn:Kroneckergenusone}. Both Bernard's formula and the representation \eqref{eqn:kronecker_g1_z_w} of the Kronecker function rely on a Schottky parametrization of the surface. Recall that for a complex torus of modulus $\tau$, with $q= e^{2\pi\iunit \tau}$, one can consider the Schottky cover constructed out of a pair of \textit{concentric circles} of radii $|q| = e^{-2 \pi \mathrm{Im}(\tau)}$ and $1$, as described and pictured in \secref{sec:concentricSchottky}. For this configuration, the Schottky group is generated by the scaling transformation $\gamma_1 z \coloneq q z$. 

It is instructive for the generalization to be considered below to explicitly match expressions \eqref{eqn:Bernard_connection} and \eqref{eqn:kronecker_g1_z_w} \cite{Chan:2022}. To do so, we set the twist homomorphism of \eqn{eqn:Bernard_connection} to be
\begin{equation}\label{eqn:twisthomomorphism}
	g(\gamma_1) = w,  
\end{equation}
where $w$ is a formal variable; this uniquely determines $g(\gamma)$ for every~$\gamma$. Then, using the coordinate~$z$ and pole location~$x$ on the concentric Schottky cover, Bernard's Kronecker function \eqref{eqn:Bernard_connection} can be related to the expression \eqref{eqn:kronecker_g1_z_w} of the Kronecker function by 
\begin{equation}\label{eqn:genus1-schottky-bernard}
	\begin{aligned}
	\sum_{n \in \Integers} \frac{q^n}{q^n z-x} w^{-n} 
	& =-\frac{1}{z} \left( \frac{(z/x)}{1-(z/x)} - \frac{w}{1-w}  + 
	\sum_{n,m > 0} (w^{-n} (z/x)^{m}-w^{n} (z/x)^{-m}) \,q^{m n}\right)\\
	&=\frac{1}{2\pi \iunit z}F(\xi-\chi,\alpha|\tau),
	\end{aligned}
\end{equation}
where the first equality can be formally\footnote{The sum defining the left-hand side is not absolutely convergent, one needs to make sense of it for the parameters in an appropriate region (see \cite[Sec.~3]{ZagierJacobi}, as well as the discussions around eqs.~(2.9) and (3.7) of \cite{Bernard:1988yv}).} obtained by splitting the sum into positive and negative terms, and expanding the denominators as power series, and the second equality follows from \eqn{eqn:kronecker_g1_z_w} after identifying $\xi$ and $\chi$ on the fundamental parallelogram with $z$ and $x$ on the concentric Schottky cover via $z = \exp(2\pi \iunit \xi)$, $x = \exp(2\pi \iunit \chi)$, and setting $w=\exp(-2\pi \iunit \alpha)$ (cf.~\eqn{eqn:Schottky_conventions}). Accordingly, the ratio $(z/x)$ translates into the difference $\xi-\chi$ corresponding to shifting the pole on the fundamental parallelogram, and we can set $x = 1$ to match the common convention $\chi=0$  in the genus-one case. The overall factor of $(2\pi \iunit z)^{-1}$ in \eqn{eqn:genus1-schottky-bernard} compensates for the change of variables in the one-form $\der z = 2 \pi \iunit z \der \xi$, yielding 
\begin{equation}
	K_{\mathrm{B}}(z,x=1|g)=
	F(\xi,\alpha|\tau) \,\dd \xi\,,
\end{equation}
where the twist homomorphism $g$ was defined in \eqn{eqn:twisthomomorphism}. Let us emphasize again that the above equivalence hinges on the concentric configuration of the two Schottky circles described in \secref{sec:concentricSchottky}, which implies the Schottky generator $\gamma_1$ to be a scaling transformation. 

Being based on placing pairs of circles on the Riemann sphere, the Schottky parametrization should be invariant under the M\oe{}bius action of $\sigma\in\mathrm{PSL}(2,\mathbb C)$: two Schottky parametrizations given in terms of circles and generators $\{C_i, C_i', \sigma_i\}$ and $\{\sigma(C_i), \sigma(C_i'), \sigma \gamma_i \sigma^{-1}\}$ will describe the same Riemann surface. Correspondingly, we would like the Schottky–Kronecker form~$S$ mentioned in the previous section to satisfy
\begin{equation}
	\skron(z,x,w|\SGroup)=\skron(\sigma z,\sigma x,\sigma w|\sigma\SGroup\sigma^{-1})
\end{equation}
for any M\oe{}bius transformation $\sigma$.

Still remaining at genus one, let us introduce the manifestly M\oe{}bius-invariant generalization of Bernard's form:
\begin{equation} \label{eqn:kronecker_schottky}
  \skron(z,x,w|\SGroup) 
  \coloneq 
  \sum_{\gamma \in \SGroup} \frac{\der(\gamma z)}{\gamma z - x} 
  \frac{P_1 - x}{P_1 - \gamma z} W(\gamma^{-1})\,,
\end{equation}
where $P_1$ is one of the fixed points\footnote{The definition is independent of which of the two fixed points is used.} of the Schottky generator $\gamma_1$. Furthermore, $W : \gamma_1^n \mapsto w^n$ maps Schottky generators to (in general: non-commuting) variables to keep track of the monodromies. The Schottky–Kronecker form is therefore a formal power series in $w$ and $w^{-1}$. 
The Kronecker form in \eqn{eqn:kronecker_schottky} can be shown to be M\oe{}bius invariant for \textit{any} configuration of two circles on the Riemann sphere parametrizing a genus-one Schottky uniformization by using the transformation rules \eqref{eqn:transformation_properties}. When choosing the concentric arrangement from above, one of the fixed points of the generator $\gamma_1 z = q z$ is $P_1 = \infty$ (cf.~\secref{sec:concentricSchottky}), which lets the additional term compared to \eqn{eqn:Bernard_connection} disappear.

Below, we will customarily use an alternative version of \eqn{eqn:kronecker_schottky}, where no Schottky group elements $\gamma\in\SGroup$ appear within the differential:
\begin{equation}\label{eqn:kron_schottky_convenient}
  \skron(z,x,w|\SGroup)  = \sum_{\gamma \in \SGroup} 
  \frac{\der z}{z - \gamma x} \frac{\gamma P_1 - \gamma x}{\gamma P_1 - z} W(\gamma)\,.
\end{equation}
\Eqns{eqn:kronecker_schottky}{eqn:kron_schottky_convenient} are related by relabeling $\gamma \mapsto \gamma^{-1}$ and applying a M\oe{}bius transformation, thereby moving the explicit dependence on Schottky generators $\gamma$ from $z$ to $x$. 

The M\oe{}bius-invariant expression \eqref{eqn:kron_schottky_convenient} can be generalized to genus $\genus > 1$ as follows: let $\SGroup$ be the Schottky group corresponding to a genus-\genus{} Riemann surface, which is generated by $\gamma_j$, $j=1,\ldots,\genus$. For each $j$ we introduce a \emph{Schottky–Kronecker form}
\begin{equation} \label{eqn:schottky_kronecker_gen}
    \skron_j(z,x,\vec w | \SGroup) 
	\coloneq 
	\sum_{\gamma \in \SGroup} \frac{\der(\gamma z)}{\gamma z - x} 
    \frac{P_j - x}{P_j - \gamma z} W(\gamma^{-1})
	= 
    \sum_{\gamma \in \SGroup} \frac{\der z}{z - \gamma x} 
    \frac{\gamma P_j - \gamma x}{\gamma P_j - z} W(\gamma)\,,
\end{equation}
where each $P_j$ is a fixed point of the $j$-th Schottky generator $\gamma_j$. In analogy to \eqn{eqn:kronecker_schottky}, $W : \gamma_{i_1}^{n_1} \cdots \gamma_{i_s}^{n_s} \mapsto w_{i_1}^{n_1} \cdots w_{i_s}^{n_s}$ maps Schottky generators to non-commutative variables $w_j$ and thus implements how the monodromies are kept track of by non-commutative variables.

\subsection{Properties of Schottky–Kronecker forms} \label{sec:properties_schottky_kronecker}
In this section we discuss properties of the Schottky–Kronecker forms $\skron_j$ defined in \eqn{eqn:schottky_kronecker_gen} and show how they can be expressed as a particular average of genus-one Kronecker forms. Along the way we are going to comment on degenerations of the Kronecker form bridging between different genera. 
\paragraph{Residue and quasiperiodicity.}

In order to compare our construction with the connection form of Enriquez (see \secref{ssec:Enriquezconn}), we (formally) compute the residues and the quasiperiodicity, i.e.~the monodromies around the \Btxt-cycles, of the Schottky–Kronecker forms.

The only summand contributing to the computation of the residue of $\skron_j$ at $z=x$ is $\gamma = \mathrm{id}$, so that one finds
\begin{equation}
    \Res_{z=x} \skron_j(\gamma_k z,x,\vec w | \SGroup) = 1\,.
\end{equation}
If $\gamma_k$ is the generator of the Schottky group $\SGroup$ which corresponds to the cycle $\mathfrak B_k$, then the corresponding quasiperiodicity is given by
\begin{equation}\label{eqn:z-monodromy}
	\skron_j(\gamma_k z,x,\vec w | \SGroup) = w_k\, \skron_j(z,x,\vec w | \SGroup)\,,
\end{equation}
which is obtained by relabeling $\gamma \mapsto \gamma \gamma_k^{-1}$ in \eqn{eqn:kron_schottky_convenient}, so that $\gamma \gamma_k z \mapsto \gamma z$.

When moving the pole location $x$ along a \Btxt-cycle, one instead finds 
\begin{equation}\label{eqn:auxiliary-monodromy}
    \skron_j(z,\gamma_k x,\vec w | \SGroup) 
    = \skron_k(z,x,\vec w | \SGroup) (w_k^{-1}-1) + 
    \skron_j(z,x,\vec w | \SGroup)\,.
\end{equation}
as shown in \appref{app:auxiliary-quasiperiodicity}. The difference between \eqns{eqn:z-monodromy}{eqn:auxiliary-monodromy} can be traced back to the action of the generator $\gamma_k$ on the fixed point $P_j=\lim_{n\to\infty}\gamma_j^nz$ when using \eqn{eqn:schottky_kronecker_gen}. At genus one, however, $\gamma_1 P_1 = P_1$, which leads to $\skron_1(z,\gamma_1 x,w_1|\SGroup_1) = \skron_1(z,x,w_1 | \SGroup_1) w_1^{-1}$ as expected from setting $j=k=1$ in \eqn{eqn:auxiliary-monodromy}. 

\paragraph{Averaging of lower genus objects.}
Using partial fractioning, the components \eqref{eqn:schottky_kronecker_gen} of the Schottky–Kronecker forms can be rewritten as
\begin{equation} \label{eqn:kronfun_expansion_genus_0}
    \skron_j(z,x,\vec w | \SGroup) 
    =
    \sum_{\gamma \in \SGroup} 
    \left(\frac{\der z}{z-\gamma x} - \frac{\der z}{z - \gamma P_j}\right) W(\gamma)\,,
\end{equation}
which can be seen as averages of genus-zero integration kernels over the Schottky group.

Formally\footnote{This formula can be given an analytic meaning at genus one, and its specialization to the concentric Schottky cover, with $z_0=x=P_1^{-1}=0$, coincides with Levin's generating function of depth-one elliptic polylogarithms \cite{Levin} (see also~\cite[Sec.~6.1]{BrownLevin}).}, the above form can be integrated, yielding 
\begin{equation} 
    \int_{z_0}^z \skron_j(z,x,\vec w| \SGroup) 
    = 
    \sum_{\gamma \in \SGroup} 
    \log\left(\frac{(z - \gamma x)(z_0 - \gamma P_j)}{(z - \gamma P_j)(z_0 - \gamma x)}\right) W(\gamma)\,.
\end{equation}

At genus one, the analytic behaviour of the Kronecker function in the variable~$w$ can be controlled: recall that, for $w=\exp(2\pi \iunit\alpha)$, one has the asymptotic behaviour $F(z,\alpha | \tau) \sim 1/\alpha$ as $\alpha\to 0$. Upon canceling this pole,~$F$ can be Taylor-expanded to generate the integration kernels $g^{(n)}$ as in \eqn{eqn:genus1expansion}. Due to the non-commutative nature of the variables~$w_i$ for genus $h>1$, we cannot give a complex-analytic meaning to~$\skron_j$ as a function of these variables, and it is therefore not clear from the representation \eqref{eqn:kronfun_expansion_genus_0} how to replace the pole-subtraction operation from genus one and define a generating series of higher-genus integration kernels.
To surmount this obstacle, we formally manipulate the defining equation of the Schottky–Kronecker form in order to express it in terms of genus-one objects. This can conveniently be done by writing Schottky group elements $\gamma \in \SGroup$ as $\gamma = \gamma_j^n \tilde \gamma$ with $\tilde \gamma \in \SGroup_j \setminus \SGroup$ and $n \in \mathbb Z$ (cf.~\secref{sec:Schottky_Poincare_series} for the definition of $\SGroup_j$). The resulting expression for the Kronecker forms reads
\begin{align}
    \skron_j(z,x,\vec w | \SGroup) 
    & = 
    \sum_{\tilde \gamma \in \SGroup_j \setminus \SGroup} \,\,
    \sum_{n \in \mathbb Z} \,\,W(\tilde \gamma^{-1}\gamma_j^{-n}) 
    \frac{\der (\gamma_j^n \tilde \gamma z)}{\gamma_j^n \tilde \gamma z - x} 
    \frac{P_1 - x}{P_1 - \gamma_j^n \tilde \gamma z} 
    \notag\\
    & = 
    \sum_{\tilde \gamma \in \SGroup_j \setminus \SGroup} W(\tilde \gamma^{-1}) 
    \skron( \tilde \gamma z,x,w_j | \SGroup_j)\,,\label{eqn:schottky-genus1-average}
\end{align}
where $\skron(z,x,w_j | \SGroup_j)$ denotes the (only) Schottky–Kronecker form defined on the Schottky cover with group~$\SGroup_j$ of a genus-one Riemann surface. 
This shows that in the same way as we identified the Kronecker form as a weighted average of genus-zero integration kernels, it can also be seen as an average of genus-one Kronecker forms. By combining this observation with the analytic control that we have on the genus-one Kronecker form, we will be able to define higher-genus integration kernels in \secref{sec:expansionschottky} below.

\begin{figure}[t]
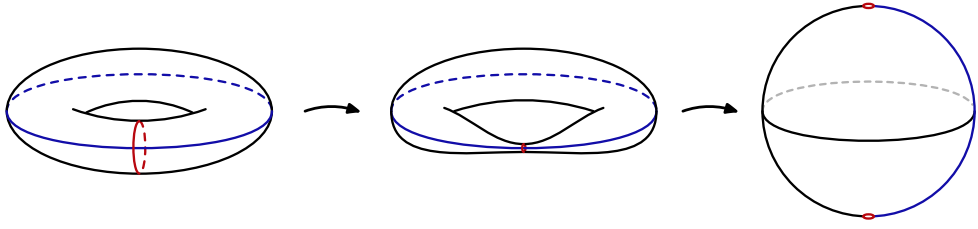

\begin{center}
	\mpostuse[width=0.8\textwidth]{morphing}
\end{center}
\caption{Degeneration of a torus into a Riemann sphere with two marked points by pinching the \Atxt-cycle of the torus, corresponding to the limit $\tau\rightarrow \iunit\infty$.}
\label{fig:morphing}
\end{figure}

\begin{figure}[ht]
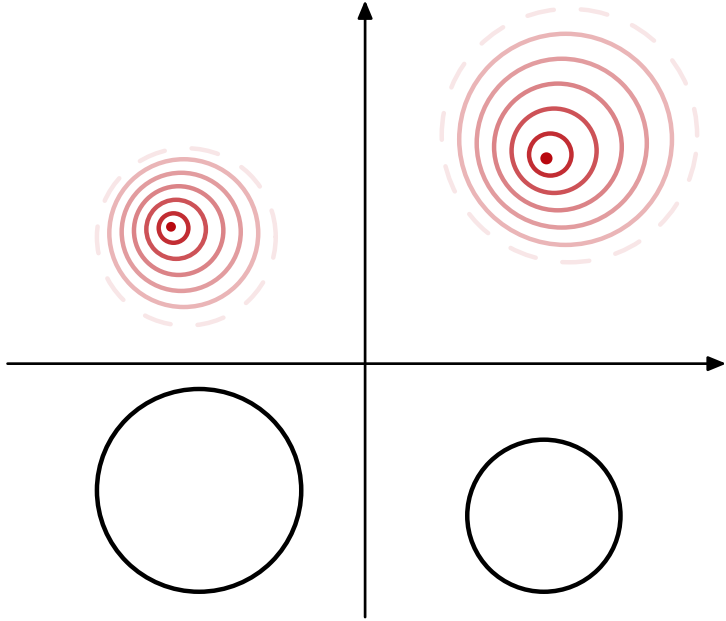

	\begin{center}
		\mpostuse[width=0.6\textwidth]{schottky-degeneration}
	\end{center}
	\caption{A surface of genus two in the Schottky parametrization degenerates into a genus one surface by shrinking a pair of Schottky circles corresponding to an \Atxt-cycle to the two fixed points of their associated M\oe{}bius transformation.}
	\label{fig:schottky-degeneration}
\end{figure}

\paragraph{Degeneration limits.} 
An important feature of the Schottky cover is a prescription for understanding the degeneration of the Kronecker form when shrinking an \Atxt-cycle to a point. At genus one, this process corresponds to taking $\tau \rightarrow i\infty$, which implies $\gamma_1 : z \mapsto 0 = P_1$ and $\gamma_1^{-1}: z \mapsto \infty = P_1'$ (cf.~\secref{sec:concentricSchottky}).

When considering degenerations of higher-genus surfaces, this procedure gets more involved. On the Schottky cover, however, such a prescription is readily available: degenerating the surface by shrinking the cycle $\acyc_j$ corresponds to taking the limit where $\gamma_j : z \mapsto P_j$ and $\gamma_j^{-1} : z \mapsto P_j'$ for every $z$, while all other generators remain inert. This means that we shrink the $j$-th pair of Schottky circles to their fixed points $P_j,\,P_j'$, as illustrated in \figref{fig:schottky-degeneration} for a degeneration from genus two to genus one.

When testing the behaviour of the Kronecker form $\skron_k$ when degenerating along the cycle~$\acyc_j$, we have to distinguish between the situations $j=k$ and $j\neq k$:
\begin{itemize}
	\item If $j \neq k$, we find that the term $\gamma P_k - \gamma x$ vanishes if $\gamma$ contains the Schottky generator~$\gamma_j$, since it would map both terms to the same point. As a result, defining~$\tilde \SGroup=\{\gamma_{i_1}^{n_1}\cdots\gamma_{i_k}^{n_k}\,|\,n_\ell\in\zZ,\,i_\ell\neq j\}$, we find for the degeneration of a genus-\genus{} Kronecker form
	\begin{equation}
	    \skron_k(z,x,\vec w | \SGroup_{\gamma_j \rightarrow P_j}) = \sum_{\gamma \in \tilde \SGroup} \frac{\dd z}{z - \gamma x} \frac{\gamma P_k - \gamma x}{\gamma P_k - z} W(\gamma) = \skron_k(z,x,\{w_i\}_{i \neq j}| \tilde \SGroup)\,,\label{eqn:degeneration-unmatched}
	\end{equation}
	recovering the corresponding Kronecker form at genus $\genus-1$ with the Schottky group to the degenerated surface.
\item In the case $j = k$, we expect the result to be different, knowing that the $\genus-1$ Kronecker forms available at genus $\genus-1$ are already accounted for. Indeed, by writing $\gamma=\tilde\gamma\gamma_j^n$ with $\tilde \gamma \in \SGroup / \SGroup_j$ and by using properties of the fixed points we find the degeneration
	\begin{align}
	    &\skron_j(z,x,\vec w | \SGroup_{\gamma_j \rightarrow P_j})  = \sum_{\tilde \gamma \in \SGroup / \SGroup_j} \sum_{n \in \mathbb Z} \frac{\dd z}{z - \tilde \gamma \gamma_j^n x} \frac{\tilde \gamma \gamma_j^n P_j - \tilde \gamma \gamma_j^n x}{\tilde \gamma \gamma_j^n P_j - z} W(\tilde \gamma \gamma_j^n) \nonumber\\
	    & = \sum_{\tilde \gamma \in \SGroup / \SGroup_j} \frac{\dd z}{z - \tilde \gamma P_j'}\frac{\tilde \gamma P_j - \tilde \gamma P_j'}{\tilde \gamma P_j - z} W(\tilde \gamma)\,\,\sum_{n < 0} w_j^{n}\,\,+ \sum_{\tilde \gamma \in \SGroup / \SGroup_j} \frac{\dd z}{z - \tilde \gamma x}\frac{\tilde \gamma P_j - \tilde \gamma x}{\tilde \gamma P_j - z} W(\tilde \gamma)\,.\label{eqn:degeneration-matched}
	\end{align}
	In the above expression, the first term corresponds to the summation over $n < 0$, which turned $\gamma_j^{n} x \mapsto P_j'$. The second term corresponds to $n=0$, and all contributions with $n>0$ vanish due to the appearance of $\tilde\gamma P_j-\tilde\gamma P_j=0$ in the numerator. Furthermore, for both terms, we find that any element $\tilde \gamma$ containing an appearance of $\gamma_j$ will make the numerator vanish, allowing us to replace the sum over $\SGroup / \SGroup_j$ by a sum over $\tilde \SGroup$, where $\tilde \SGroup$ is the Schottky group of the degenerated surface. Thus, the degeneration limit becomes
	\begin{align}
		\skron_j(z,x,\vec w | \SGroup_{\gamma_j \rightarrow P_j}) &= \sum_{\tilde \gamma \in \tilde \SGroup} \frac{\dd z}{z - \tilde \gamma P_j'}\frac{\tilde \gamma P_j - \tilde \gamma P_j'}{\tilde \gamma P_j - z} W(\tilde \gamma)\,\,\sum_{n < 0} w_j^{n}\notag\\
		&\quad+ \sum_{\tilde \gamma \in \tilde \SGroup} \frac{\dd z}{z - \tilde \gamma x}\frac{\tilde \gamma P_j - \tilde \gamma x}{\tilde \gamma P_j - z} W(\tilde \gamma)\,.
	\end{align}
	Both of these terms resemble the Kronecker forms themselves, although they both involve a fixed point which is now at one of the punctures resulting from shrinking the cycle. 
\end{itemize}

\paragraph{Fay identity at genus one.}

At genus one, the Fay identity \eqref{eqn:KroneckerFayGenusOne} for the Kronecker function can be deduced by formally\footnote{This manipulation ignores the fact that the Poincar\'e series considered are not absolutely convergent.} partial-fractioning the representation on a concentric Schottky cover given by \eqn{eqn:genus1-schottky-bernard}:
\begin{equation}\label{eqn:schottky-fay-proof}
	\begin{aligned}
	&F(\xi-\chi,\alpha|\tau) F(\tilde \xi-\tilde\chi, \tilde \alpha|\tau) = \sum_{m \in \mathbb Z} \sum_{\tilde m \in \mathbb Z} \frac{1}{z-q^m x} w^m \frac{1}{\tilde z - q^{\tilde m} \tilde x}  \tilde w^{\tilde m} \\
	& = \sum_{m \in \mathbb Z} \sum_{\tilde m \in \mathbb Z} \frac{1}{z - q^m x} (w \tilde w)^m \frac{1}{\tilde z - q^{\tilde m - m} z \tilde x / x} \tilde w^{\tilde m - m} + ( \cdot \leftrightarrow \tilde \cdot) \\
	& = F(\xi-\chi,\alpha+\tilde\alpha | \tau) F(\tilde \xi-\tilde \chi- \xi+\chi , \tilde \alpha | \tau) + F(\tilde \xi-\tilde \chi,\alpha+\tilde\alpha | \tau) F(\xi -\chi-\tilde\xi+\tilde\chi , \alpha | \tau)\,,
	\end{aligned}
\end{equation}
where $(\cdot \leftrightarrow \tilde \cdot)$ corresponds to the term where variables with and without tilde are swapped. 

Unfortunately, this approach to deriving a Fay identity is not successful for higher genera: at genus one, the fixed points are invariant under the action of the Schottky group, so it can be ignored how they transform by going to the concentric cover. Furthermore, since there is only a single generator involved at genus one, the actions thereof are commutative and can be combined between the two Kronecker forms. These properties are lost once one goes to higher genera, as fixed points of one Schottky generator are not invariant under the action of another Schottky generator and because successive application of different Schottky generators is not commutative anymore.

\subsection{Expansion into kernels}\label{sec:expansionschottky}

We would like now to extract from each Kronecker form $\skron_j$, which is a formal object, a family of meromorphic differentials with at most simple poles on the Schottky cover, i.e.~analytic objects, which will serve as integration kernel for higher-genus analogues of polylogarithms. We will succeed in doing this in the case of hyperelliptic polylogarithms on real hyperelliptic curves. Similar to the conventions of refs.~\cite{EnriquezHigher,EZ1}, we set $w_i = \exp(b_i)$, with non-commutative variables~$b_i$. At genus one, the variable $b=\log(w)$ differs from the $\alpha$ notation  of~\eqn{eqn:genus1expansion} only by a factor of $-2\pi \iunit$. The idea then is that one can formally cancel the poles in the variables $b_i$ from the genus-one Kronecker forms in the expression \eqref{eqn:schottky-genus1-average}, and obtain formal series in non-commutative variables, whose coefficients will be the sought-for integration kernels. 

Let us be more precise here: as pointed out in \secref{sec:eMPLs}, the classical Kronecker form $F(\xi,\alpha|\tau)$ has a pole at both $\xi=0$ and $\alpha=0$. Since in \eqn{eqn:schottky-genus1-average} we express the higher-genus generalizations of the Kronecker form as averages over a Schottky group of the (genus-one) Kronecker form, formally the pole in the formal variable $b_j$ is inherited by each $\skron_j(z, x, \vec{w} | \SGroup)$.
We can then therefore (formally) cancel this pole by multiplying the Schottky–Kronecker form by~$b_j$ from the right\footnote{Multiplication must be from the right because of our convention that $(\gamma_1\gamma_2)(z)=\gamma_1(\gamma_2(z))$.}, obtaining the following expansion in the non-commutative variables~$b_i$:
\begin{subequations}
\begin{align}\label{eqn:schottky-kronecker-expansion-definitionb}
    \skron_j(z,x,\vec w| \SGroup) b_j 
    &= 
    \sum_{\gamma \in \SGroup / \SGroup_j} W(\gamma) \skron(\gamma^{-1} z, x,w_j|\SGroup_j)\, b_j
     \\
    &\eqdef -2\pi \iunit \sum_{s = 0}^\infty \sum_{i_1,\cdots,i_s = 1}^h 
    \omega_{i_1\cdots i_s j}(z,x |\SGroup) \,
    b_{i_1} \cdots b_{i_s}\,,\label{eqn:schottky-kronecker-expansion-definition}
\end{align}
\end{subequations}
where the first equality is obtained from \eqn{eqn:schottky-genus1-average} by 
relabeling $\gamma \mapsto \gamma^{-1}$ which also changes the summation domain $ \SGroup_j\! \setminus \!\SGroup  \mapsto \SGroup / \SGroup_j$.

It remains now to determine the coefficient of a particular word $b_{i_1} \cdots b_{i_s}$ in the formal expansion \eqref{eqn:schottky-kronecker-expansion-definition}. In order to do this, it is helpful to write the word as well as the Schottky element $\gamma$ in \eqn{eqn:schottky-kronecker-expansion-definitionb} in a way that groups repeated consecutive entries:
\begin{subequations}
\begin{align}
	\text{word} & = b_{i_1}^{n_1} \cdots b_{i_s}^{n_s}\label{eqn:groupword} \\ 
	\gamma & = \gamma_{j_1}^{m_1} \cdots \gamma_{j_l}^{m_l} \label{eqn:Schottkyword}
\end{align}
\end{subequations}
where $n_k > 0$ label the lengths of the individual groups of consecutive repeated letters within the words in \eqn{eqn:groupword}, $m_k \neq 0$ label the lengths of the individual groups of consecutive repeated generators in \eqn{eqn:Schottkyword}, and $s$ and $l$ denote the numbers of groups. In particular, one is allowed to have $m_k < 0$, to account for inverses that appear in the element of the Schottky group. Uniqueness of the notation is ensured by requiring $i_k \neq i_{k+1}$ and $j_k \neq j_{k+1}$.

By Taylor-expanding a genus-one Schottky–Kronecker forms, we obtain integration kernels $\skern{k}(z, x| \SGroup_j)$
on the Schottky cover of the genus-one Riemann surface obtained from the Schottky subgroup ${\SGroup}_j \subset {\SGroup}$ generated by the $j$-th generator $\gamma_j$:
\begin{equation}
	\skron(z, x, w_j |\SGroup_j)\, b_j \eqdef \sum_{k = 0}^\infty b_j^k \,\skern{k}(z , x | \SGroup_j).
\end{equation}
These differentials are related to the standard genus-one kernels \eqref{eqn:genus1expansion} via\footnote{The normalization factor of $(-2 \pi \iunit)^{1-n}$ is due to the expansion in $b \sim -2 \pi \iunit \alpha$. } 
\begin{equation} \label{eqn:genus-one-kernels-relation}
	\skern{n}(z, x | {\SGroup}_j) = (-2 \pi \iunit)^{1-n} \gkern{n}(\abel(z,x|\SGroup_j), \tau(\SGroup_j))\, \omega(z|\SGroup_j),
\end{equation}
where $\abel$ and $\omega$ are Abel's map and the holomorphic differential on the Schottky cover of a genus-one Riemann surface as in \secref{sec:Schottky_Poincare_series}, and $\tau(\SGroup_j)$ is the modular parameter of the corresponding complex torus.

For a formal power series $\sum_w a_w\cdot w$ over words $w$ in non-commutative letters, with corresponding coefficients $a_w$, denote by $[w']$ the operation which extracts the coefficient $a_{w'}$ of $w'$ from the series, i.e.~$a_{w'}=[w']\sum_w a_w\cdot w$. 
Then, by placing \eqn{eqn:schottky-kronecker-expansion-definition} on the LHS, and plugging in \eqn{eqn:genus-one-kernels-relation} for \eqn{eqn:schottky-kronecker-expansion-definitionb} on the RHS, we can isolate the word $b_{i_1}^{n_1} \cdots b_{i_s}^{n_s}$ to find\footnote{Notice that we may have contributions to $b_{i_s}^{n_s}=b_j^{n_s}$ also from the summation over $\SGroup / \SGroup_j$, because the fact that the last letter of a word in $\SGroup / \SGroup_j$ cannot be $w_j$ does not imply that $b_j$ cannot be the last letter after expanding the exponentials, as $w_k=1+b_k+\cdots$. This explains the summation over $k$ on the right-hand side.}
\begin{equation}
	\omega_{i_1\cdots i_s j}(z,x |\SGroup) \,
    = -\frac{1}{2\pi i} [ b_{i_1}^{n_1} \cdots b_{i_s}^{n_s}] \sum_{\gamma \in \SGroup / \SGroup_j} \sum_{k=0}^{\delta_{ji_s} n_s} W(\gamma)\, b_{i_s}^k \, \skern{k}(\gamma^{-1} z, x | \SGroup_j).\label{eqn:expansion-setup}
\end{equation}
What remains is to find the coefficient of $b_{i_1}^{n_1} \cdots b_{i_s}^{n_s}$ in $W(\gamma)$. We can do this recursively, writing
\begin{align} \label{eqn:higher-genus-kerels-preexp}
	&C(b_{i_1}^{n_1} \cdots b_{i_s}^{n_s} , \gamma_{j_1}^{m_1} \cdots \gamma_{j_l}^{m_l}) \nonumber\\[3pt]
	&\hspace{28pt}=
	\begin{cases} s = 0 : & 1 \quad \text{(empty word)}, \\
		s \neq 0 = l : & 0 \quad  \text{(non-empty word, identity element)}, \\
		i_1 \neq j_1 : & C(b_{i_1}^{n_1} \cdots b_{i_s}^{n_s} , \gamma_{j_2}^{m_2} \cdots \gamma_{j_l}^{m_l}) \quad \text{(first letters don't match)}, \\
		i_1 = j_1 : & \sum_{k=0}^{n_1} \frac{(m_1)^k}{k!} C(b_{i_1}^{n_1-k} \cdots b_{i_s}^{n_s} , \gamma_{j_2}^{m_2} \cdots \gamma_{j_l}^{m_l}) \quad \text{(first letters match)},
	\end{cases}
\end{align}
where using the base cases for empty words and elements, one can parse the groups of repeated generators in the element one at a time, using the coefficients coming from the expansion of the exponential when the indices match. A detailed derivation of \eqn{eqn:higher-genus-kerels-preexp} is given in \appref{app:coef-ker-derivation}. Putting everything together, we finally find
\begin{equation} \label{eqn:schottky_kernels}
	\omega_{i_1\cdots i_s j}(z,x |\SGroup) = -\frac{1}{2\pi i}\sum_{\gamma \in \SGroup / \SGroup_j} \sum_{k=0}^{\delta_{ji_s} n_s} C(b_{i_1}^{n_1} \cdots b_{i_s}^{n_s-k} , \gamma) \,\skern{k}(\gamma^{-1} z, x | \SGroup_j).
\end{equation}
The first two simplest examples are 
\begin{subequations}
	\begin{align}\label{eq:weight-0-example}
			\omega_{j}(z,x | \SGroup) 
			&=
			-\frac{1}{2\pi i}\sum_{\gamma \in \SGroup/\SGroup_j} \skern{0}(\gamma^{-1}z, x| \SGroup_j)\,,
			\\ \label{eq:weight-1-example}
			\omega_{ij}(z,x | \SGroup) 
			&= 
			-\frac{1}{2\pi i}\bigg[\sum_{\gamma \in \SGroup / \SGroup_j} \!\! C(b_i , \gamma) \skern{0}(\gamma^{-1} z,x | \SGroup_j) \, + \delta_{ij} \sum_{\gamma \in \SGroup / \SGroup_j} \skern{1}(\gamma^{-1} z,x | \SGroup_j)\bigg],		
	\end{align}
\end{subequations}
where the coefficients of the terms in the second sum of \eqref{eq:weight-1-example} are $\delta_{ij} C(1,\gamma) = \delta_{ij}$, since the right-hand side of \eqn{eqn:expansion-setup} suggests that we find $[b_i] W(\gamma) b_j$. Notice that the formula for the kernels $\omega_j(z, x | \SGroup)$ in \eqn{eq:weight-0-example} matches the Poincaré series representation \eqref{eqn:decomp_hol_diff} of the normalized holomorphic differentials.

From the above formula one can immediately deduce that the only genus-$\genus$ kernels that have a pole at $z=x$ are $\{\omega_{jj}(z,x| {\SGroup})\}_{j=1,\ldots,\genus}$. Only those kernels which have $\skern{1}(z,x|\SGroup_j)$ as a summand can have a pole at $x$. In order for this term to have a non-zero coefficient, the identity generator $\id \in {\SGroup} / {\SGroup}_j$ must contribute to the term with $k = 1$ in \eqref{eqn:higher-genus-kerels-preexp}. This can only happen when $s=1$ and $i_s = j$: kernels with $s > 1$ will not have a contribution from the identity element, and kernels with $i_s \neq j$ will not have a contribution with $k=1$. The rest of the kernels are holomorphic in the fundamental domain $\funddom$, and have at most simple poles on the Schottky cover $\Omega({\SGroup})$.

\subsection{Relation to Enriquez' connection}\label{sec:KZB}
This subsection is dedicated to translating and connecting our findings to the connection of Enriquez \cite{EnriquezHigher}. Let us combine the Schottky–Kronecker forms $\skron_j$ defined in \eqn{eqn:schottky_kronecker_gen}  into one object which generates all the differentials by making use of~$h$ extra formal variables $a_1,\ldots ,a_h$:
\begin{equation}\label{eqn:schottkyK}
    \conn(z,x) 
    \coloneq 
    -\sum_{j=1}^h \skron_j(z,x,\vec w | \SGroup)\, b_j a_j / 2\pi \iunit
    = 
    \sum_{s=0}^{\infty} \sum_{i_1,\cdots,i_s=1}^h \sum_{j=1}^h 
    \omega_{i_1\cdots i_s j}(z,x |\SGroup)\,  b_{i_1} \cdots b_{i_s} a_j\,.
\end{equation}

It follows from the quasiperiodicity and the residue formulas satisfied by the Schottky–Kronecker forms, discussed in \secref{sec:properties_schottky_kronecker}, that $\conn(z,x) $ satisfies
\begin{subequations}
\begin{align}
    \conn(\gamma_k z,x) &= e^{b_k} \conn(z,x)\,,\\ 
    \Res_{z=x} \conn(z,x) &= -\sum_{j=1}^h b_j a_j / 2\pi \iunit \,,
\end{align}
\end{subequations}
which are precisely the conditions \eqref{eqn:connection_constraint1}, \eqref{eqn:connection_constraint2} 
that uniquely determine the connection of Enriquez, and which imply the following conditions coefficient-wise: 
\begin{subequations}
\begin{align}
    \omega_{i_1\cdots i_s j}(\gamma_k z,x|\SGroup) 
    &= \sum_{l=0}^s \delta_{i_1 \cdots i_{l} k}\, \omega_{i_{l+1} \cdots i_s j}(z,x|\SGroup) \,,
    \\
    \Res_{z=x}\, \omega_{i_1\cdots i_s j}(z,x |\SGroup) 
    &= -\delta_{s1} \delta_{i_1 j}/2\pi \iunit\,.
\end{align}
\end{subequations}
These conditions uniquely determine the family $\{\omega_{i_1\cdots i_s j}(z,x)\}$ of meromorphic differentials on the Schottky cover with at most simple poles at $z=\gamma x$ defined in Enriquez' original paper \cite{EnriquezHigher} via a non-constructive method (cf.~\eqn{eq:diffenriq}), whose iterated integrals should be higher-genus analogues of the elliptic polylogarithms~$\Gt$.

In other words, we have established that the Enriquez differentials $\omega_{i_1\cdots i_s j}(z,x)$ can be identified with our $\omega_{i_1\cdots i_s j}(z,x|{\SGroup})$, provided that the latter can be promoted from formal objects to meromorphic differentials on the Schottky cover, i.e.~that their defining series converge. This is possible in special cases, such as that of real hyperelliptic curves, as argued in \secref{sec:numerics-convergence}.


\subsection{Hyperelliptic polylogarithms from Schottky–Kronecker forms}
\label{sec:hepolyschottky}
Let us introduce the shorthand notation $\omega_{\mindx{i}_n}(z,x | \SGroup)$, where $n$ indicates the length of the multi-index $\mindx{i}_n$. Whenever $\omega_{\mindx{i}_n}(z,x | \SGroup)$ define analytic objects, one can consider the iterated integrals
\begin{equation} \label{eq:higher-genus-polylogs}
  \Gtargzg{\mindx{i}_{n_1}, \ldots ,\mindx{i}_{n_k}}{x_1, \ldots, x_k} 
  \coloneq
  \int_{z_0}^z \omega_{\mindx{i}_{n_1}}(t, x_1|\SGroup) \Gtargtg{\mindx{i}_{n_2}, \ldots ,\mindx{i}_{n_k}}{x_2, \ldots, x_k},\quad\quad \Gtargzg{}{}\coloneq 1  \,,
\end{equation}
where the integration basepoint $z_0$ is a point on the universal cover of the surface, which may also be chosen to be a preimage of $x$, in which case one needs to regularize the integral. As already mentioned, we restrict our attention to real hyperelliptic curves, in which case these integrals make sense, and we call them \emph{hyperelliptic polylogarithms}. They generalize the elliptic polylogarithms $\Gt$ from \cite{Broedel:2017kkb}.

As shown in \eqn{eqn:schottky_kernels}, the kernels $\omega_{\mindx{i}_n}(z,x|\SGroup)$ can be expressed as averages over the Schottky group of genus-one kernels. Therefore, the above iterated integrals of the higher-genus kernels can be expressed as similar sums of iterated integrals over genus-one kernels of the form
\begin{equation} \label{eqn:iterated_integral_higher_genus}
\int_{z_0}^{z} \skern{n_1}(\gamma_{1} z_1, x_1 | \SGroup_{j_1}) 
\int_{z_0}^{z_1} \skern{n_2}(\gamma_{2} z_2, x_2 | \SGroup_{j_2}) \int_{z_0}^{z_2} \ldots
\int_{z_0}^{z_k-1} \skern{n_k}(\gamma_{k} z_k, x_k | \SGroup_{j_k})\,.
\end{equation}
We want now to make use of the relation \eqref{eqn:genus-one-kernels-relation} to rewrite \eqn{eqn:iterated_integral_higher_genus} in terms of the standard genus-one kernels $\gkern{k}$. To do so, let us introduce the notations $\tau_{j_i}:=\tau(G_{j_i})$ and $r_{ik} \coloneq \abel_{j_i} \circ \gamma_i \circ \gamma_k^{-1} \circ \abel_{j_k}^{-1}$, where $\abel_{j_i}$ is a shorthand for Abel's map on the Schottky cover of group $\SGroup_j$
\begin{equation}
\abel(z,x_i|\SGroup_{j_i})
= 
\frac{1}{2\pi \iunit} \log 
\left( \frac{z - P_{j_i}}{z - P'_{j_i}}\frac{x_i - P'_{j_i}}{x_i - P_{j_i}} \right),
\end{equation}
where $P_{j_i}$ and $P'_{j_i}$ are the fixed points of the $j_i$-th Schottky generator. Then \eqn{eqn:iterated_integral_higher_genus} can also be written (up to the factor $(-2\pi\iunit)^{k-n_1-\ldots-n_k}$) as
\begin{equation}  \label{eqn:iterated_integral_higher_genus_std}
	\int_{t_0}^{t} \der t_1\, \gkern{n_1} (r_{1k}(t_1)|\tau_{j_1})\, r_{1k}'(t_1) 
	\int_{t_0}^{t_1} \der t_2\, \gkern{n_2} (r_{2k}(t_2) | \tau_{j_2})\, r_{2k}'(t_2) \int_{t_0}^{t_2}
\ldots \int_{t_0}^{t_{k-1}}\der t_k \,\gkern{n_k}(t_k |\tau_{j_k})\,,
\end{equation}
where we used that $r_{kk}$ reduces to the identity in the innermost integral.
The end points $t$ and $t_0$ are given by $t = \abel(\gamma_k z, x_k | \SGroup_{j_k}),$ and $t_0 = \abel(\gamma_k z_0, x_k | \SGroup_{j_k})$.

Due to the presence of the maps $r_{ik}$ in the integration, the given expression is not an elliptic polylogarithm, but rather its generalization, where the endpoints of the intermediate integrations are shifted by the M\oe{}bius transformations $\gamma_i$. Summation of these functions over (cosets of) the Schottky group with appropriate combinatorial factors then gives rise to hyperelliptic polylogarithms. Notice that such an expression could be taken as a formal definition of \eqn{eq:higher-genus-polylogs} for any Riemann surface. 

In \secref{sec:numerics-convergence} we argue that the Poincar\'e series \eqref{eqn:schottky_kernels} converges under certain assumptions. More precisely, under these assumptions we expect absolute convergence, uniformly in the compact subsets, so that the series define meromorphic differentials, and all the formal manipulations on the series are justified. Moreover, uniform convergence implies that one can interchange integration and summation, so that also the series of genus-one iterated integrals \eqref{eqn:iterated_integral_higher_genus} which compute hyperelliptic polylogarithms are convergent away from the singularities. These series can be seen as generalizations of the series which define elliptic polylogarithms as averages of genus-zero polylogarithms in the classical approach of \cite{Bloch, Zagier, Levin, BrownLevin}.

Since we express hyperelliptic polylogarithms as Schottky averaging sums of elliptic polylogarithms, the question of the regularization when the basepoint $z_0$ is placed at a pole of the kernels may be tackled via the regularization mentioned in \secref{sec:genus-one-polylogs}. Applying the tangential basepoint regularization prescription directly on \eqn{eq:higher-genus-polylogs} should yield the same result. 


\section{Numerical evaluation of hyperelliptic polylogarithms} \label{sec:practicabilities}

Representing the Kronecker form in the Schottky parametrization opens up the possibility to evaluate the resulting hyperelliptic polylogarithms numerically. In this section we discuss a possible approach towards a numerical implementation of the objects constructed so far.

\subsection{A numerical approach}

\paragraph{From an algebraic curve to the Schottky parametrization.}
Many possible applications of hyperelliptic polylogarithms -- foremost Feynman integral calculations -- rely on representations of higher-genus Riemann surfaces as algebraic curves. In order to be able to apply our results to this context, one therefore needs to be able to find a Schottky uniformization of a given algebraic curve. In general, the question of existence of a uniformization with a classical Schottky group is not settled. However, there exists an algorithm 
to obtain a non-classical (i.e.~based on Jordan loops that are not necessarily circles) Schottky uniformization of a given algebraic curve \cite{Bobenko:2011}. As it was mentioned in \secref{sec:schottky_intro}, the convergence of the Poincar\'e series generally is an open question, but it is guaranteed at least for circle decomposable Schottky groups. In \appref{sec:curve_to_schottky} we review a numerical algorithm for real hyperelliptic curves with real roots that results in a manifestly circle decomposable Schottky group. This algorithm can be extended to arbitrary real hyperelliptic curves \cite{Seppala:2004}. 

\paragraph{Functions on a curve.} 
Provided that we have a description of a (real) algebraic curve in terms of the Schottky parametrization, it is then straightforward to implement various objects defined on the curve. 
The expressions \eqref{eqn:schottky-holomorphic-basis}, \eqref{eq:abel-map-schottky} and \eqref{eq:period-matrix-schottky} can be exploited to numerically implement the canonical basis of holomorphic differentials, Abel's map and the period matrix. The convergence of these formulas all rely on the convergence of the Poincar\'e series expression of the holomorphic differentials. We will take a closer look at the convergence aspects in~\secref{sec:numerics-convergence}.

\paragraph{Hyperelliptic polylogarithms.}
One of the main results of this article is an expression for higher-genus integration kernels of hyperelliptic polylogarithms in terms of Schottky averaging sums of the genus-one kernels \eqref{eqn:schottky_kernels}. Convergence of these expression is argued, and illustrated with numerical tests, in \secref{sec:numerics-convergence}.

Having at disposal a tool to numerically evaluate the higher-genus kernels, one may straightforwardly numerically evaluate the hyperelliptic polylogarithms \eqref{eq:higher-genus-polylogs} defined as iterated integrals of these kernels. However, it is in practice more efficient to use the fact that the kernels can be expressed as Schottky averaged sums of the genus-one kernels, which leads to computing hyperelliptic polylogarithms as Schottky averaged sums of the elliptic ones from \eqn{eqn:iterated_integral_higher_genus_std}. The convergence of this series is guaranteed by the convergence of the series defining the underlying kernels. In this way, one can exploit all sorts of known efficient methods and tricks to compute elliptic polylogarithms, such as modular properties (that allow to shift the modular parameter~$\periodmatrix$ to the fundamental domain and make the $q$-expansion converge quickly) and regularization schemes.

\paragraph{An implementation.}
In the subsequent sections the numerical evaluations of the functions on a Schottky parametrization is performed. We use a numerical implementation that goes along the lines of the previous paragraphs using \textsc{Python}, in particular \textsc{SciPy} (a scientific library for \textsc{Python}), and -- for time-critical computations -- \textsc{C} with \textsc{GSL} (GNU scientific library). All the plots are produced using \textsc{matplotlib}, a mathematical drawing library for \textsc{Python}.

\subsection{Convergence of the Poincar\'e series} \label{sec:numerics-convergence}
In this section we discuss the question of convergence of various functions defined on a Schottky uniformized Riemann surface earlier. We choose a particular genus-two (real) hyperelliptic curve  
\begin{equation} \label{eq:example-curve}
	y^2 = x (x - 1) (x - 6) (x - 7) (x - 12)
\end{equation}
as an example for the following computations. In order to discuss the convergence of the series we define an ordering in the Schottky group such that for any group element of length $l$ any other element of length $l-1$ has smaller index.

\begin{figure}
	\centering
	\begin{minipage}[t]{0.45\textwidth}
		\centering
		\includegraphics[width=\textwidth]{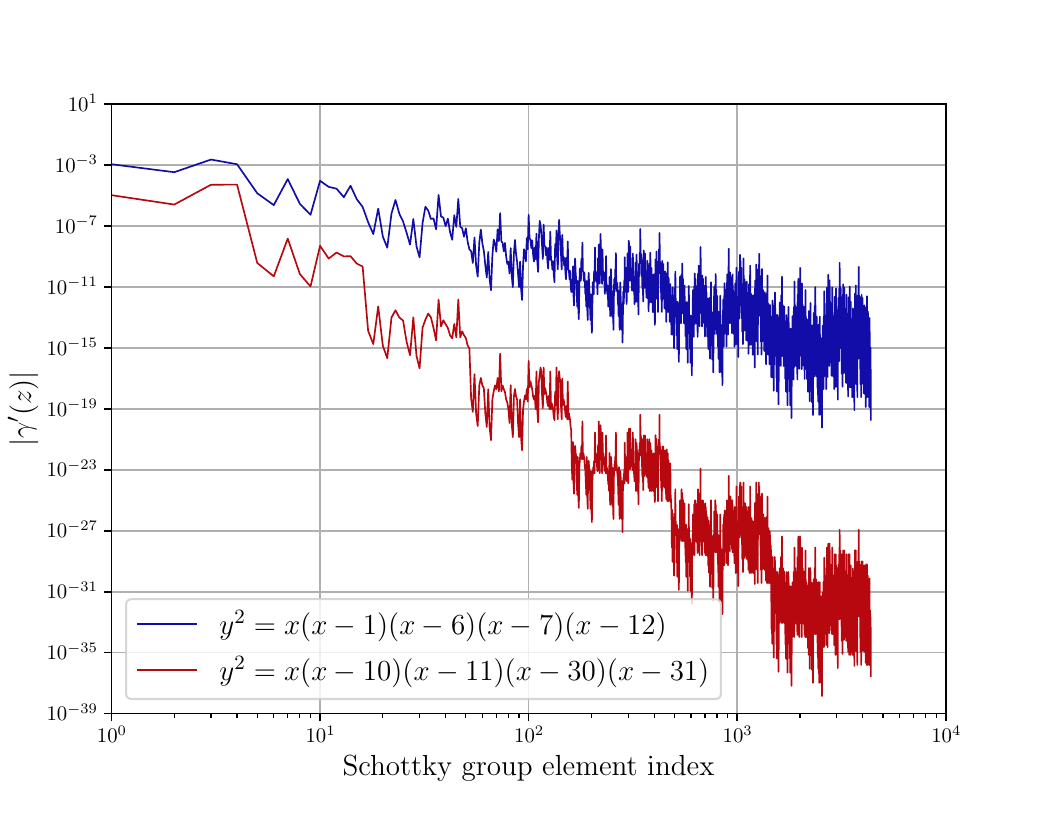}
	  \end{minipage}
	\hspace*{0.02\textwidth}
	\begin{minipage}[t]{0.45\textwidth}
		\centering
		\includegraphics[width=\textwidth]{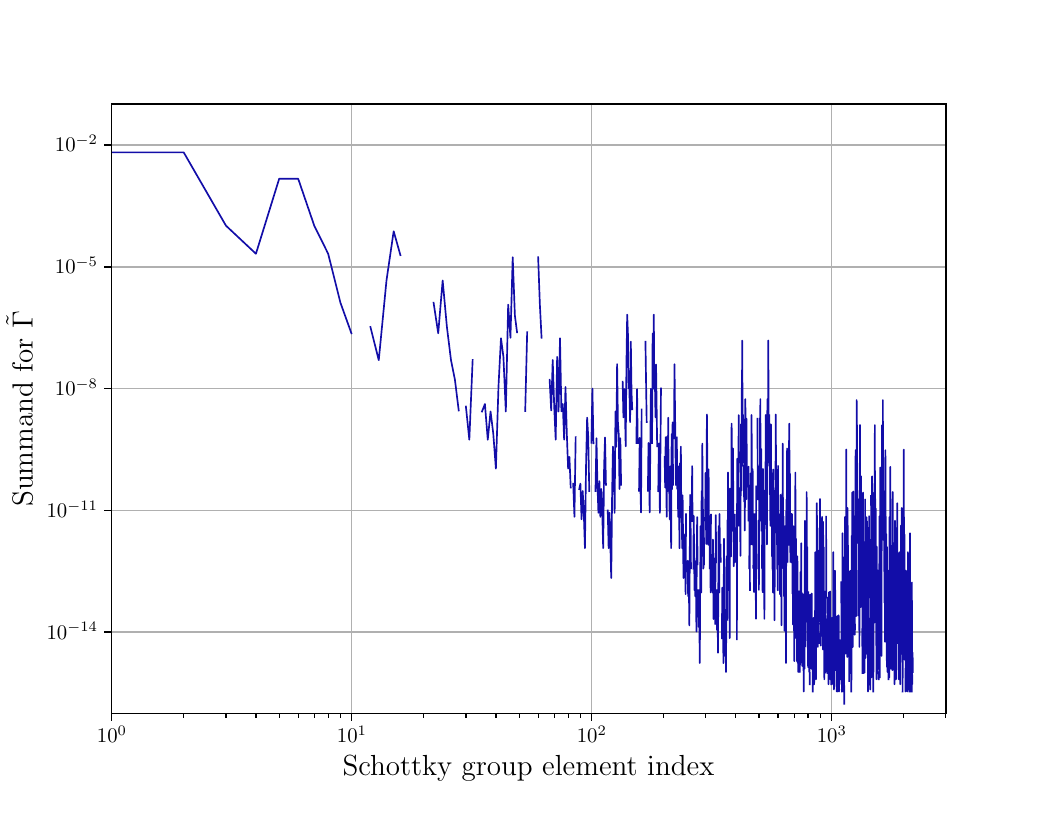}
	\end{minipage} 
	\caption{Illustration of convergence for the holomorphic differentials and polylogarithms. The horizontal axis on both graphs enumerates the Schottky group elements. In the left plot the vertical axis measures the absolute values of $\gamma'$ evaluated at a generic point. The blue line corresponds to the example curve \protect\eqref{eq:example-curve}, whereas the red line is plotted for an alternative curve (see the legend). In the right figure the vertical axis corresponds to the absolute value of \protect\eqref{eqn:iterated_integral_higher_genus} for the polylogarithm $\Gtargzg{(2\ 1)}{x}$ at depth 1 evaluated at a generic point with the basepoint $z_0 = 0$ and the pole $x=-\iunit$. }
	\label{fig:conv_genus_2}
\end{figure}

\paragraph{Holomorphic differentials.}
As it was discussed earlier, the convergence of the series \eqref{eqn:schottky-holomorphic-basis} for the holomorphic differentials is guaranteed for real hyperelliptic curves, and relies on the behavior of the factor $\gamma'(z)$ in the series, which decreases exponentially with the length of the Schottky words. Since the number of Schottky group elements of the fixed length $l$ grows exponentially as $2\genus(2\genus -1)^{l-1}$, the convergence of the series is polynomial.

This behavior in the case of the curve \eqref{eq:example-curve} is depicted in \figref{fig:conv_genus_2} (left) by the blue line. While this is a general feature for circle decomposable Schottky groups, the rate of the convergence varies significantly for different Schottky groups. The red line represents another Schottky uniformized curve, whose convergence is much faster. Generally, the speed of convergence depends on the relative positions and radii of the isometric circles of the classical Schottky group: the smaller the circles and the bigger the distances between them, the faster the convergence. In a sense, one may view the Poincar\'e series as an analogue of $q$-expansions in the genus-one case. There, the modular transformations $\grp{SL}(2,\Integers)$ can be used to shift the modular parameter~$\tau$ such that $q = \eunit^{2 \pi \iunit \tau}$ has a smaller absolute value, thus boosting the convergence of the expansion. This gives efficient numerical algorithms to resum the $q$-expansions of functions on elliptic curves with a nice modular behaviour. For higher-genus Schottky uniformized curves one may try to relate Poincar\'e series over different Schottky groups, enhancing the convergence of the numerical methods. This is particularly important to establish, because an improvement of the precision by an order comes with a price of exponentially more terms in the Poincar\'e series to evaluate.

\paragraph{Kronecker form.}
When considering the Schottky–Kronecker forms of \eqref{eqn:schottky_kronecker_gen}, one may notice that if we strip the $W(\gamma)$ term, the resulting series is not a Poincar\'e series: the meromorphic function in the summand contains a pole in the singular set of ${\SGroup}$, over which the summation is performed. This makes the stripped series diverge. 

Convergence can be restored in the genus-one case by an appropriate summation prescription and choice of the variable $w$.  
However, in the higher genera cases we replace the single complex variable $w$ by formal non-commuting variables $w_i$, and we treat the series \eqref{eqn:schottky_kronecker_gen} as a formal series, for which the problem of convergence is not well-posed.

\paragraph{Hyperelliptic polylogarithms.}
Although the Schottky–Kronecker forms can only be treated as a formal series, it makes sense to ask whether the series defining the higher-genus kernels \eqref{eqn:schottky_kernels} of the hyperelliptic polylogarithms, which are produced by formal manipulations on the Schottky–Kronecker forms, do converge. 
Formally, the resulting Schottky averaging sums for the kernels do not form Poincar\'e series, which does not allow us to draw a conclusion about the convergence right away. We argue that the convergence in the real hyperelliptic case still holds as follows. 

In \eqref{eqn:schottky_kernels} the summation over $k$ occurs over a finite domain, so this summation does not affect the convergence. The remaining series over the coset ${\SGroup}/{\SGroup}_j$ is similar to a general $(-2)$-dimensional Poincar\'e series \eqref{eqn:poincare-series}, for which convergence holds for any circle-decomposable Schottky group, and the one-form $\skern{n}(z, x | {\SGroup}_j)$ has only a single pole for $n = 1$ at $z = x$, which does not belong to the limiting set of the coset. However, there are two features that make the series non-Poincar\'e: due to the combinatorial factor \eqref{eqn:higher-genus-kerels-preexp} some of the terms in the series vanish and there is an additional factor in the sum that grows polynomially with the length of the Schottky words.  We argue that neither of the features spoils the convergence as the term $\gamma'(z)$ within the differentials $\skern{n}(z, x | {\SGroup}_j)$ gets smaller exponentially with the length of the Schottky words, which makes the series absolutely convergent.

Now that the convergence of the series defining the higher-genus kernels is established, the resulting hyperelliptic polylogarithms are well-defined, and can be computed by the series over the terms \eqref{eqn:iterated_integral_higher_genus_std}, which is also convergent. While we refrain from writing explicit expressions as sums of the genus-one polylogarithms due to complicated combinatorics, a numerical test of the convergence is given in \figref{fig:conv_genus_2} (right) for the particular polylogarithm $ \Gtargzg{(2\ 1)}{x}$ on the curve \eqref{eq:example-curve}. The summand indeed exhibits an expected exponential decay and the missing parts of the plot represent the exact 0 terms due to combinatorial factors.

\subsection{An example}

\begin{figure}
	\centering
	\includegraphics[width=0.45\textwidth]{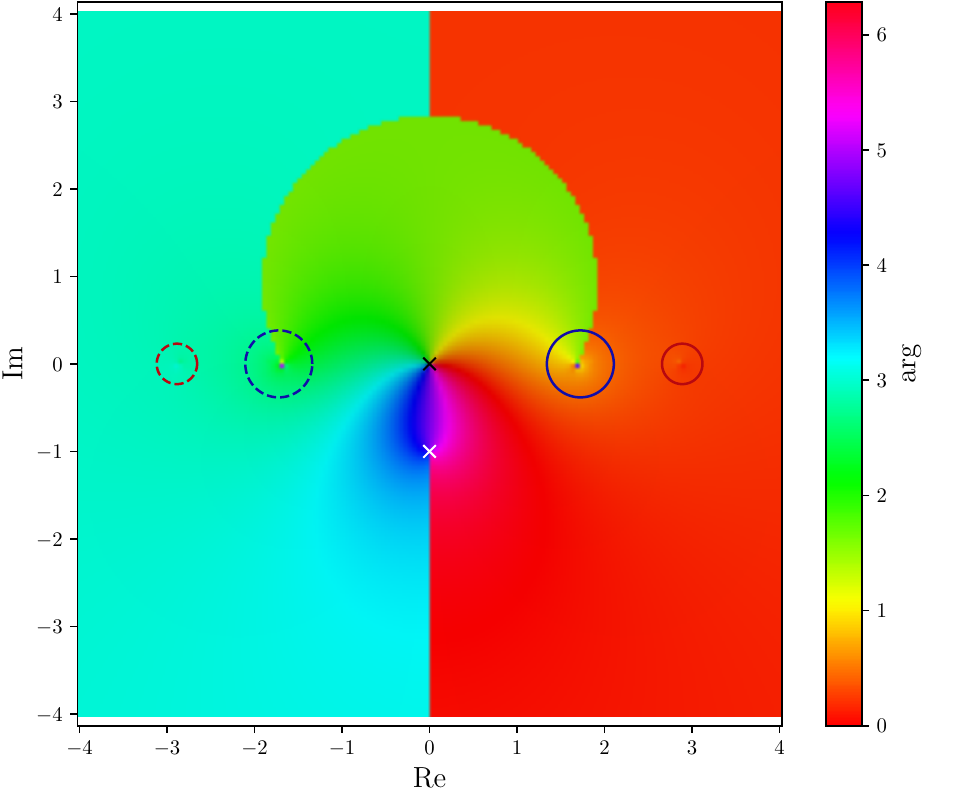}
	\includegraphics[width=0.45\textwidth]{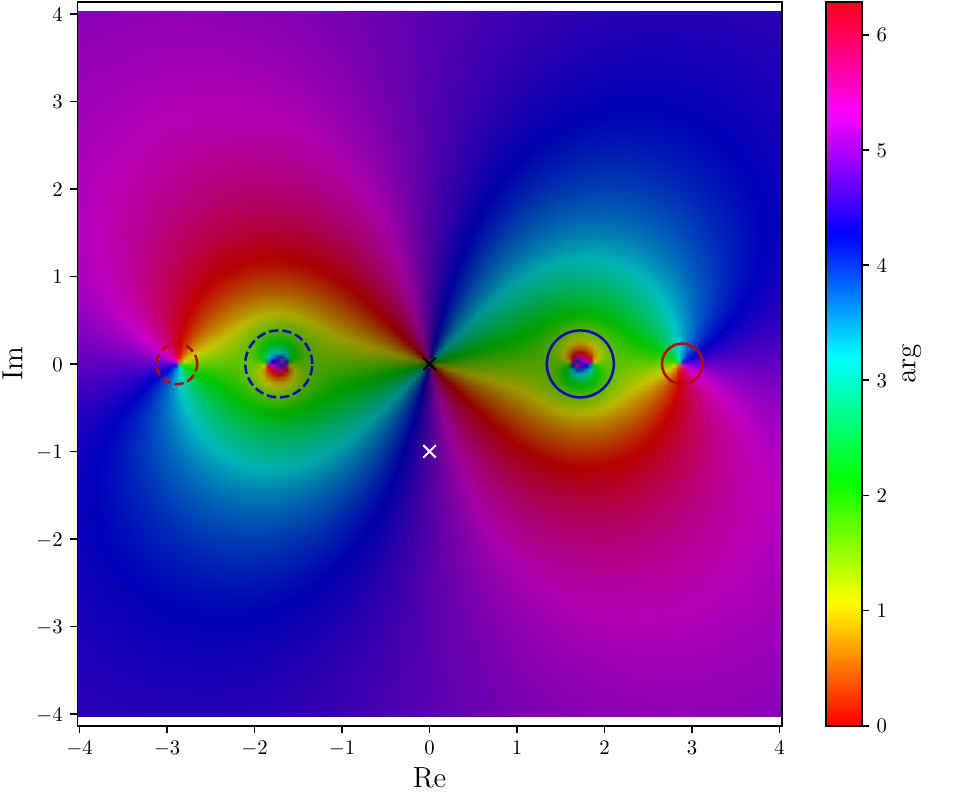}
	\caption{Plots of the polylogarithms $\Gtargzg{(1\ 1)}{x}$ (left) and $\Gtargzg{(2\ 1)}{x}$ (right) with the pole $x = - \iunit$ and the basepoint $z_0 = 0$. The color and brightness represent the argument and absolute value of the polylogarithms evaluated at a given point. The circles of the same color and different line style represent a pair of the isometric Schottky circles, the black point is located at the base point $z_0$ and the white point denotes the point $x$.}
	\label{fig:polylog_plot}
\end{figure}

\begin{table}
	\centering
	\begin{tabular}{|c|c|c|c|c|}
		\hline
		$C_i$ & $c_i$ & $r_i$ & $\gamma_i$ & $P_i$ \\
		\hline
		$C_1$ & 1.723 & 0.383 & $\begin{pmatrix}
			\phantom{-}4.503 & -7.377 \\
			-2.613 & \phantom{-}4.503
		\end{pmatrix}$
		& 1.680 \\
		\hline
		$C_2$ & 2.887 & 0.231 & $\begin{pmatrix}
			\phantom{-}12.505 & -35.876 \\
			-4.331 & \phantom{-}12.505
		\end{pmatrix}$
		& 2.878 \\
		\hline
	\end{tabular}
	\caption{Numerical approximation for Schottky group generators and isometric circle parameters for the curve \protect\eqref{eq:example-curve}.}
	\label{tab:schottly-group-numerically}
\end{table}

As a proof of concept, we depict two depth-one polylogarithms $\Gtargzg{(1\ 1)}{x}$ and $\Gtargzg{(2\ 1)}{x}$ defined on the curve \eqref{eq:example-curve} with the pole $x = - \iunit$ and the basepoint $z_0 = 0$ in \figref{fig:polylog_plot} left and right respectively using our numerical implementation. The Schottky uniformization of the curve is given by the algorithm from \appref{sec:curve_to_schottky} that yields two circles $C_1$ and $C_2$ (blue and red solid lines respectively) and their images under the Schottky generators $\gamma_1$ and $\gamma_2$ given by the dashed lines of the corresponding color. The centers $c_i$ and radii $r_i$ of the circles $C_i$ as well as the Schottky generators $\gamma_i$ in the fundamental representation and the corresponding fixed points $P_i$ are given in \tabref{tab:schottly-group-numerically}. 
 
As it was already mentioned in \secref{sec:expansionschottky}, only the kernels $\omega_{jj}(z,x|{\SGroup})$ have a pole in the fundamental domain at $z=x$. This can be observed in \figref{fig:polylog_plot}: in the left plot the polylogarithm $\Gtargzg{(1\ 1)}{x}$ demonstrates a branch cut stretching from the point $x = -\iunit$, the pole of the underlying kernel, since it involves the genus one kernel $\skern{1}$ in the sum \eqref{eqn:schottky_kernels} that contains the pole. Due to the quasi-periodicity, the same singularity is also observed inside the Schottky circles, which are images of the branch cut under the translations along \Btxt-cycles. On top of that one observes a second branch cut going between two fixed points, which represents a non-trivial monodromy of the polylogarithm along the \Atxt-cycles. On the contrary, in the right figure the plot for $\Gtargzg{(2\ 1)}{x}$ is given, the kernel for which has the last weight $j=1$ repeated only once. Therefore the corresponding kernel essentially consists of only $\skern{0}$ genus-one kernels, which are proportional to the holomorphic forms, thus, there are no branch cuts stretching from the point $x$ as it is a regular point. Moreover, the branch cut between the fixed points is also absent rendering the polylogarithm single-valued on the Schottky cover. 

For the numerical evaluation we truncated the series after all the Schottky group elements of length $l$ if the norm of each of the element was smaller than $10^{-7}$, which gives a rough estimation of the precision. For the given curve this precision requires an order of $10^3$ Schottky group elements. The evaluation time of a single point on the plot is of order of $10^{-2}$ seconds on a single thread on a modern CPU. Our implementation serves the purpose of a prototype of a software to numerically evaluate the hyperelliptic polylogarithms, which, to the best of our knowledge, has not been done before in the literature. Being the prototype, we do not use the most efficient methods to evaluate the genus-one polylogarithms, the Poincar\'e series, neither do we make any exact statements about error estimates. Therefore, further work towards implementing more efficient algorithms \cite{Bobenko:2011,bogatyrev2012extremal,lyamaev2022summation} and taking the precision under control is needed for practical use of the numerical implementation and in order to reduce the evaluation time.

Being under heavy development, the source code is not planned to be published in the near future. Interested readers are hereby encouraged to contact the authors for code or more information on the implementation.


\section{Conclusions}\label{sec:summary}
\subsection{Summary}
In this article we explore a generalization of the genus-one Kronecker function \eqref{eqn:Kroneckergenusone} to a genus-\genus{} Riemann surface using the language of Schottky uniformization. We introduce a set of~\genus{} Schottky–Kronecker forms, and for real hyperelliptic curves we expand them into integration kernels and relate them to the connection of Enriquez. In addition we provide numerical access to genus-\genus{} hyperelliptic polylogarithms. 

In particular:  
\begin{itemize}
	\item Inspired by Bernard's connection form \eqref{eqn:Bernard_connection}, we define higher-genus Schottky–Kronecker form at higher genera in \eqn{eqn:schottky_kronecker_gen} via Poincaré series, which (formally) satisfy the analytical constraints \eqns{eqn:KroneckerResidue}{eqn:KroneckerQuasiper} that uniquely determine Enriquez' connection form. 

		The Schottky formulation permits easy access to finding relationships that the Schottky–Kronecker forms have to other genera, discussed among other properties in \secref{sec:properties_schottky_kronecker}. In \eqn{eqn:schottky-genus1-average}, we suggest through a formal manipulation that the Schottky–Kronecker form at higher genera may be identified as an average of genus-one Kronecker forms. In \eqn{eqn:degeneration-unmatched} and \eqn{eqn:degeneration-matched}, we sketch how the Schottky–Kronecker form degenerates when an \Atxt-cycle of the underlying surface is shrunk down to a point.

\item The higher-genus Schottky–Kronecker forms can be expanded into formal integration kernels given by Poincar\'e series, using an alphabet of non-commuting letters. Under technical assumptions which ensure the convergent of the series, which include the case of real hyperelliptic curves, we use our kernels to construct higher-genus polylogarithms in \secref{sec:hepolyschottky}. 
The sum of the genus-\genus{} Schottky–Kronecker forms (multiplied with appropriate letters) was linked to the connection from Enriquez \cite{EnriquezHigher} in \secref{sec:KZB}.

\item Combining with the averaging result from \eqn{eqn:schottky-genus1-average}, the kernels can  be expressed recursively in terms of known genus-one kernels as shown in \eqn{eqn:schottky_kernels}. Based on this recursive formulation, we can numerically evaluate hyperelliptic polylogarithms in \secref{sec:practicabilities}. 
\end{itemize}

\subsection{Outlook}
While our approach gives an explicit construction of hyperelliptic polylogarithms allowing for numerical evaluations of these functions, there are several questions for future research: 
\begin{itemize}
\item \textbf{Beyond real hyperelliptic curves.} The construction of the higher-genus Kronecker forms and hence the higher-genus polylogarithms in this paper is constrained to the case of Riemann surfaces whose Schottky cover is ``circle-decomposable'', which in particular includes all real hyperelliptic curves: for this subset of Riemann surfaces, the convergence of the Poincar\'e series which define the integration kernels can be guaranteed. It is important to explore to what extent this condition can be relaxed, allowing for the construction via rapidly converging Poincaré series of higher-genus polylogarithms for a wider class of the Riemann surfaces.
	\item \textbf{Relationship to theta ratio.} 
		A higher-genus generalization of the Kronecker form could have started from the genus-one representation as a ratio of $\theta$ functions in \eqn{eqn:Kroneckergenusone} as well. Taking the naive/straightforward approach results in the formula 
		\begin{equation}\label{eqn:theta-kronecker-defn}
		\Fth(z,x,-\vec b / 2 \pi \iunit|\tau) = \frac{\theta\big(\abel(z,p_0|\tau) - \abel(x,p_0|\tau) - \vb / 2\pi \iunit\big|\tau\big)}{\theta\big(\abel(z,p_0|\tau) - \abel(x,p_0|\tau)\big|\tau\big) \theta\big(-\vb / 2\pi \iunit\big|\tau\big)} \sum_{j=1}^\genus \partial_j \theta(\vec 0|\tau) \omega_j(z),
		\end{equation}
		where $\abel(z,p_0|\tau)$ has been defined in \eqn{eqn:abel-universal} and $\theta$ is a higher-genus theta-function of odd charateristics, $\abel$ denotes Abel's map and $\omega_j(z)$ are the holomorphic differentials on the genus-\genus{} surface. \\
		As a consequence of the Fay trisecant equation \cite{mumford1984tata}, this ratio satisfies a Fay identity \cite{Lisitsyn:2024}
		\begin{equation} \label{eq:FayHigher}
		\Fth(z,x,\val|\tau)\Fth(y,x,\vbe|\tau) = \Fth(z,y,\val|\tau)\Fth(y,x,\val + \vbe|\tau) + \Fth(z,x,\val + \vbe|\tau)\Fth(y,z,\vbe|\tau)
		\end{equation}
		for $\vec \alpha = \mathbb C^\genus$ and $\vec \beta = \abel(p) - \abel(y)$ for some point $p \in \RSurf$.

		Expanding this ratio into kernels, however, lacks the required non-commuting structure that we saw for the expansion letters $b_i$ in \secref{sec:schottky_kronecker} as well as in Enriquez construction \cite{EnriquezHigher}. It remains to be settled, whether a different prescription for the expansion will allow to remedy this drawback. Once a valid expansion is identified, one could hopefully translate the ratio's Fay identity at each genus into what is known as Fay-like identities in \cite{DHoker:2023vax}. The precise relation between the $\theta$-function approach and the Schottky kernels formulated in this paper is under investigation.

Let us mention in addition that the ``twisted Green function'' constructed in \cite{enriquez2000solutions} in the context of the KZB equations precisely matches our proposal. Furthermore, \eqn{eqn:theta-kronecker-defn} is also remarkably similar to a function considered by Tsuchiya in \cite[Eq.~2.13]{Tsuchiya:2022lqv}; in particular, the cyclic products used by Tsuchiya are equivalent to cyclic products of the ratio in \eqn{eqn:theta-kronecker-defn}.
However, understanding of the precise relations and possible implications needs further examination.

\item \textbf{Fay identities.} So far we have not been investigating the Fay-like identities for the Schottky kernels analytically.  However, numerical experiments lead to a series of candidate relations for higher-genus Enriquez kernels. In what way they are related to the Fay trisecant equation, like \eqref{eq:FayHigher}, remains unsettled. 
\item \textbf{Hyperelliptic MZVs.} Special values of (elliptic) polylogarithms are of particular interest when computing physical quantities such as scattering amplitudes in field or string theory as reviewed in \secref{sec:polylogreview}. The natural question then is to study special values of hyperelliptic polylogarithms, i.e.~hyperelliptic multiple zeta values. For instance, relations among them could be investigated by exploiting the numerical approximation methods proposed in this article. Moreover, it is known that elliptic MZVs can be written as iterated integrals of Eisenstein series, and have therefore an interesting modular behaviour. It would be interesting to investigate the dependence of hyperelliptic MZVs on the moduli of the surface via numerical approximation.
In fact, numerical approximation itself would be boosted by having control on the modular properties of hyperelliptic MZVs. Similarly to the genus-one case, where the convergence of the $q$-expansions of various objects can be improved by choosing a modular parameter $\tau$ with larger imaginary part, for the higher genera the sums over Schottky group exhibit better convergence if the circles are small and far from each other. Just as the $\grp{SL}(2, \Integers) = \grp{Sp}(2, \Integers)$ transformations relate different values of $\tau$, the higher modular group $\grp{Sp}(2\genus, \Integers)$ should relate different Schottky groups, allowing us to choose those that have better convergence.
\item \textbf{Application to Feynman calculations.} For Feynman calculations it is convenient to formulate the integration kernels as differential forms on an algebraic curve, as already remarked in the case of Feynman integrals involving genus-one polylogarithms \cite{Broedel:2017kkb}. Recursive formulas for Enriquez' kernels in terms of algebraically defined objects where given in \cite{EZ1}. It would be interesting to apply our numerical setup to evaluating a higher-genus Feynman integral like the one considered in \rcite{Marzucca:2023gto}.

\end{itemize}

\subsection*{Acknowledgments}
We are grateful to Yannis Moeckli and Zhi Cong Chan for various discussions.  The work of K.B., J.B.~and E.I.~is partially supported by the Swiss National Science Foundation through the NCCR SwissMAP. The research of F.Z. has been partly funded by the European Union’s Horizon 2020 research and innovation programme under the Marie Sklodowska-Curie grant agreement No 843960 for the project “HIPSAM”.


\vspace*{2cm}
\noindent{\LARGE\textbf{Appendix}}
\addcontentsline{toc}{section}{Appendix}
\appendix


\section{From the algebraic curve to the Schottky cover} \label{sec:curve_to_schottky}

In this appendix we review a numerical algorithm to obtain a Schottky group and corresponding isometric circles for a given real hyperelliptic curve. The methods described in the following are a collection from the references \cite{Bobenko:2011, BBE+94, OSBORNE2018112}, where the latter reference explains more technical details about numerical calculations when going from an algebraic curve to a Schottky cover.

We focus on hyperelliptic curves with real roots, given -- for genus \genus{} -- by a polynomial equation of degree $2\genus+1$: 
\begin{equation}
	\label{eqn:polycurve}
	y^2=(x-a_1)(x-a_2)\cdots(x-a_{2\genus+1})
\end{equation}
where the $a_i\in\mathbb{R}$ are the real roots of the polynomial and we assume them to be ordered, i.e.~$a_1<a_2<\ldots<a_{2\genus+1}$. Technical aspects become more cumbersome once one has a hyperelliptic curve with non-real roots. 

\begin{figure}[t]
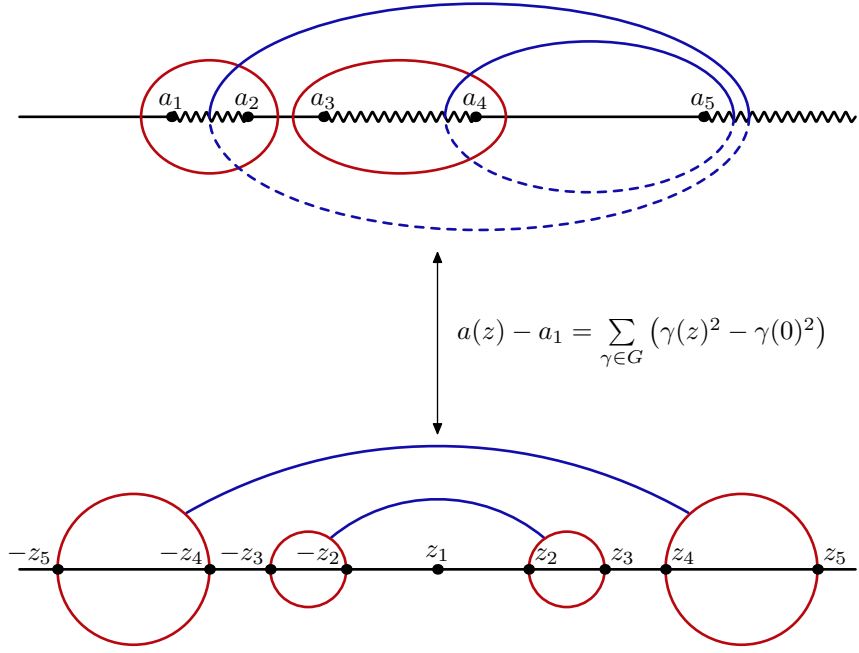

	\centering
	\mpostuse{algcurve-schottky}
	\caption{Schematics of going from an algebraic curve (upper part) to a Schottky uniformization (lower part) for the case of a genus two surface. \Atxt-cycles are drawn in red and \Btxt-cycles in blue. Note that the dashed line in the upper figure represents the change of sheet when passing through the branch cuts of the algebraic curve (the branch cuts are indicated by zig-zag lines in the upper part of the figure). Figure similar as in \cite{OSBORNE2018112}.}
	\label{fig:curveschottky}
\end{figure}

A visualization of the relation between an algebraic curve with real branch points and the corresponding Schottky cover is shown in \figref{fig:curveschottky}, which is useful to refer to when reading the following explanations.

The starting point is an algebraic curve with real roots, as given in \eqn{eqn:polycurve} and the goal is to have a procedure yielding the parameters of the Schottky uniformization corresponding to this algebraic curve. At the end of the process, we will have a Schottky uniformization which is symmetric w.r.t.~both the real and the imaginary axis of the complex plane, i.e.~the circles of the uniformization are cut exactly in half by the real axis and the knowledge of the intersection points $z_2,\ldots,z_{2\genus+1}$ (with $z_1=0$) yields all the information about the Schottky uniformization.

Let $\SGroup$ be the Schottky group corresponding to the desired Schottky uniformization, and let~$z$ be a point on the Schottky cover with its counterpart $a(z)$ on the algebraic curve. Then the relation between the two formulations of the Riemann surface is
\begin{equation}
	\label{eqn:curveschottky}
	a(z)-a_1=\sum_{\gamma\in\SGroup}\left(\gamma(z)^2-\gamma(0)^2\right)\,.
\end{equation}
When coming from the algebraic curve, one does not know the Schottky group $\SGroup$ yet, so one starts off with the first approximation of \eqn{eqn:curveschottky} (only using the term of the sum with $\gamma=\mathrm{id}$), i.e.
\begin{equation}
	a(z)-a_1\approx z^2 \iff z\approx \pm \sqrt{a(z)-a_1}\,. 
\end{equation}

From this first approximation, the corresponding Schottky group elements can be found (see below). The approximation and the Schottky group can then be used to solve \eqn{eqn:curveschottky} iteratively to the desired precision. \Eqn{eqn:curveschottky} should be solved for the mapping
\begin{equation}
	a_n=a(z_n),
\end{equation}
so that each branch point of the algebraic curve has a counterpart on the Schottky cover corresponding to an intersection point of a circle with the real axis (from this it follows immediately that $z_1=0$).

The following shows the relations of the intersection points $\pm z_i$ with the other parameters and quantities of the Schottky uniformization of genus \genus{}:
\begin{itemize}
	\item The radii of the $2\genus$ circles (where each circle has exactly one other circle it is identified with, leaving \genus{} different circles after the identification) are (for $i=1,\ldots,\genus$) 
	\begin{equation}
		r_i=\frac12(z_{2i+1}-z_{2i})\,,
	\end{equation}
	and their centers are at
	\begin{equation}
		\pm c_i=\frac12(z_{2i}+z_{2i+1})\,.
	\end{equation}
	\item The expressions for the parameters $P_i$ (i.e.~the fixed points of the Schottky generators $\gamma_i$) and $\lambda_i$ of the Schottky cover are obtained from the intersection points $z_k$ as follows (for $i=1,\ldots,\genus$) 
	\begin{align}
		P_i&=\sqrt{z_{2i}z_{2i+1}}\,,\\
		\lambda_i&=\left(\frac{\sqrt{z_{2i}}-\sqrt{z_{2i+1}}}{\sqrt{z_{2i}}+\sqrt{z_{2i+1}}}\right)^2.
	\end{align}
	\item Using these parameters, the generators $\gamma_i$ of the Schottky group $\SGroup$ are (for $i=1,\ldots,\genus$) 
	\begin{equation}
		\gamma=\frac{1}{2\sqrt{\lambda_i}}
		\left(\!\!
		\begin{array}{cc} 
			1+\lambda_i & -P_i(1-\lambda_i) \\ 
			-\frac1{P_i}(1-\lambda_i) & 1+\lambda_i 
		\end{array}\!\!\right),
	\end{equation}
	with their corresponding inverses
	\begin{equation}
		\gamma^{-1}=\frac{1}{2\sqrt{\lambda_i}}\left(\!\!\begin{array}{cc} 1+\lambda_i & P_i(1-\lambda_i) \\ \frac1{P_i}(1-\lambda_i) & 1+\lambda_i \end{array}\!\!\right).
	\end{equation}
\end{itemize}


\section{Extra details for Schottky–Kronecker form}

\subsection{Quasiperiodicity in auxiliary variable}\label{app:auxiliary-quasiperiodicity}

We start by expressing the terms in $F_j$ as differences of fractions, as in \eqn{eqn:kronfun_expansion_genus_0},
\begin{equation}
    F_j(z,x) = \sum_{\gamma \in \SGroup} \dd z \left(\frac{1}{z-\gamma x} - \frac{1}{z - \gamma P_j}\right)W(\gamma)\,.
\end{equation}
As we plug in $x \mapsto \gamma_k x$, one may naively assume that one may be able to relabel the sum on only the first term to absorb the new generator. However, each fraction independently is not M\oe{}bius invariant, and their partial sums are not absolutely convergent, which would make such a relabeling change the value of the form. Instead, we must insert additional terms to recover M\oe{}bius invariant cross-ratios,
\begin{align}
    F_j(z,\gamma_k x) & = \sum_{\gamma \in \SGroup} \dd z \left( \frac{1}{z-\gamma \gamma_k x} - \frac{1}{z - \gamma P_j} \right) W(\gamma) \nonumber\\
    &  = \sum_{\gamma \in \SGroup} \dd z \left( \frac{1}{z - \gamma \gamma_k x} - \frac{1}{z-\gamma x} + \frac{1}{z - \gamma x} - \frac{1}{z - \gamma P_j}\right)W(\gamma) \nonumber\\
    &  = \sum_{\gamma \in \SGroup} \dd z \left(\frac{1}{z - \gamma \gamma_k x} - \frac{1}{z-\gamma x}\right) W(\gamma) + F_j(z,x)\,.
\end{align}
Now, we can use a similar strategy to identify the first term on the right hand side,
\begin{align}
	\sum_{\gamma \in \SGroup} \dd z \bigg(\frac{1}{z - \gamma \gamma_k x} &- \frac{1}{z-\gamma x}\bigg) W(\gamma) \nonumber \\
    & = \sum_{\gamma \in \SGroup} \dd z \left(\frac{1}{z - \gamma \gamma_k x} - \frac{1}{z - \gamma \gamma_k P_k}- \frac{1}{z-\gamma x} + \frac{1}{z - \gamma P_k}\right) W(\gamma) \nonumber\\
    & = \sum_{\gamma \in \SGroup} \dd z  \left(\frac{1}{z - \gamma \gamma_k x} + \frac{1}{z - \gamma \gamma_k P_k}\right) W(\gamma) - F_k(z,x) \nonumber\\ 
    & = F_k(z,x) (w_k^{-1} - 1)\,.
\end{align}
Putting the results together, we find
\begin{equation}
    F_j(z,\gamma_k x) = F_k(z,x) (w_k^{-1} - 1) + F_j(z,x)\,,
\end{equation}
as in \eqn{eqn:auxiliary-monodromy}.


\subsection{Derivation for the coefficients of the kernels} \label{app:coef-ker-derivation}
In this appendix we provide details on the derivation of formula \eqref{eqn:higher-genus-kerels-preexp} for the coefficients $C(b_{i_1}^{n_1} \cdots b_{i_s}^{n_s} , \gamma_{j_1}^{m_1} \cdots \gamma_{j_l}^{m_l}) $ of the higher-genus kernels. 
The element $\gamma_{j_1}^{m_1} \cdots \gamma_{j_l}^{m_l}$ corresponds to the expression $e^{m_1 b_{j_1}} \cdots e^{m_l b_{j_l}}$ under the action of the map $W : \gamma_{i_1}^{n_1} \cdots \gamma_{i_l}^{n_l} \mapsto w_{i_1}^{n_1} \cdots w_{i_l}^{n_l}$.
Consider the procedure of extracting the coefficient of $b_{i_1}^{n_1} \cdots b_{i_s}^{n_s}$:
\begin{equation}
C\left(b_{i_1}^{n_1} \cdots b_{i_s}^{n_s}, \gamma_{j_1}^{m_1} \cdots \gamma_{j_l}^{m_l}\right)=\left[b_{i_1}^{n_1} \cdots b_{i_s}^{n_s}\right] e^{m_1 b_{j_1}} \cdots e^{m_l b_{j_l}}.
\end{equation}
\begin{itemize}
	\item If $s=0$, then
	\begin{equation}
	C\left(1, \gamma_{j_1}^{m_1} \cdots \gamma_{j_l}^{m_l}\right)=[1] e^{m_1 b_{j_1}} \cdots e^{m_l b_{j_l}}=1.
	\end{equation}
	\item If $l=0$ and $s \neq 0$, then
	\begin{equation}
	C\left(b_{i_1}^{n_1} \cdots b_{i_s}^{n_s}, \id \right)=\left[b_{i_1}^{n_1} \cdots b_{i_s}^{n_s}\right] 1=0.
	\end{equation}
\end{itemize}
This covers the base cases of the formula.
Then, we expand the first exponential:
\begin{equation}
C\left(b_{i_1}^{n_1} \cdots b_{i_s}^{n_s}, \gamma_{j_1}^{m_1} \cdots \gamma_{j_l}^{m_l}\right)=\left[b_{i_1}^{n_1} \cdots b_{i_s}^{n_s}\right] \sum_{k=0}^{\infty} \frac{\left(m_1 b_{j_1}\right)^k}{k!} e^{m_2 b_{j_2}} \cdots e^{m_l b_{j_l}}.
\end{equation}
\begin{itemize}
	\item If $j_1 \neq i_1$, then the only contributing term is when $k=0$, so
	\begin{equation}
	C\left(b_{i_1}^{n_1} \cdots b_{i_s}^{n_s}, \gamma_{j_1}^{m_1} \cdots \gamma_{j_l}^{m_l}\right)=\left[b_{i_1}^{n_1} \cdots b_{i_s}^{n_s}\right] e^{m_2 b_{j_2}} \cdots e^{m_l b_{j_l}}=C\left(b_{i_1}^{n_1} \cdots b_{i_s}^{n_s}, \gamma_{j_2}^{m_2} \cdots \gamma_{j_l}^{m_l}\right).
	\end{equation}
	\item If $j_1=i_1$, then we must include the terms where $0 \leq k \leq n_1$, so
	\begin{align}
	C\left(b_{i_1}^{n_1} \cdots b_{i_s}^{n_s}, \gamma_{j_1}^{m_1} \cdots \gamma_{j_l}^{m_l}\right)
	&=
	\sum_{k=0}^{n_1} \frac{(m_1)^k}{k!}\left[b_{i_1}^{n_1-k} \cdots b_{i_s}^{n_s}\right] e^{m_2 b_{j_2}} \cdots e^{m_l b_{j_l}}
	\notag \\
	&=
	\sum_{k=0}^{n_1} \frac{(m_1)^k}{k!} C\left(b_{i_1}^{n_1-k} \cdots b_{i_s}^{n_s}, \gamma_{j_2}^{m_2} \cdots \gamma_{j_l}^{m_l}\right). 
	\end{align}
\end{itemize}
This covers the recursive cases where in each case the length of the Schottky element gets shorter, so the computation terminates.


\bibliographystyle{HigherGenusKronecker}
\bibliography{HigherGenusKronecker}

\end{document}